\documentclass[12pt, a4paper]{article}
\pdfoutput=1

\usepackage[height=25cm,width=16cm]{geometry}
\usepackage{amsmath}
\usepackage{dcolumn}
\usepackage{bm,here}
\usepackage{subfig}
\usepackage{comment}
\usepackage{ifpdf}
\usepackage{slashed}
\usepackage{colortbl}
\usepackage{color}
\usepackage[mathscr]{eucal}
\usepackage[sort&compress, numbers, merge]{natbib}

\usepackage{cancel}

\ifpdf
  \usepackage{graphicx}     
  \usepackage[bookmarksopen,colorlinks=true,linkcolor=bblue,citecolor=ppink,urlcolor=ppink]{hyperref}
\else     
  \usepackage[dvipdfmx]{graphicx}     
  \usepackage[dvipdfmx,bookmarksopen,colorlinks=true,linkcolor=bblue,citecolor=ppink,urlcolor=ppink]{hyperref}
\fi

\usepackage{multicol}
\definecolor{red}{rgb}{1,0,0}
\definecolor{ppink}{rgb}{0.921545,0.440586,0.687243}
\definecolor{bblue}{rgb}{0.400000,0.400000,1.000000}
\usepackage[charter]{mathdesign}

\newcommand{\gev}{\ensuremath{\,\mathrm{GeV}}}
\newcommand{\tev}{\ensuremath{\,\mathrm{TeV}}}

\begin{document}

\begin{titlepage}

\begin{flushright}
\hfill IPMU18-0180 \\
\end{flushright}

\begin{center}

\vskip 2.5cm
{\Large \bf Light Fermionic WIMP Dark Matter \\[.3em] with Light Scalar Mediator}

\vskip 1.5cm
{\large
Shigeki Matsumoto$^{(a)}$,
Yue-Lin Sming Tsai$^{(b,c)}$
and
Po-Yan Tseng$^{(a)}$
}

\vskip 1.5cm
$^{(a)}$ {\sl Kavli IPMU (WPI), UTIAS, University of Tokyo, Kashiwa, 277-8583, Japan}\\[.3em]
$^{(b)}$ {\sl Institute of Physics, Academia Sinica, Nangang, Taipei 11529, Taiwan}\\[.1em]
$^{(c)}$ {\sl Key Laboratory of Dark Matter and Space Astronomy,  
Purple Mountain Observatory, Chinese Academy of Sciences, Nanjing 210008, China}\\[.1em]

\vskip 3.5cm
\begin{abstract}
\noindent
A light fermionic weakly interacting massive particle (WIMP) dark matter is investigated by studying its minimal renormalizable model, where it requires a scalar mediator to have an interaction between the WIMP and standard model particles. We perform a comprehensive likelihood analysis of the model involving the latest but robust constraints and those will be obtained in the near future.
In addition, we pay particular attention to properly take the kinematically equilibrium condition into account. 
It is shown that near-future experiments and observations such as low-mass direct dark matter detections, 
flavor experiments and CMB observations play important roles to test the model. 
Still, a wide parameter region will remain even if no WIMP and mediator signals are detected there. 
We also show that precise Higgs boson measurements at future lepton colliders will significantly 
test this remaining region.

\end{abstract}

\end{center}

\end{titlepage}

\tableofcontents
\newpage
\setcounter{page}{1}

\section{Introduction}
\label{sec: intro}

In past decades, people tried to develop particle physics based on the electroweak naturalness\,\cite{Ellis:1986yg, Barbieri:1987fn}, namely how the electroweak scale should be naturally explained. Many new physics scenarios such as supersymmetry, extra-dimension and composite Higgs have been proposed in this context, and those have been and still are being tested by various experiments including Large Hadron Collider (LHC). However, people recently start doubting this guiding principle, for new physics signals predicted by the scenarios have not been detected at all. 
In other words, 
though the electroweak scale should be naturally explained, it may be achieved by some other mechanisms (or ideas) which are totally different from what we have thought about so far. 
Particle physicists are seeking new mechanisms based on this consideration, but 
none of candidate models can successfully explain electroweak naturalness yet.
Under this circumstance, one starts 
taking another strategy: developing particle physics by solving the dark matter problem. Once the nature of dark matter is clarified, it 
may launch out into the exploration of new physics beyond the standard model (SM).

The thermal dark matter, 
often called the weakly interacting massive particle (WIMP), is known to be one of influential dark matter candidates among others, for the dark matter abundance observed today is naturally explained by the so-called freeze-out mechanism~\cite{Bernstein:1985th, Srednicki:1988ce}, 
which can also successfully explain the history of 
the Big Band Nucleosynthesis (BBN) and recombination in the early universe. Though the WIMP is in general predicted to be in the mass range between ${\cal O}(1)$\,MeV\,\cite{Boehm:2002yz, Boehm:2003bt} and ${\cal O}(100)$\,TeV\,\cite{Murayama:2009nj, Hambye:2009fg, Hamaguchi:2009db, Antipin:2014qva, Antipin:2015xia, Huo:2015nwa}, those with the mass around the electroweak scale have been intensively studied because of a possible connection to new physics models for the electroweak naturalness. However, present negative experimental results, not only from the LHC experiment but also from direct dark matter detection experiments, start eroding the parameter space of the WIMP with the electroweak mass.
Thus, it motivates us to consider other WIMPs with a lighter ($\lesssim {\cal O}(10)\gev$) or heavier ($\gtrsim {\cal O}(1)\tev$) mass. 
We focus on the former case in this paper.

Light WIMP must be singlet under the SM gauge group, otherwise it 
would be discovered already.
Concerning the spin of the WIMP, we take one-half, namely a light fermionic WIMP in this paper.\footnote{Light scalar WIMP in its minimal model (Higgs-portal dark matter) is already excluded by the constraint from the invisible Higgs decay\,\cite{Kanemura:2010sh, LopezHonorez:2012kv, Djouadi:2012zc, Cheung:2012xb, Falkowski:2015iwa}. Light scalar WIMP still survives in next-to-minimal models\,\cite{Barger:2010yn, Gonderinger:2012rd}.} In the minimal (renormalizable) model to describe such a light fermionic WIMP, a new additional particle called the mediator must be introduced to have an interaction between the WIMP and SM particles.
In addition, such a mediator is required to be as light as the WIMP to explain the 
dark matter abundance observed today, to be singlet under the SM gauge group to avoid constraints from 
the current performed collider experiments, 
to be bosonic being consistent with the Lorentz symmetry, and to be even under the $Z_2$ symmetry 
in order to make the WIMP stable. Such a light fermionic dark matter with a light bosonic mediator recently 
receives many attentions, as it has a potential to have a large and velocity-dependent scattering cross section between WIMPs and solve the so-called small scale crisis of the universe\,\cite{Vogelsberger:2012ku, Rocha:2012jg, Peter:2012jh}. Among two possibilities of the bosonic mediator, 
either a scalar or a vector~\cite{Pospelov:2007mp,Pospelov:2008jd,Bertolami:2007wb,Matsumoto:2010bh,Kanemura:2011nm}, 
we take the scalar one in this paper.\footnote{A careful model-building is required to have a light fermionic WIMP with a light vector mediator\,\cite{Yanagida new}, because such a WIMP has a s-wave annihilation and tends to be excluded by the cosmic microwave background (CMB) observation. Such a constraint can be avoided for the scalar mediator case\,\cite{Kamada:2018zxi}, as seen in section\,\ref{sec: constraints}.}

To investigate the present status and future prospects of a light fermionic WIMP with a light scalar mediator, we perform a comprehensive analysis of its minimal (renormalizable) model\,\cite{Patt:2006fw, Kim:2006af, Djouadi:2011aa, Pospelov:2011yp, Baek:2011aa, Esch:2013rta, Ghorbani:2014qpa, Freitas:2015hsa, Dutra:2015vca, Ghorbani:2016edw, Beniwal:2018hyi, Kanemura:2010sh, LopezHonorez:2012kv}, where our likelihood involves all robust constraints obtained so far and those will be obtained in the near future (if no WIMP and mediator signals are detected) from particle physics experiments as well as cosmological and astrophysical observations. We carefully involve a kinematical equilibrium condition assuming that the freeze-out (chemical decoupling) of the WIMP occurs when it is in kinematically equilibrium with SM particles. We pay particular attention to a possible case that the light WIMP can be in the kinematical equilibrium via existent mediators at the freeze-out even if the WIMP does not have an interaction to SM particles with enough magnitude\,\cite{Kamada:2016ois, Evans:2017kti}. We find that a very wide parameter region is surviving at present in the certain mass region of the WIMP. We also show quantitatively how near-future experiments and observations such as low-mass direct dark matter detections, flavor experiments and CMB observations play important roles, by comparing the results of analyses for the present status and future prospects of the model. Moreover, we see that a wide parameter region will still 
 remain even if neither WIMP nor mediator signal 
 is detected in the near future, and show that precise Higgs boson measurements at future lepton colliders will play a significant role to test the region.
Such comprehensive analysis by the global scanning is a recent trend in WIMP studies\,\cite{Workgroup:2017lvb, Athron:2017kgt, Athron:2018hpc,Athron:2017qdc,Athron:2017yua,Athron:2018ipf}.

The rest of the paper is organized as follow. In section\,\ref{sec: model}, we show our setup, the minimal renormalizable model to describe a light fermionic WIMP with a scalar mediator. We will give all interactions predicted by the model and discuss physics of the mediator. All constraints that we have involved in our likelihood are discussed in section\,\ref{sec: constraints}.\footnote{For readers who are not very much interested in the detail of the constraints and want to see the results of our analysis quickly, please skip this section (section\,\ref{sec: constraints}) and go to the next section (section\,\ref{sec: results}) directly.} Results of our likelihood analysis are given and discussed in section\,\ref{sec: results}, including several implications of the result to (near) future projects for the WIMP search. Section\,\ref{sec: summary} is devoted to the summary of our discussion. There are several appendices at the end of this paper, where preselection criteria we have involved in our analysis (Appendix\,\ref{app: initial parameter region}), the kinematical equilibrium condition (Appendix\,\ref{app: equilibrium condition}) and results of our analysis (Appendix\,\ref{app: supplemental figures}) are shown in details.

\section{The minimal model}
\label{sec: model}

\subsection{Lagrangian}

Two general properties of the WIMP field are considered before constructing a simplified model: the spin and weak-isospin(s) of the field(s).\footnote{Several dark matter fields with a different weak-charge are introduced for the well-tempered WIMP\,\cite{Banerjee:2016hsk}.} Since we are interested in a light fermionic WIMP with its mass of ${\cal O}(1)$\,GeV or less, we focus on a Majorana WIMP, namely the simplest one among spin half WIMPs. On the other hand, the weak-isospin of the WIMP must be fixed to be zero, because a light WIMP carrying a non-zero weak-charge is excluded by collider experiments (LEP, etc.) performed so far. Such a singlet Majorana WIMP is well-motivated by e.g. a neutralino (Bino, Singlino, etc.) in supersymmetric models.

An additional new mediator has to be introduced, 
otherwise the WIMP cannot have any renormalizable interaction with SM particles due to SM gauge symmetry, Lorentz symmetry and $Z_2$ symmetry making the WIMP stable. The mediator is required to be as light as the WIMP to satisfy the relic abundance condition, 
as seen in following sections. If we consider a $Z_2$-odd mediator, 
it must be charged under the SM gauge interactions due to the SM gauge symmetry and 
such a light charged particle is also excluded by the collider experiments~\cite{Matsumoto:2014rxa, Matsumoto:2016hbs}. 
Hence, the mediator must be even under the $Z_2$ symmetry. 
Moreover, the mediator is either a scalar or a vector boson, 
because it must have a renormalizable interaction with the WIMP (the WIMP-WIMP-mediator interaction). 
We consider the case of the scalar mediator in this paper, and leave the vector mediator case for future work.

We assume that the mediator is described by a real singlet 
from the viewpoint of minimality. 
Then, the Lagrangian involving all possible renormalizable interactions of the singlet Majorana WIMP $\chi$, the mediator $\Phi$ and SM particles is given as follows\,\cite{Krnjaic:2015mbs}:
\begin{equation}
	\mathcal{L} =
	\mathcal{L}_{\rm SM} + \frac{1}{2} \bar{\chi} (i\slashed{\partial} - m_{\chi}) \chi + \frac{1}{2} (\partial \Phi)^2
	- \frac{c_s}{2} \Phi \bar{\chi} \chi - \frac{c_p}{2} \Phi \bar{\chi} i \gamma_5 \chi
	-V(\Phi,H),
	\label{eq: L}
\end{equation}
with $\mathcal{L}_{\rm SM}$ and $H$ being the SM Lagrangian and the SM Higgs doublet, respectively. 
In order to make the WIMP stable as mentioned above, 
a $Z_2$ symmetry is imposed, where $\chi$ is odd under the symmetry 
but other particles are charged even. 
The scalar potential of the model is composed of $V(\Phi, H) \equiv V_\Phi(\Phi) + V_{\Phi H}(\Phi, H)$ and $V_H(H)$, where $V_H(H)$ is the potential of the SM doublet $H$ involved in $\mathcal{L}_{\rm SM}$. Its explicit form is written as follows:
\begin{eqnarray}
	V_H(H) &=& \mu^2_H H^{\dagger} H + \frac{\lambda_H}{2} (H^{\dagger} H)^2,
	\nonumber \\
	V_{\Phi}(\Phi) &=& \mu^3_1 \Phi + \frac{\mu^2_\Phi}{2} \Phi^2 + \frac{\mu_3}{3!} \Phi^3 + \frac{\lambda_\Phi}{4!} \Phi^4,
	\nonumber \\
	V_{\Phi H}(\Phi, H) &=& A_{\Phi H} \Phi H^{\dagger} H + \frac{\lambda_{\Phi H}}{2} \Phi^2 H^{\dagger} H.
	\label{eq: V}
\end{eqnarray}
Here $\lambda_i$s are dimensionless coupling constants of quartic interactions, while others ($\mu_i$s and $A_{\Phi H}$) are mass dimension one coupling constants for cubic and quadratic interactions.

We take $v_H = (-2\mu_H^2/\lambda_H)^{1/2} \simeq 246$\,GeV and $v_\Phi$ as vacuum expectation values of $H$ and $\Phi$. 
Taking the unitary gauge, the fields are expressed as $H = [0, (v_H + h')/\sqrt{2}]^T$ and $\Phi = v_\Phi + \phi'$, where $v_\Phi$ can be fixed to be zero without a loss of generality. Mass eigenstates of the scalars are then obtained by diagonalizing the quadratic terms of the potential, 
\begin{eqnarray}
	\mathcal{L} \supset
	-\frac{1}{2} (h', \phi')
	\left( \begin{matrix} m^2_{h' h'} & m^2_{h'\phi'} \\ m^2_{h'\phi'} & m^2_{\phi'\phi'} \end{matrix} \right)
	\left( \begin{matrix} h' \\ \phi' \end{matrix} \right)
	=
	-\frac{1}{2} (h, \phi)
	\left( \begin{matrix} m_h^2 & 0 \\ 0 & m_\phi^2 \end{matrix} \right)
	\left( \begin{matrix} h \\ \phi \end{matrix} \right),
	\label{eq: mass matrix}
\end{eqnarray}
where $m^2_{h' h'} = \lambda_H v_H^2$, $m^2_{h'\phi'} = A_{\Phi H} v_H$ and $m_{\phi'\phi'}^2 = \mu_\Phi^2 + \lambda_{\Phi H} v_H^2/2$. The diagonalization matrix is described by the mixing angle $\theta$, which controls the strength of interactions between $\phi$ and SM fermions or $\phi$ and SM gauge bosons. This angle is defined through the relation:
\begin{eqnarray}
	\left( \begin{matrix} h \\ \phi \end{matrix} \right) =
	\left( \begin{matrix} \cos \theta & - \sin \theta \\ \sin \theta & \cos \theta \end{matrix} \right)
	\left( \begin{matrix} h' \\ \phi' \end{matrix} \right).
\end{eqnarray}
Shown in following sections, the mixing angle $\sin \theta$ is severely constrained, and hence its absolute value is phenomenologically required to be much smaller than one, namely $|\theta| \ll 1$. As a result, the mass eigenstates and the mixing angle are expressed as 
\begin{eqnarray}
	m^2_{h\,(\phi)} = \frac{ m_{h' h'}^2 + m_{\phi'\phi'}^2 \pm \sqrt{ (m_{h' h'}^2 - m_{\phi'\phi'}^2)^2 + 4 m_{h'\phi'}^4 } }{2},
	\qquad 
	\tan 2\theta = - \frac{ 2 m_{h'\phi'}^2 }{ (m_{h' h'}^2 - m_{\phi'\phi'}^2 )}.
	\label{eq: masses and mixing}
\end{eqnarray}
Note that the solution of the second equation for the angle $\theta$ has a multi-fold ambiguity within the domain of definition, $-\pi \leq \theta \leq \pi$. However, we do not have to worry about this ambiguity because the angle is phenomenologically required to be suppressed as $|\theta| \ll 1$.

\subsection{Interactions}
\label{subsec: interactions}

Because of the mixing between the singlet $\Phi$ and the SM doublet $H$, the model predicts various interactions. First, interactions between the Higgs boson $h$ and SM fermions \& SM gauge bosons are suppressed slightly by $\cos \theta$ compared to the SM prediction. On the other hand, the mixing introduces interactions between the mediator $\phi$ and the SM particles, where $\phi$ behaves like a light Higgs boson with the couplings suppressed by $\sin \theta$ compared to the SM prediction. Both scalars have interactions with the WIMP as follows:
\begin{equation}
	\mathcal{L}_{\rm int} \supset
	- \frac{\cos\theta}{2} (c_s \phi \bar{\chi} \chi + c_p \phi \bar{\chi} i \gamma_5 \chi)
	+ \frac{\sin\theta}{2} (c_s h \bar{\chi} \chi + c_p h \bar{\chi} i \gamma_5 \chi).
\end{equation}
Other interactions among SM fermions and SM gauge bosons in $\mathcal{L}_{\rm SM}$ are not changed.

We next consider interactions among the scalars $h$ and $\phi$. From the scalar potential, four kinds of triple scalar interactions are obtained. Their explicit forms are given by
\begin{eqnarray}
	\mathcal{L} &\supset& - \frac{c_{h h h}}{3!} h^3 - \frac{c_{\phi h h}}{2} \phi h^2 - \frac{c_{\phi \phi h}}{2} \phi^2 h
	- \frac{c_{\phi \phi \phi}}{3!} \phi^3,
	\label{eq: cubic terms}
	\\
	c_{h h h} &=& 3 \lambda_H v_H c_\theta^3 - 3 A_{\Phi H} c_\theta^2 s_\theta - \mu_3 s_\theta^3
	+ 3 \lambda_{\Phi H} v_H s_\theta^2 c_\theta,
	\nonumber \\
	c_{\phi h h} &=& 3 \lambda_H v_H c_\theta^2 s_\theta + A_{\Phi H} (c_\theta^3 - 2 c_\theta s_\theta^2 )
	+ \mu_3 c_\theta s_\theta^2 + \lambda_{\Phi H} v_H ( s_\theta^3 - 2c_\theta^2 s_\theta),
	\nonumber \\
	c_{\phi \phi h} &=& 3 \lambda_H v_H c_\theta s_\theta^2 + A_{\Phi H } (2c_\theta^2 s_\theta - s_\theta^3)
	- \mu_3 c_\theta^2 s_\theta + \lambda_{\Phi H} v_H (c^3_\theta - 2 c_\theta s_\theta^2),
	\nonumber \\
	c_{\phi \phi \phi} &=& 3 \lambda_H v_H s^3_\theta + 3 A_{\Phi H} c_\theta s^2_\theta + \mu_3 c_\theta^3
	+ 3 \lambda_{\Phi H} v_H c_\theta^2 s_\theta,
	\nonumber
\end{eqnarray}
where we define $s_\theta \equiv \sin\theta$ and $c_\theta \equiv \cos\theta$, respectively. In addition to the cubic scalar interactions, the model predicts five quartic interactions from the scalar potential:
\begin{eqnarray}
	\mathcal{L} &\supset& - \frac{c_{h h h h}}{4!} h^4 - \frac{c_{\phi h h h}}{3!} \phi h^3 - \frac{c_{\phi \phi h h}}{4} \phi^2 h^2
	- \frac{c_{\phi \phi \phi h}}{3!} \phi^3 h - \frac{c_{\phi \phi \phi \phi}}{4!} \phi^4,
	\\
	c_{h h h h} &=& 3 \lambda_H c_\theta^4 + 6 \lambda_{\Phi H} c_\theta^2 s_\theta^2 + \lambda_\Phi s_\theta^4,
	\nonumber \\
	c_{\phi h h h} &=& 3 \lambda_H c_\theta^3 s_\theta - 3 \lambda_{\Phi H} (c_\theta^3 s_\theta - c_\theta s_\theta^3)
	- \lambda_\Phi c_\theta s_\theta^3,
	\nonumber \\
	c_{\phi \phi h h} &=& 3 \lambda_H c_\theta^2 s_\theta^2
	+ \lambda_{\Phi H} (c_\theta^4 - 4 c_\theta^2 s_\theta^2 + s_\theta^4 ) + \lambda_\Phi c_\theta^2 s_\theta^2,
	\nonumber \\
	c_{\phi \phi \phi h} &=& 3 \lambda_H c_\theta s_\theta^3 + 3 \lambda_{\Phi H} ( c_\theta^3 s_\theta - c_\theta s_\theta^3 )
	- \lambda_\Phi c_\theta^3 s_\theta,
	\nonumber \\
	c_{\phi \phi \phi \phi} &=& 3 \lambda_H s_\theta^4 + 6 \lambda_{\Phi H} c_\theta^2 s_\theta^2 + \lambda_\Phi c_\theta^4.
	\nonumber
\end{eqnarray}
As seen from all interactions discussed in this subsection, when the mixing angle $\theta$ is very suppressed, the scalar $h$ becomes almost the SM Higgs boson, and it couples to the WIMP weakly. On the other hand, $\phi$ couples to the WIMP without any suppression, while interacts with SM particles very weakly.\footnote{The mediator $\phi$ seems to have unsuppressed interactions with $h$ through the couplings $A_{\phi H}$ and $\lambda_{\phi H}$ even if the mixing angle is very close to zero. The coupling $A_{\phi H}$ is, however, suppressed when the mixing angle is small, as can be deduced from eq.\,(\ref{eq: masses and mixing}). On the other hand, the coupling $\lambda_{\Phi H}$ can be still large, though interactions between $\phi$ and SM fermions, which plays an important role in cosmology, are suppressed by small Yukawa couplings especially when the dark matter mass (hence, the freeze-out temperature) is small.} As a result, the WIMP $\chi$ and the mediator $\phi$ form a dark sector coupling to the SM sector weakly through the mixing between $H$ and $\Phi$. Hence, precision measurements of particle physics experiments are expected to play important roles to search for the dark sector particles, as we will see in following sections.

\subsection{Model parameters}

There are eight parameters in the scalar potential: $\mu_H^2$, $\mu_1^3$, $\mu_\Phi^2$, $\mu_3$, $\lambda_H$ $\lambda_\Phi$, $\lambda_{\Phi H}$ and $A_{\Phi H}$. Since the two vacuum expectation values of the SM doublet $H$ and the singlet $\Phi$ are fixed and the Higgs mass is determined to be $m_h \simeq 125$\,GeV thanks to the LHC experiment, the number of free parameters in the potential becomes five. Adding the WIMP mass $m_\chi$ as well as two couplings between the WIMP and the singlet $\Phi$, namely $c_s$ and $c_p$, 
the total number of free model parameters is then reduced to be eight 
in the end.

The condition of the two vacuum expectation values, $v_H \simeq 246$\,GeV and $v_\Phi = 0$, gives two relations among the model parameters, $\mu_H^2 + \lambda_H v_H^2/2 = 0$ and $\mu_1^3 + A_{\Phi H} v_H^2/2 = 0$, so that we can drop the two parameters $\mu_H^2$ and $\mu_1^3$ from the set of independent parameters. Moreover, as 
seen in eq.\,(\ref{eq: mass matrix}), the Higgs mass is determined mainly by the parameter $\lambda_H$ when the mixing angle $\theta$ is suppressed, 
and hence it can also be dropped from the parameter set. 
On the other hand, when $\theta \ll 1$, the mediator mass $m_\phi$ is determined mainly by the combination of the two parameters $\mu_\Phi^2$ and $\lambda_{\Phi H}$, while the mixing angle is given mainly by the parameter $A_{\Phi H}$, as can be seen in eq.\,(\ref{eq: mass matrix}). 
Thus, we adopt
$m_\phi$ and $\sin \theta$ as independent parameters instead of $\lambda_{\phi H}$ and $A_{\phi H}$. As a result, we have the following eight parameters, $m_{\chi}$, $c_s$, $c_p$, $m_\phi$, $\sin \theta$, $\mu_\phi^2$, $\mu_3$ and $\lambda_\Phi$, 
as free model input parameters.

\subsection{Mediator decay}
\label{subsec: phi decay}

Because the decay of the mediator $\phi$ plays an important role in phenomenology of the model, we discuss some details of the decay in this subsection. As already mentioned in section\,\ref{subsec: interactions}, $\phi$ behaves like a light Higgs boson, and its decay into SM particles is from the mixing between $H$ and $\Phi$. Its partial decay width into a specific SM final state is
\begin{equation}
	\Gamma(\phi \to {\rm SMs}) =
	\sin^2 \theta \times \left. \Gamma(h_{\rm SM} \to {\rm SMs}) \right|_{m_{h_{\rm SM}}^2 \to m_\phi^2}.
	\label{eq: phi decay}
\end{equation}
We will discuss each decay mode below, and present how we estimate the decay width.

When the mediator is lighter than the electron-positron threshold, namely $m_\phi \leq 2m_e$, it decays mainly into two photons. We use the formula in Ref.\,\cite{shifman} to calculate its partial decay width. Above the threshold, though this channel is not a dominant one anymore, 
the same formula is still used up to $m_\phi = 0.6$\,GeV. 
If $m_\phi \geq 2$\,GeV, the decay width is computed by 
using the \texttt{HDECAY} code\,\cite{hdecay}. In the region of 0.6\,GeV $\leq m_\phi \leq$ 2\,GeV, the formula in Ref.\,\cite{gunion} is used to connect the regions $m_\phi \leq 0.6$\,GeV and $m_\phi \geq 2$\,GeV smoothly.

At the region of $2 m_e \leq m_\phi \leq 2 m_\mu$ with $m_\mu$ being the muon mass, it decays mainly into a electron and a positron. Its partial decay width is given by the following formula,
\begin{equation}
	\Gamma(\phi \to e^+ e^-) = \sin^2 \theta \times \frac{m_e^2 m_\phi}{8 \pi v_H^2} 
	\left( 1 - \frac{4 m_e^2}{m_\phi^2} \right)^{3/2}.
	\label{eq: electrons}
\end{equation}
When the mediator mass is above the muon threshold but below the pion threshold, namely $2 m_\mu \leq m_\phi \leq 2 m_\pi$, the mediator decays mainly into a muon pair. Its partial decay width is computed by the same formula as above with $m_e$ being replaced by $m_\mu$. The formulae for the two channels are used up to $m_\phi = 2$\,GeV. In the mass region of $m_\phi \geq 2$\,GeV, the \texttt{HDECAY} code is used to compute the partial decay widths of the two channels.

When $\phi$ is heavier than the pion threshold but lighter than a few GeV, it decays mainly into a pair of pions (and a pair of K mesons if $m_\phi \geq 2 m_K$). Concerning the decay channel into $\pi \pi$, we use the result in Ref.\,\cite{pipikk} to compute its partial decay width in the mass region of $2 m_\pi \leq m_\phi \leq 1.4$\,GeV,
which is also consistent with the latest result in Ref.\,\cite{Winkler:2018qyg}.
On the other hand, the width becomes negligibly small compared to other channels when $m_\phi \geq 2$\,GeV, so that we set $\Gamma(\phi \to \pi\pi) = 0$ in this region. The width is evaluated by a linear interpolation in the mass range of 1.4\,GeV $\leq m_\phi \leq$ 2\,GeV. On the other hand, the result in Ref.\,\cite{pipikk} is used again to compute the partial decay width of the $\phi \to KK$ channel at $2 m_K \leq m_\phi \leq 1.4$\,GeV. Then, the width is evaluated using a linear interpolation at 1.4\,GeV $\leq m_\phi \leq$ 2\,GeV by connecting the width to that of the $\phi \to s \bar{s}$ channel continuously, where $s$ is the strange quark. The partial decay width of the $\phi \to s \bar{s}$ channel is computed by the \texttt{HDECAY} code at $m_\phi \geq 2$\,GeV.\footnote{In order to take the effect of the K meson threshold into account, we multiply the partial decay width of the $\phi \to s \bar{s}$ channel (computed in the \texttt{HDECAY} code) by the phase space factor of $(m_\phi^2 - 4 m_K^2)^{3/2}/(m_\phi^2 - 4 m_s^2)^{3/2}$. A similar prescription is also applied for the other channel $\phi \to c \bar{c}$ with $c$ being the charm quark.} Moreover, in order to take other hadronic decay channels into account, we also consider the $\phi \to g g$ channel with $g$ being the gluon in the mass range of $m_\phi \geq 1.4$\,GeV. When $m_\phi \geq 2$\,GeV, we use the result in the \texttt{HDECAY} code to compute its partial decay width, while the width is evaluated at 1.4\,GeV $\leq m_\phi \leq$ 2\,GeV by a linear interpolation with the boundary condition of $\Gamma(\phi \to g g) = 0$ at $m_\phi = 1.4$\,GeV.

Other decay channels open in the mass range of $m_\phi \geq 2$\,GeV (e.g. decays into a tau lepton pair, a charm quark pair, a bottom quark pair, etc.). We consider all possible decay channels in this region and compute their partial decay widths using the \texttt{HDECAY} code.

Here, we consider the $\phi$ decay into $\pi \pi$ ($K K$) in more details, because it is known to have a large theoretical uncertainty in the range of $2 m_\pi\,(2 m_K) \leq m_\phi \leq 1.4$\,GeV\,\cite{charm_eq} due to non-perturbative QCD effects. For instance, the total width of the SM Higgs boson in this mass range is predicted in other literature to be larger\,\cite{Duchovni:1989ii} or smaller\,\cite{Truong:1989my} than what we have estimated based on Ref.\,\cite{pipikk}. 
In order to make our analysis conservative, we introduce a nuisance parameter $\sigma$ to take this uncertainty into account.
Then, the partial decay widths of $\phi \to \pi \pi$ and $\phi \to K K$ channels are computed according to the equations
\begin{eqnarray}
	\Gamma(\phi \to \pi \pi) &\equiv&
	\frac{\Gamma_{\pi \pi}}{\Gamma_{\pi \pi} + \Gamma_{K K}}
	\left[ \sigma \Gamma_+ + (1 - \sigma) \Gamma_- \right],
	\nonumber \\
	\Gamma (\phi \to KK) &\equiv&
	\frac{\Gamma_{K K}}{\Gamma_{\pi \pi} + \Gamma_{K K}}
	\left[ \sigma \Gamma_+ + (1 - \sigma) \Gamma_- \right],
\end{eqnarray} 
where $\Gamma_{\pi \pi}$ and $\Gamma_{K K}$ are the partial decay widths into $\pi \pi$ and $K K$ computed based on Ref.\,\cite{pipikk}. On the other hand, $\Gamma_+$ and $\Gamma_-$ are the sum of the widths computed based on Ref.\,\cite{Duchovni:1989ii} and Ref.\,\cite{Truong:1989my}, respectively. The nuisance parameter $\sigma$ varies between $0 \leq \sigma \leq 1$.

In addition to the decay channels into various SM particles, 
the mediator particle $\phi$ can also decay into a pair of WIMPs when the mediator mass is larger than twice the WIMP mass, namely $m_\phi \geq 2 m_\chi$. Its partial decay width is described by 
\begin{eqnarray}
	\Gamma(\phi \to \chi \chi) = \cos^2 \theta \frac{m_\phi}{16 \pi}
	\left[ c_s^2 \left( 1 - \frac{4 m_\chi^2}{m_\phi^2} \right)^{3/2} + c_p^2 \left(1 - \frac{4 m_\chi^2}{m_\phi^2} \right)^{1/2} \right].
	\label{eq: phi to chi chi}
\end{eqnarray}

\begin{figure}[t!]
	\centering
	\includegraphics[height=2.2in, angle=0]{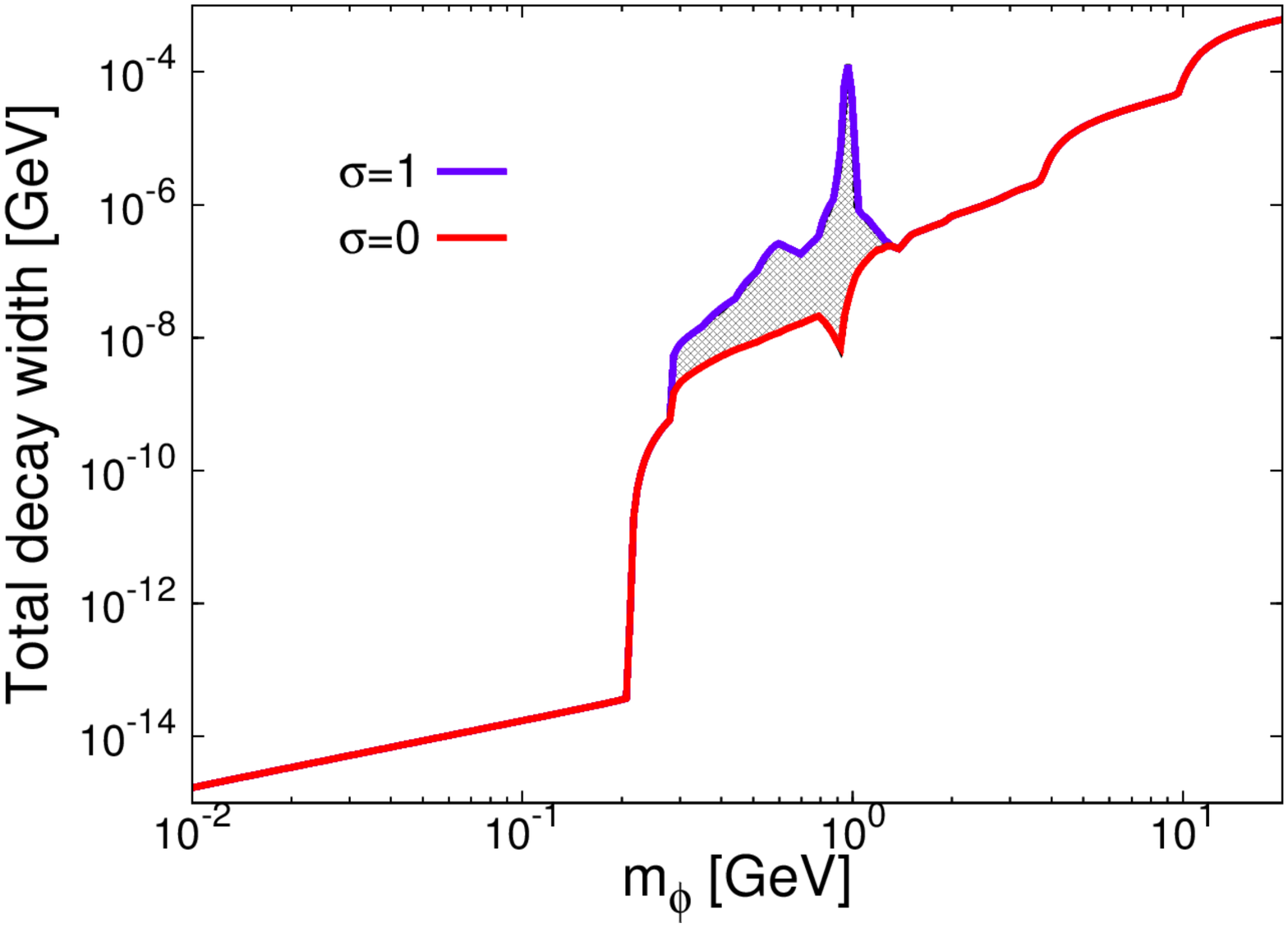}
	\includegraphics[height=2.2in, angle=0]{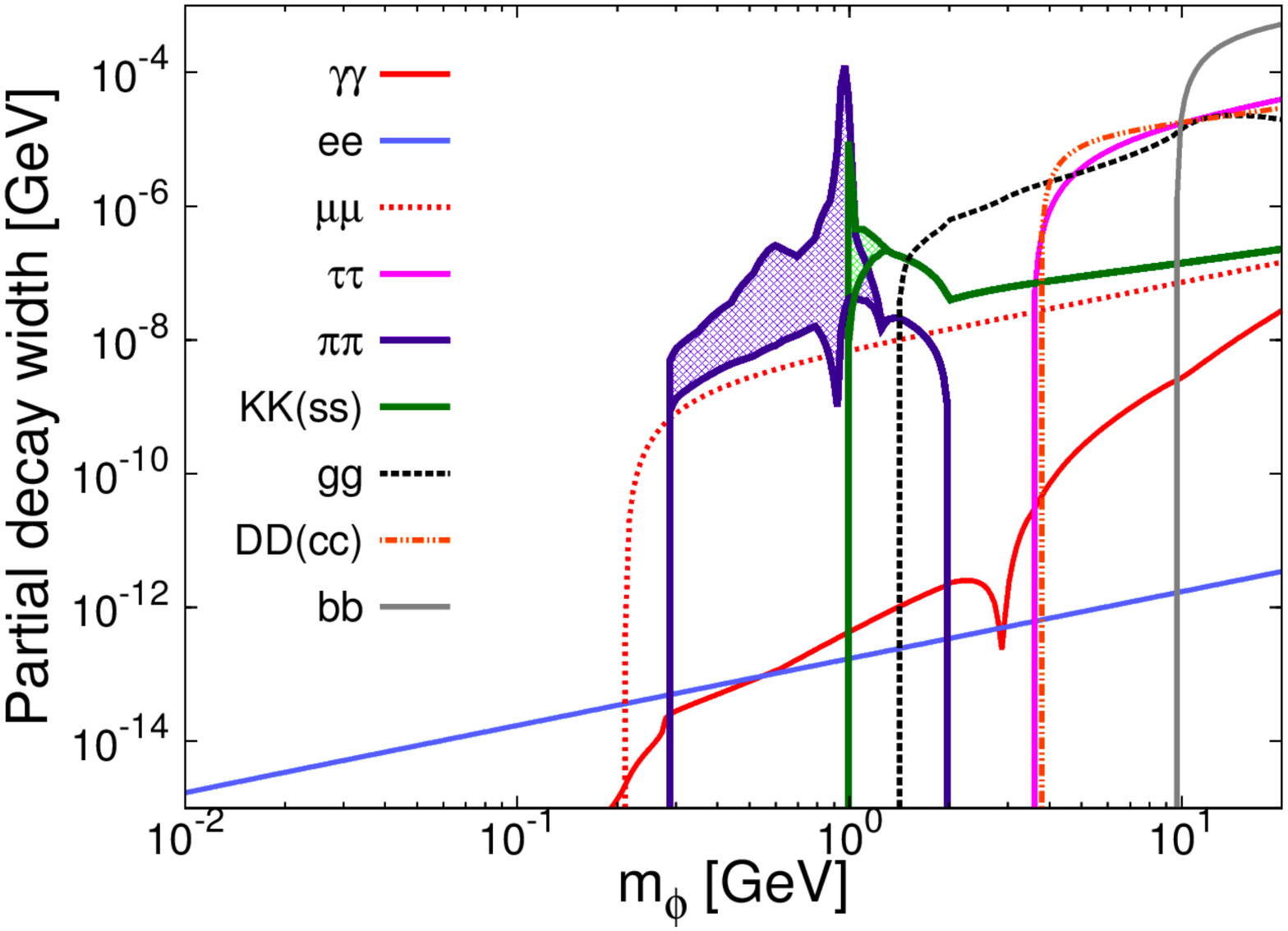}
	\caption{\small \sl (Left panel) The total decay width of the mediator $\phi$ assuming that $\sin \theta = 1$ and $\phi$ does not decay into a WIMP pair. The gray band indicates the theoretical uncertainty due to non-perturbative QCD effects. (Right panel) Partial decay widths contributing to the total width.}
	\label{fig: width}
\end{figure}

All results obtained so far in this subsection are summarized in Fig.\,\ref{fig: width}, where the total decay width of $\phi$ is shown in the left panel of the figure assuming that $\sin \theta = 1$ and $\phi$ does not decay into a pair of WIMPs. The gray band indicates the theoretical uncertainty due to non-perturbative QCD effects, which are taken into account by the nuisance parameter $\sigma$. In the right panel, the partial decay widths of several channels of $\phi$ are depicted.

\subsection{Higgs decay}

In this model, the partial decay width of the Higgs boson into SM particles is slightly modified from the SM prediction because of the mixing between $H$ and $\Phi$. Its partial decay width into a SM final state (a fermion or a gauge boson pair) is given by the following formula:
\begin{equation}
	\Gamma(h \to {\rm SMs}) =
	\cos^2 \theta \times \Gamma(h_{\rm SM} \to {\rm SMs}),
	\label{eq: phi decay}
\end{equation}
where the Higgs mass is fixed to be 125\,GeV. In addition to these channels, the Higgs boson decays into a pair of WIMPs, and its partial decay width is given by the same formula as the one shown in eq.\,(\ref{eq: phi to chi chi}) with $\cos \theta$ and $m_\phi$ 
replaced by $\sin \theta$ and $m_h$, respectively:
\begin{eqnarray}
	\Gamma(h \to \chi \chi) = \sin^2 \theta \frac{m_h}{16 \pi}
	\left[ c_s^2 \left( 1 - \frac{4 m_\chi^2}{m_h^2} \right)^{3/2} + c_p^2 \left(1 - \frac{4 m_\chi^2}{m_h^2} \right)^{1/2} \right].
\end{eqnarray}

The Higgs boson can also decay into several mediators when the mediator $\phi$ is light enough. Considering the fact that the mixing angle $\theta$ is phenomenologically required to be much less than one and multi-$\phi$ channels are suppressed by their final state phase spaces, only the decay channel $h \to \phi \phi$ can be potentially comparable to the other decay channels into SM particles (and a WIMP pair). The explicit form of its partial decay width is
\begin{equation}
	\Gamma(h \to \phi \phi ) = \frac{c_{\phi \phi h}^2}{32 \pi m_h} \left( 1 - \frac{4 m_\phi^2}{m_h^2} \right)^{1/2},
\end{equation}
where the coefficient $c_{\phi \phi h}$ is given in eq.\,(\ref{eq: cubic terms}). This decay channel gives distinctive signals at the LHC experiment, depending on the mass and the decay length of the mediator. For instance, when the mediator decays mainly into leptons and its decay length is enough shorter than 1\,mm, this Higgs decay channel gives a signal of $h \to 4 \ell$. On the other hand, if the decay length is much longer than the detector size, the channel gives a signal of the invisible $h$ decay, regardless of the decay channel of the mediator. When the decay length is around the detector size, we can expect various signals at displaced vertex searches.

\section{Constraints}
\label{sec: constraints}

We introduce all constraints used in our comprehensive analysis. The likelihood of the constraints will be modeled in various functions which will be further discussed in section\,\ref{subsec: simulation framework}. The usage and information of the constraints are discussed in the following in details.

\subsection{Preselection criteria}
\label{subsec: apriori constraints}

Before performing the likelihood analysis,
we apply preselection criteria on the model parameter space 
as one of our prior distribution. 
The following three criteria are imposed: CP conserving criterion, the criterion on the maximal value of the mixing angle, and the vacuum stability criterion. Those are implemented by a $1/0$ logical cut in our analysis. All the preselection criteria 
involved in our analysis are summarized in 
Table~\ref{tab: apriori constraints}.

\begin{table}[t!]
	\centering
	\begin{tabular}{ l | l r r }
	& Present & Future & Section \\
	\hline
	CMB distortion & Planck\,\cite{Slatyer:2015jla} & -- -- -- & \ref{subsubsec: CP conserving} \\
	Higgcision & LHC\,\cite{Khachatryan:2016vau} & -- -- -- & \ref{subsubsec: Higgcision} \\
	Vacuum stability & See the text & -- -- -- & \ref{subsubsec: vacuum stability} \\
	\hline
	\end{tabular}
	\caption{\small \sl Preselection criteria that we have imposed in our analysis. The second and third columns are for present and near future experiments/observations used to apply the criteria. The last column is for the section where each criterion is discussed in detail. The third column is now blanked, as the preselection criteria are not stronger than other constraints we discuss in following subsections.}
	 \label{tab: apriori constraints}
\end{table}

\subsubsection{CP conserving criterion}
\label{subsubsec: CP conserving}

The pseudo-scalar interaction between the WIMP $\chi$ and the singlet $\Phi$ in eq.\,(\ref{eq: L}) breaks the CP symmetry under the general scalar potential of $V(\Phi, H)$. Then, the existence of the non-zero pseudo-scalar coupling $c_p$ induces the so-called s-wave WIMP annihilation into SM particles, namely the WIMP annihilates into SM particles without any velocity suppression at present universe. 
Moreover, when both the scalar and pseudo-scalar couplings, $c_s$ and $c_p$, are non-zero, the WIMP annihilates into a pair of the mediators $\phi$ without the velocity suppression. In such cases, the WIMP annihilation cross section at present universe 
is almost the same as the one during the freeze-out process in the early universe, and this fact means that the cross section at present universe is expected to be about 1\,pb.

WIMP annihilation cross section of ${\cal O}(1)$\,pb at present universe is, however, not favored by the CMB observation, when the WIMP mass is less than ${\cal O}(1)$\,GeV\,\cite{Slatyer:2015jla}. This is because the annihilation distorts the recombination history of the universe, while the observational result is consistent with the standard predication without the contribution from the WIMP. Thus, in order to avoid the CMB constraint, we set the pseudo-scalar coupling constant to be zero, namely \mbox{\boldmath $c_p = 0$}, in our analysis. In other words, we impose the CP symmetry on the interactions relevant to the WIMP. After switching off the coupling $c_p$, the WIMP annihilates into SM particles as well as a pair of the mediators with a velocity suppression at present universe, which enables us to avoid the CMB constraint easily.

\subsubsection{Criterion on $|\theta|$}
\label{subsubsec: Higgcision}

As already mentioned in previous section, the mixing angle $\theta$ is severely constrained by various experiments. In order to reduce a computational cost in our numerical analysis, we impose a condition on the angle as \mbox{\boldmath $|\theta| \leq \pi/6$}, as a preselection criterion. This condition comes from the precision measurement of Higgs boson properties at the LHC experiment\,\cite{Khachatryan:2016vau}. The measurement of the Higgs boson production process $gg \to h$ followed by its decay into a pair of gammas or $W$ bosons gives a constraint on $\theta$ as $\cos \theta \geq 0.9$, which is translated into the constraint on the angle $\theta$ as above. On the other hand, much severe constraints are eventually obtained from other particle physics experiments and cosmological observations, and those are taken into account in the subsequent likelihood analysis.

\subsubsection{Vacuum stability criterion}
\label{subsubsec: vacuum stability}

The minimal WIMP model in eq.\,(\ref{eq: L}) has an extended scalar sector composed of $\Phi$ and $H$. In order to guarantee the stability of our electroweak vacuum, we impose the following condition on the scalar potential. We first define the fields $\xi$ and $\eta$ as $\Phi = \eta$ and $H = (0, \xi/\sqrt{2})^T$, respectively, after taking the unitary gauge. Then, we impose a condition on the potential as \mbox{\boldmath $V(\eta, \xi) \geq V(v_\Phi, v_H)$} in the range of the fields \mbox{\boldmath $|\xi| \leq 1$}\,TeV and \mbox{\boldmath $|\eta| \leq 1$}\,TeV. Philosophy behind the condition is as follows: We consider the minimal model as an effective theory of the WIMP defined at the energy scale of 1\,TeV, and require our electroweak vacuum to be absolutely stable within the range where the effective theory is applied. One might think that it is even possible to put a more conservative constraint on the potential by using the meta-stability condition, where our vacuum is required to be, at least, meta-stable with its lifetime much longer than the age of the universe. On the other hand, other constraints from particle physics experiments and cosmological observations give severer constraints on the potential, as we will see in the following subsections.\footnote{Radiative corrections to the potential are also not taken into account because of the same reason.} Hence, we adopt the absolute stable condition as a preselection criterion applied before the likelihood analysis.

\subsubsection{Region of the parameters scanned}
\label{subsubsec: parameters}

In addition to the the preselection criteria mentioned above, 
we scan the parameter space ($m_{\chi}$, $c_s$, $m_\phi$, $\theta$, $\mu_\phi^2$, $\mu_3$ and $\lambda_\Phi$) 
over the following ranges in our numerical analysis:
\begin{eqnarray}
	0 \leq &m_\chi& \leq 30\,{\rm GeV}, \nonumber \\
	-1 \leq &c_s& \leq 1, \nonumber \\
	0 \leq & m_\phi & \leq 1\,{\rm TeV}, \nonumber \\
	- \pi/6 \leq &\theta& \leq \pi/6, \nonumber \\
	- 1\,{\rm TeV}^2 \leq &\mu_\Phi^2 & \leq 1\,{\rm TeV}^2, \nonumber \\
	-1 \,{\rm TeV} \leq &\mu_3& \leq 1\,{\rm TeV}, \nonumber \\
	-1 \leq &\lambda_\Phi& \leq 1. 
	\label{eq: the parameters}
\end{eqnarray}
The dimensionless coupling constants $c_s$ and $\lambda_\Phi$ vary between $-1$ to $+1$, assuming that UV completion behind the minimal WIMP model is described by a weak interacting theory. On the other hand, dimensionful coupling constants basically vary within the energy scale of 1\,TeV due to the validity of the minimal WIMP model. Upper limit on the WIMP mass $m_\chi$ is fixed to be 30\,GeV, simply because we are interested in the light WIMP region.

The surviving parameter space after applying the three preselection criteria are shown in appendix\,\ref{app: initial parameter region}.
Though the seven parameters in eq.\,(\ref{eq: the parameters}) are used to put the preselection criteria,\footnote{Five parameters ($m_\phi$, $\theta$, $\mu_\phi^2$, $\mu_3$ and $\lambda_\phi$) among seven parameters are relevant to the preselection criteria.} two of them, $\mu_\Phi^2$ and $\lambda_\Phi$, are not very much relevant to the result of our likelihood analysis. This is because the coupling constant of the quadratic term, $\mu_\Phi^2$, appears only in the mass matrix of eq.\,(\ref{eq: mass matrix}), but the mass matrix is already parameterized by two parameters, $m_\phi$ and $\theta$.\footnote{Since $\mu_\Phi^2$ determines the value of the parameter $\lambda_{\Phi H}$ via eq.\,(\ref{eq: mass matrix}), it is scanned as a nuisance to be honest.} The other parameter $\lambda_\Phi$ appears in scalar quartic interactions, however no significant constraints on the quartic interactions are obtained so far and even in the near future. As a result, we will present our results of the likelihood analysis in terms of the following five parameters \mbox{\boldmath $m_{\chi}$}, \mbox{\boldmath $c_s$}, \mbox{\boldmath $m_\phi$}, \mbox{\boldmath $\theta$} and \mbox{\boldmath $\mu_3$} in subsequent sections.

\subsection{Conditions/Constraints from cosmology and astrophysics}
\label{subsec: cosmological constraints}

In this subsection, we summarize cosmological and astrophysical conditions/constraints used to figure out the present status and future prospects of the minimal WIMP model. Those are taken into account through the likelihood analysis, unless otherwise stated. All the conditions/constraints involved in our analysis are summarized in Table\,\ref{tab: cosmological and astrophysical constraints}.

\begin{table}[t!]
	\centering
	\begin{tabular}{ l | l l r }
	& Present & Future & Section \\
	\hline
	Relic abundance
	& Planck\,\cite{Adam:2015nya} & -- -- -- & \ref{subsubsec: relic abundance} \\
	Equilibrium
	& See the text & -- -- -- & \ref{subsubsec: kinematical equilibrium} \\
	Direct detection
	& XENON1T\,\cite{Aprile:2018dbl}, CRESST\,\cite{Petricca:2017zdp}, & NEWS-SNOLAB\,\cite{news1, news2},
	& \ref{subsubsec: direct detection} \\
	&PANDAX\,\cite{Tan:2016zwf}, SuperCDMS\,\cite{Agnese:2015nto}, & SuperCDMS\,\cite{Agnese:2016cpb}, & \\
	& NEWS-G\,\cite{Arnaud:2017bjh}, Darkside-50\,\cite{Agnes:2018ves} & LZ\,\cite{Akerib:2015cja, Szydagis:2016few} & \\
	DOF $\left( \Delta N_{\rm eff} \right)$
	& PLANCK\,\cite{Ade:2013zuv} & CMB-S4\,\cite{Abazajian:2016yjj} & \ref{subsubsec: neff} \\
	BBN & See the text & -- -- -- & \ref{subsubsec: bbn} \\
	\hline
	\end{tabular}
	\caption{\small \sl Cosmological and astrophysical conditions/constraints imposed in our likelihood analysis. See the caption of Table\,\ref{tab: apriori constraints}, for the meaning of all the columns are the same as those in the table.}
	 \label{tab: cosmological and astrophysical constraints}
\end{table}

\subsubsection{Relic abundance condition}
\label{subsubsec: relic abundance}

The WIMP should satisfy the so-called relic abundance condition. On observational side, the relic abundance of the WIMP, or in other words, the averaged mass density of the WIMP at present universe, is very precisely measured by the PLANCK collaboration\,\cite{Adam:2015nya}:
\begin{equation}
	\Omega h^2 = 0.1193 \pm 0.0014,
\end{equation}
with $h$ being the normalized Hubble constant. The uncertainty of the observation is less than 2\%, which is comparable to precise measurements at collider experiments.

On theoretical side, we calculate the relic abundance using the \texttt{MicrOMEGAs} code\,\cite{Belanger:2014hqa} based on the minimal WIMP model. The code first calculates the (thermal-averaged) annihilation cross section of the WIMP, and next compute the abundance by solving the Boltzmann equation numerically. Concerning the cross section, when $m_\chi \geq m_\phi$, the WIMP annihilates mainly into a pair of the mediators, where the mediators eventually decay into SM particles. On the other hand, when $m_\chi \leq m_\phi$, the WIMP annihilates into SM particles through the exchange of the mediator (or the Higgs boson) in the s-channel. Because the process is suppressed by small Yukawa couplings and the mixing angle, only the resonant region with $m_\chi \sim m_\phi/2$ satisfies the relic abundance condition. Concerning the Boltzmann equation, it is known to have a theoretical uncertainty originating in the massless degrees of freedom of the universe during the freeze-out process. Since we are interested in the light WIMP scenario, the freeze-out temperature can be as low as the energy scale of the QCD phase transition, which makes the estimate of the massless degrees of freedom uncertain due to several non-perturbative QCD effects. Its uncertainty is reported to be at most 10\%\,\cite{Drees:2015exa}, and we adopt this maximum value to make our analysis conservative. Here, it is also worth mentioning that
the mediator may contribute to the massless degrees of freedom 
and affect the Hubble expansion rate if it is light enough.
However, we do not include this contribution in this analysis, because it is negligibly small compared to the above QCD uncertainty.

Note that it may have 
the Sommerfeld effect on the WIMP annihilation process\,\cite{Hisano:2003ec, Hisano:2004ds, Hisano:2006nn}, for the mediator is sometimes much lighter than the WIMP in our setup\,\cite{Tulin:2012wi, Tulin:2013teo, Kaplinghat:2013yxa, kimmo, Kahlhoefer:2017umn}. 
We found that the effect is not sizable for model parameter sets passing all conditions and constraints adopted in this paper. This is verified by computing the quantity, $c_s^2/(4\pi)\cdot(m_\chi/m_\phi)$, and confirmed that it is always enough smaller than one for the sets.\footnote{This quantity is known to be the one for testing whether the Sommerfeld effect becomes sizable or not\,\cite{Hisano:2004ds, Hisano:2006nn, Tulin:2012wi}. If it is smaller than one, the effect only gives a small correction to the thermally averaged cross section.} We therefore do not take the effect into account to calculate the WIMP annihilation cross section.

\subsubsection{Kinematical equilibrium condition}
\label{subsubsec: kinematical equilibrium}

We also impose the kinematical equilibrium condition on the model, where the WIMP and SM particles are required to be in thermal equilibrium during the freeze-out process. Though the condition is automatically satisfied for a typical WIMP with its mass of $\mathcal{O}$(100)\,GeV, it should be imposed independently for the light WIMP, because both WIMP and mediator connect to SM with small couplings, and the condition is not automatically satisfied. It is worth emphasizing that we adopt this condition to figure out a very conventional WIMP parameter region in our setup. On the other hand, the condition can be relaxed by requiring that the WIMP is in the equilibrium at some temperature of the universe before the freeze-out, because it still allows us to make a quantitative prediction on its abundance. We will discuss in appendix\,\ref{app: relaxing KEC} how the result of our analysis alters by relaxing the condition.

The WIMP annihilates directly into SM particles when $m_\chi \leq m_\phi$. Its reaction rate at the freeze-out temperature $T_f$ is estimated to be the product of the thermally averaged annihilation cross section (times the relative velocity) and the number density of the WIMP, namely $\Gamma_{\chi \chi} \sim \langle \sigma_{\chi \chi} v \rangle_{T_f} n_{\rm WIMP}(T_f)$. The reaction rate $\Gamma_{\chi \chi}$ becomes the same order as the expansion rate of the universe $H(T_f)$ due to the relic abundance condition. The existence of the annihilation process guarantees that of the scattering process between the WIMP and the SM particles because of the crossing symmetry, and its reaction rate is $\Gamma_{\chi {\rm SM}} \sim \langle \sigma_{\chi {\rm SM}} v \rangle_{T_f} n_{\rm SM}(T_f)$ with $\sigma_{\chi {\rm SM}}$ and $n_{\rm SM}(T_f)$ being the scattering cross section and the number density of the SM particles, respectively. Since the number density of the SM particles is much larger than that of the WIMP at the freeze-out temperature, this fact gives $\Gamma_{\chi {\rm SM}} \gg \Gamma_{\chi \chi} \sim H(T_f)$ unless the annihilation cross section is significantly boosted compared to the scattering cross section. Hence, the equilibrium condition is usually automatically satisfied as far as $m_\chi \leq m_\phi$.\footnote{When $m_\chi$ is very close to $m_\phi/2$ and the decay width of the mediator is very suppressed, $\langle \sigma_{\chi \chi} v \rangle_{T_f}$ is indeed significantly boosted, and it requires a special treatment to calculate the correct relic abundance\,\cite{Binder:2017rgn}.}

On the other hand, the WIMP annihilates mainly into two mediators when $m_\chi \geq m_\phi$, so that $\Gamma_{\chi {\rm SM}} \geq H(T_f)$ is not always guaranteed. However, even if $\Gamma_{\chi {\rm SM}} \leq H(T_f)$, the WIMP can be in the kinematical equilibrium with SM particles when the equilibrium is maintained between the WIMP and the mediator and between the mediator and the SM particles simultaneously. It means that, even if the reaction rate between the WIMP and SM particles is smaller than the expansion rate of the universe, the WIMP has a possibility to be in the kinematical equilibrium via the mediator in the universe. We thus take the following strategy to impose the kinematical equilibrium condition to the minimal WIMP model.

At each set of the input model parameters to define the model, we first calculate the freeze-out temperature $T_f$ by the \texttt{MicrOMEGAs} code. We next calculate the reaction rate $\Gamma_{\chi {\rm SM}}$ and compare it with the expansion rate of the universe $H(T_f)$. We accept the set if $\Gamma_{\chi {\rm SM}} \geq H(T_f)$. If it is not, we further calculate two reaction rates between the WIMP and the mediator and between the mediator and SM particles. The former reaction rate is estimated to be $\Gamma_{\chi \phi} \sim \langle \sigma_{\chi \phi} v \rangle_{T_f} n_\phi(T_f)$ with $\sigma_{\chi \phi}$ and $n_\phi(T_f)$ being the scattering cross section and the number density of the mediator at the freeze-out temperature. The latter reaction rate has a more complicated form than the former one. In fact, three different processes contribute to the reaction rate; decay, scattering and absorption processes. The rate is estimated to be $\Gamma_{\phi {\rm SM}} \sim \langle \Gamma_\phi \rangle_{T_f} + \langle \sigma_{\phi {\rm SM}} v \rangle_{T_f} n_{\rm SM}(T_f) + \langle \sigma'_{\phi {\rm SM}} v \rangle_{T_f} n_\phi(T_f)$ with $\Gamma_\phi$, $\sigma_{\phi {\rm SM}}$ and $\sigma'_{\phi {\rm SM}}$ being the total decay width, the scattering and absorption cross sections of $\phi$, respectively. If both $\Gamma_{\chi \phi}$ and $\Gamma_{\phi {\rm SM}}$ are larger than $H(T_f)$, we accept the set of the model parameters. Note that, unlike the relic abundance condition, the kinematical equilibrium condition that we have discussed above is implemented by a $1/0$ logical cut in our likelihood analysis.

\begin{figure}[t!]
	\centering
	\includegraphics[height=2.2in,angle=0]{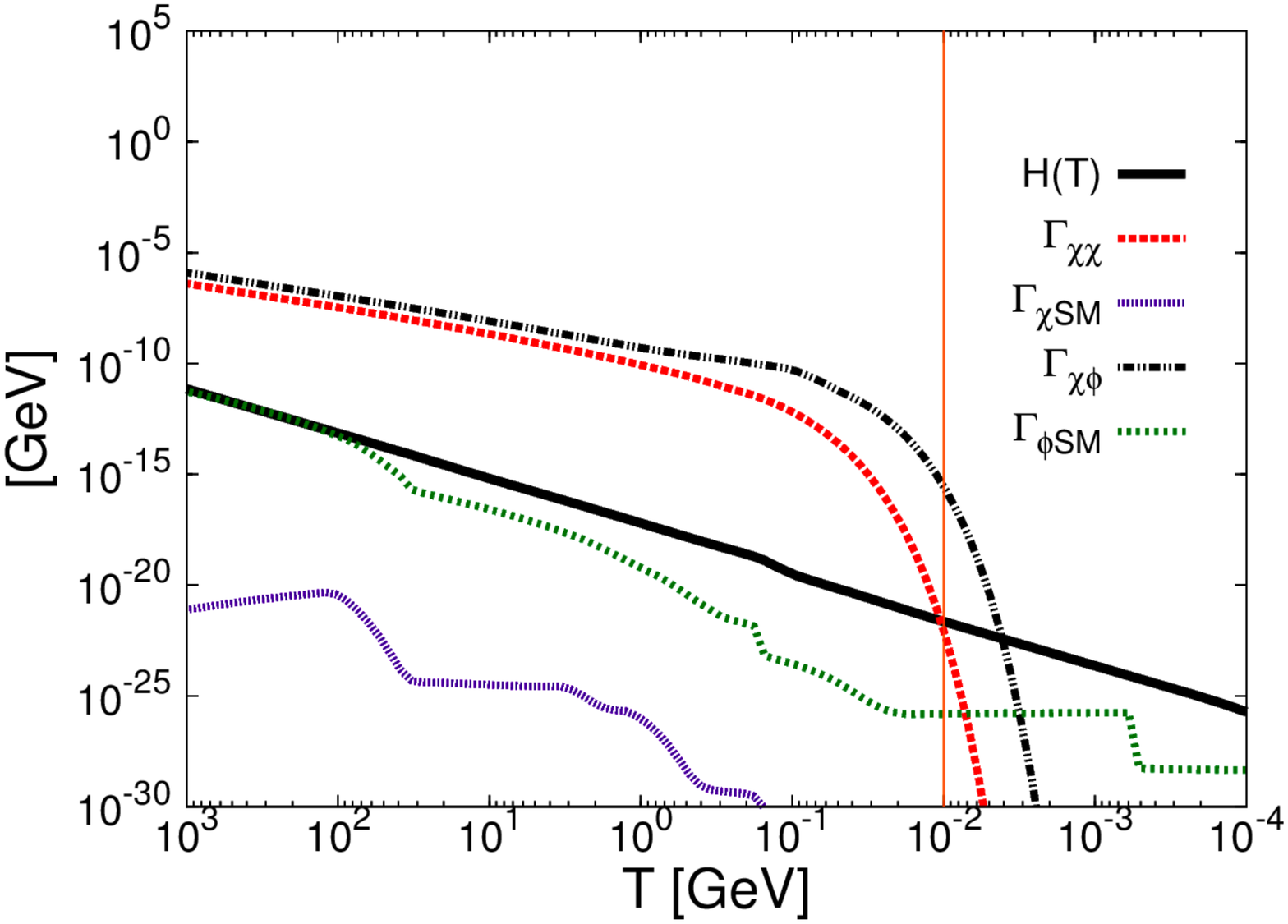}
	\includegraphics[height=2.2in,angle=0]{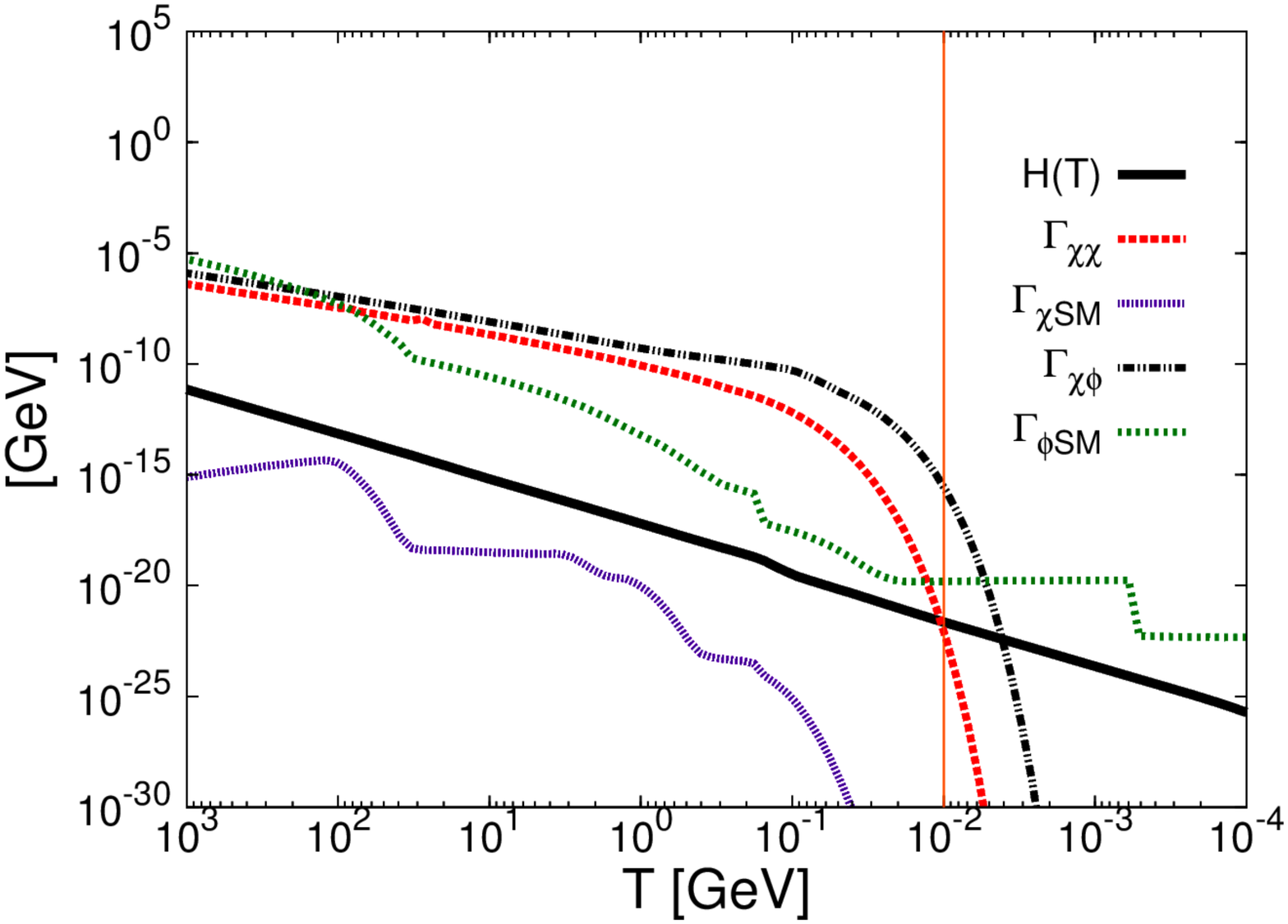}
	\caption{\small \sl Reaction rates $\Gamma_{\chi \chi}$, $\Gamma_{\chi {\rm SM}}$, $\Gamma_{\chi \phi}$ and $\Gamma_{\phi {\rm SM}}$ as well as $H(T)$ as a function of the temperature of the universe. Model parameters are fixed to be ($m_\chi$, $c_s$, $m_\phi$, $\sin \theta$, $\mu_3$) = (200\,MeV, 0.022, 100\,MeV, 10$^{-6}$, 10\,MeV) in the left panel, showing that the equilibrium condition is not satisfied. On the other hand, those are fixed to be (200\,MeV, 0.022, 100\,MeV, 10$^{-3}$, 10\,MeV) in the right panel, satisfying the condition. The freeze-out temperature is shown as a vertical (orange) line in both panels.}
	\label{fig: rates}
\end{figure}

In Fig.\,\ref{fig: rates}, the reaction rates $\Gamma_{\chi \chi}$, $\Gamma_{\chi {\rm SM}}$, $\Gamma_{\chi \phi}$ and $\Gamma_{\phi {\rm SM}}$ as well as the expansion rate of the universe $H(T_f)$ are depicted as a function of the temperature of the universe in two cases; one is for the model parameter set that does not satisfy the kinematically equilibrium condition (left panel) and the other is for that satisfying the condition (right panel). The freeze-out temperature is shown as a vertical (orange) line in both panels. See also the figure caption for the model parameters used to calculate the reaction rates. As we can see from the comparison between the results in the two panels, the equilibrium condition gives the lower limit on the mixing angle $|\theta|$. 
In order to manifest how the equilibrium condition works for the minimal WIMP model, 
the parameter region survived after imposing the equilibrium condition as well as the preselection criteria and the relic abundance condition 
is presented in appendix~\ref{app: equilibrium condition}.
All explicit forms of the reaction rates are also given there.

\subsubsection{Constraint from direct dark matter detections}
\label{subsubsec: direct detection}

Direct dark matter detection is known to be 
a stringent constraint 
for the WIMP based on the scattering between the WIMP and a nucleon. In the minimal WIMP model, the scattering occurs through the exchange of the mediator or the Higgs boson in the $t$-channel, and it contributes to the 
spin-independent scattering\,\cite{kimmo}. In our analysis, the scattering cross section is computed 
by using \texttt{MicrOMEGAs} code, where its explicit form is given by
\begin{equation}
	\sigma_{\rm SI} = c_s^2 \sin^2 \theta \cos^2 \theta
	\frac{m_\chi^2 m_N^4 f_N^2}{\pi v_H^2 (m_\chi + m_N)^2} \left( \frac{1}{m_\phi^2} - \frac{1}{m_h^2} \right)^2.
	\label{eq: sigma SI}
\end{equation}
Here, $m_N$ is the mass of a nucleon and $f_N = f_{Tu} + f_{Td} + f_{Ts} + (2/9) f_{TG}$ with $f_{Tu} \simeq 0.0153$, $f_{Td} \simeq 0.0191$, $f_{Ts} \simeq 0.0447$ and $f_{TG} \simeq 0.921$, respectively. Since $m_\phi \ll m_h$ is required to satisfy the relic abundance condition, the scattering process through the exchange of the mediator, namely the first term in the parenthesis of eq.~\eqref{eq: sigma SI}, dominates the cross section.

On experimental side, the most stringent constraint on the spin-independent scattering cross section is from XENON1T\,\cite{Aprile:2018dbl}, PANDAX\,\cite{Tan:2016zwf, Cui:2017nnn}, SuperCDMS\,\cite{Agnese:2015nto}, CRESST\,\cite{Petricca:2017zdp}, Darkside-50\,\cite{Agnes:2018ves} and NEWS-G\,\cite{Arnaud:2017bjh} experiments for the WIMP mass of our interest. On the other hand, in the near future, the constraint will be updated by LZ\,\cite{Akerib:2015cja, Szydagis:2016few}, SuperCDMS/SNOLAB\,\cite{Agnese:2016cpb} and NEWS-SNOLAB\,\cite{news1, news2} experiments, if no signal is detected. In particular, the NEWS experiment utilizes several gas detectors (e.g. Helium to Xenon) and will play an important role for the search with the mass less than a few GeV. For the present constraints (XENON1T, CRESST, and Darkside-50), we use the Poisson distribution likelihoods given by the \texttt{DDCalc} code\,\cite{Workgroup:2017lvb}. On the other hand, we involve the future constraints (SuperCDMS(SNOLAB), LZ, and NEWS-SNOLAB) assuming a half-Gaussian~\cite{Liu:2017kmx} 
form with the central value being set to be zero to figure out the future prospects of the light WIMP.

\subsubsection{Constraint from $N_{\rm eff}$ at $T_{\rm CMB}$}
\label{subsubsec: neff}

The mediator affects the expansion rate of the universe at the recombination era ($T_{\rm CMB} \simeq 4$\,eV\,\cite{Ade:2013zuv}) when it is lighter than the neutrino decoupling temperature, namely $m_\phi \lesssim T_D \simeq {\cal O}(1)$\,MeV\,\cite{Dolgov:2002wy}. This is because neutrinos are already decoupled from the thermal bath composed of photons and electrons when the temperature of the universe is below $T_D$, while a part of the entropy of the universe is still being carried by the mediator at $T \sim T_D$ and it is eventually injected into the two systems (neutrino and photon+electron systems) asymmetrically before the recombination according to the interaction of the mediator.\footnote{If the mediator is lighter than $T_{\rm CMB}$, it contributes directly to the expansion rate of the universe at the recombination era and this possibility is already excluded by the constraint on $N_{\rm eff}$ at the CMB observation.} In the minimal WIMP model, the mediator interacts only with electrons and photons when its mass is small enough, so that the entropy carried by the mediator goes into those particles. As a result, this injection contributes to the photon temperature of the universe, which makes the difference between the photon and neutrino temperatures larger, namely the expansion rate of the universe at the recombination era becomes smaller than usual.\footnote{One may think that the WIMP also affects the expansion rate if it is light. The WIMP mass is, however, constrained to be more than 10\,MeV due to the kinematical equilibrium condition and collider constraints as we will see later, so that the entropy carried by the WIMP is injected into the thermal bath at much above $T_D$.}

Such a contribution is severely constrained by the CMB observation, giving a lower limit on $m_\phi$ in order not to alter the effective number of relativistic degrees of freedom $N_{\rm eff}$\,\cite{Ade:2013zuv}. In the minimal WIMP model, the number $N_{\rm eff}$ at the recombination era is predicted as\,\cite{Ibe:2018juk}
\begin{eqnarray}
	N_{\rm eff} \simeq 3 \left( 1 + \frac{45}{11\pi^2} \frac{s_\phi (T_D)}{T_D^3} \right)^{-4/3},
	\qquad
	s_\phi(T_D) \equiv h_\phi(T_D) \frac{2\pi^2}{45} T_D^3,
	\label{eq: Neff}
\end{eqnarray}
where $h_\phi(T_D) = (15 x_\phi^4)/(4 \pi^4) \int^\infty_1 dy (4y^2-1) \sqrt{y^2 - 1}/(e^{x_\phi y} - 1)$ with $x_\phi\equiv m_\phi/T_D$ and $T_D$ being 3\,MeV\,\cite{Ibe:2018juk}. Here, we have assumed that the mediator is never chemically and kinematically decoupled from the thermal bath, 
and it is indeed satisfied in most of the parameter region of the minimal WIMP model. 
This is because kinematical equilibrium is maintained among all species at the freeze-out temperature through  the (inverse) decay process between the mediator and SM particles as well as the scattering process between the dark matter and the mediator, unless the mixing angle is vastly suppressed.\footnote{For instance, the mixing angle of ${\cal O}(10^{-5})$ is enough to maintain the kinematical equilibrium even if  the mediator is lighter than ${\cal O}(100)$\,MeV, and it is not excluded by collider constraints as seen in next section.} As a result, the mediator is in the equilibrium with SM particles below the freeze-out temperature.

On the other hand, in other parameter regions, the mediator is chemically decoupled from (but still maintains the kinematical equilibrium with) the thermal bath before the freeze-out of the WIMP and/or kinematically decoupled from (and eventually recoupled with) the thermal bath after the freeze-out. In these cases, the mediator can contribute to $N_{\rm eff}$ more than that discussed in eq.\,(\ref{eq: Neff}) and receives a severer constraint from the observation. We, however, keep using the formula\,(\ref{eq: Neff}) as a conservative constraint obtained from the CMB observation, and leave those precise computations for a future study.

The effective number of relativistic degrees of freedom is observed to be $N_{\rm eff} = 2.99 \pm 0.17$ by the Planck collaboration\,\cite{Aghanim:2018eyx}. The data could be improved by future observations, which resolves small angular scales (high multipole number $\ell$) of CMB and allows a more precise measurement of $N_{\rm eff}$ by observing its damping tail at high-$\ell$. According to their forecast, the measurement of $N_{\rm eff}$ can be improved by one order of magnitude at, for example, the CMB-S4 experiment\,\cite{Abazajian:2016yjj}, which leads to the expected limit of $\Delta N_{\rm eff} < 0.017$, if we do not see any deviation from the standard prediction. We involve these (expected) limits in our analysis for the present status and future prospects of the minimal WIMP model.

\subsubsection{Constraint from BBN}
\label{subsubsec: bbn}

If the mediator decays during or after the era of the Big Band Nuclear synthesis (BBN), it may spoil this successful scenario of generating light elements in the early universe. Hence, the mediator should be constrained so that it does not spoil the BBN scenario, which depends on its mass, lifetime, decay channels and abundance at the time that it is decaying.

For instance, if the mediator is heavier than $2m_\pi$ and 
mainly decays hadronically, it modifies the neutron-to-proton ratio and increases the primordial helium mass fraction, which gives a constraint on its lifetime as $\tau_\phi \lesssim {\cal O}(1)$ second\,\cite{Berger:2016vxi}. On the other hand, the mediator also affects the BBN scenario even when its mass is in the range of 4\,MeV $\leq m_\phi \leq 2m_\pi$ and thus decays leptonically. This is because a few MeV-electrons pass their energies to CMB photons through the inverse Compton scattering and generate MeV gammas, which disintegrate $^2$H and $^7$Be (which eventually forms $^7$Li). As a result, the leptonic decay produces less $^7$Li and more $^3$He/$^2$H than the usual case, leading to a constraint as $\tau_\phi \lesssim {\cal O}(10^5)$ seconds\,\cite{Berger:2016vxi}.\footnote{When the mediator is lighter than 4\,MeV, the energy of the electron produced by the mediator decay is not sufficient to photo-disintegrate nuclei, so that the constraint discussed here cannot be applied\,\cite{Hufnagel:2018bjp}.} The mediator may affect the BBN scenario by other mechanisms if it is very light. For instance, when $m_\phi \lesssim T_{\rm BBN} \sim 0.1$\,MeV, the mediator is relativistic during the BBN era and contributes to the expansion rate of the universe as an additional light degree of freedom. On the other hand, even when the mediator is in the mass range of $T_{\rm BBN} \lesssim m_\phi \lesssim T_D$, the mediator alters the expansion rate of the universe, because it changes the ratio between the photon and neutrino temperatures as discussed in section\,\ref{subsubsec: neff}.

In order to impose the BBN constraint on the minimal WIMP model in a rigorous way, we have to consider the cosmology of the mediator at the late universe\,\cite{Hufnagel:2018bjp, Fradette:2017sdd}\footnote{In the references, $\sin \theta$ is assumed to be so small that the absorption and decay processes do not work to maintain the chemical equilibrium. Hence, constraints derived there cannot be directly applied to our case.} and implement all the decay channels of the mediator in an appropriate BBN code, which is beyond the scope of this paper. Instead, we impose the following two constraints, $\tau_\phi \leq 1$ second when $m_\phi \geq 2m_\pi$ and $\tau_\phi \leq 10^5$ seconds when 4\,MeV $\leq m_\phi \leq 2m_\pi$, by a simple 1/0 logical cut in the likelihood analysis, which is indeed a reference constraint adopted in many literature (see, e.g. Ref.\,\cite{Krnjaic:2015mbs}).\footnote{We would also like to thank Dr. Sebastian Wild for fruitful discussion for this BBN constraint\,\cite{Hufnagel:2018bjp}.} We leave precise computations of the BBN constraints for a future study. On the other hand, we do not involve the BBN constraints concerning the expansion rate of the universe, for those in section\,\ref{subsubsec: neff}, 
$\Delta N_{\rm eff}$ had given severer constraint.

Let us briefly discuss how the BBN constraint works for the minimal WIMP model. The lifetime of the mediator is proportional to the mixing angle squared, and thus it gives the lower limit on the angle $\sin \theta$. As can be seen in Fig.\,\ref{fig: After AP RE KE constraints} of appendix\,\ref{subapp: RA and KE conditions}, the result on the $(m_\phi, \sin \theta)$-plane, no lower limit on the mixing angle is obtained by imposing preselection criteria as well as relic abundance and kinematical equilibrium conditions, so that the BBN constraint plays an important role to make the allowed model parameter space finite.

\subsubsection{Other constraints}

It is also possible to impose further cosmological and astrophysical constraints on the minimal WIMP model in addition to those discussed above. 
One of such constraints is from the distortion of the CMB spectrum due to the late time decay of the mediator. When the lifetime of the mediator is longer than $10^6$ seconds, where the double Compton scattering process ($\gamma\,e \to \gamma\,\gamma\,e$) is not active and the CMB cannot maintain the Planck distribution against an additional photon injection, the constraint on the so-called $\mu$-distortion parameter put an upper limit on the lifetime\,\cite{Hu:1992dc}. When the lifetime is longer than $10^9$ seconds, where the Compton process ($\gamma\,e \to \gamma\,e$) is also inactive and even the Bose-Einstein distribution cannot be maintained against the injection, the constraint on the $y$-distortion parameter put the upper limit on the lifetime\,\cite{Masso:1997ru}. 
These limits are, however, weaker than CMB and BBN constraints 
in Sec.~\ref{subsubsec: neff} and~\ref{subsubsec: bbn}. Hence, we do not include the constraints from the distortion of the CMB spectrum due to the late time decay of the mediator in our analysis.

The other possible constraint can be obtained by the observation of neutrinos from the supernova (SN) 1987A\,\cite{Hirata:1987hu}. Physics behind the constraint is as follows. The collapse at the SN heats up the core and its energy is estimated to be ${\cal O}(10^{59})$\,MeV, so that the cooling rate must be, at least, smaller than this value. The neutrino emission during a supernova explosion is known 
to be the only mechanism to cool down the core within the SM. On the other hand, the mediator can also contribute to the cooling in the minimal WIMP model. The dominant process is the nucleon scattering, $NN \to NN\phi$, followed by the $\phi$ decay into SM particles. It is then possible to put an upper limit on the mixing angle $\sin\theta$, 
such that the instantaneous luminosity in novel particles does not exceed the value $\mathcal{O}(10^{52})$\,erg/s. 
When $\sin\theta$ is too large, the parameter space is allowed because 
$\phi$ decays or is trapped inside the core and not contribute to the cooling. 
As a result, the SN physics has a potential to put a constraint on a certain region of $m_\phi$ and $\sin \theta$.

To include SN1987a cooling constraints into our analysis, 
it can be a highly non-trivial task because
SN is a complicated physics system and the constraint from it suffers from several uncertainties\,\cite{Chang:2016ntp, Chang:2018rso, Mahoney:2017jqk,Fischer:2016cyd,Tu:2017dhl,Sung:2019xie}. For example, the nature of the core of protoneutron star to the primary driver of the shock revival, the temperature, the density profiles, and equation of state of the progenitor star and the cross section of various QCD processes with soft radiations and environment effects to the process are known to give such uncertainties. Unfortunately, none of studies concerning the constraint is found to include all the uncertainties, and such a comprehensive study of the constraint is beyond the scope of our study. Hence, we did not include the constraint in our likelihood analysis, while we present the SN1987A exclusion contour on top of our result in the $(m_\phi,\sin\theta)$-plane in section \ref{subsec: status}.

Finally, we would like to comment on the cross section of the dark matter self-scattering process ($\chi\chi\to\chi\chi$). Since the mediator $\phi$ in this model can be a thousand lighter than the dark matter and leads to an interesting enhancement of the dark matter self-scattering cross section. 
This enhancement becomes significant and velocity-dependent when the incident dark matter velocity is as small as $10^{-3}$ because of the diagram exchanging a light mediator in the $t$-channel. On the other hand, the velocity-dependence disappears when the velocity is larger than ${\cal O}(10^{-2})$ and the self-scattering cross section is approximately given as\,\cite{Tulin:2013teo}
\begin{eqnarray}
    \sigma_T(\chi\chi \to \chi\chi) \simeq 5\times 10^{-23}\,\left( \frac{c_s^2}{0.1} \right)^2
    \left( \frac{m_\chi}{10\,{\rm GeV}} \right)^2\,\left(\frac{10\,{\rm MeV}}{m_\phi} \right)^4\,{\rm cm^2}.
\end{eqnarray}
Phenomenologically, the velocity-dependent self-scattering at $v \sim {\cal O}(10^{-3})$ could provide a solution to the ``small scale crisis''\,\cite{Weinberg:2013aya}, namely the so-called core-cusp problem of dwarf galaxies can be solved by the self-scattering. On the other hand, a robust constraint on the self-scattering with $v \sim 10^{-2}$ is obtained from various clusters of galaxies\,(see, eg., Ref.\,\cite{Tulin:2017ara}). We have checked that all parameter points satisfying all conditions and constraints imposed in this paper are also consistent with the constraint on the self-scattering, and thus we do not include this self-scattering constraint in our likelihood analysis.

\subsection{Constraints from collider experiments}
\label{subsec: collider constraints}

In this subsection, we summarize collider constraints that we have considered for our likelihood analysis. Since the WIMP couples to SM particles mainly through the mediator while the mediator couples to several SM particles, the constraints on the minimal WIMP model are mostly from mediator productions at various colliders.\footnote{Only exception comes from the interaction between the WIMP and the Higgs boson. This interaction is constrained by observing the invisible decay width of the Higgs boson at colliders, as seen in section\,\ref{subsubsec: higgs decay}.} Collider signals then depend on how the mediator is produced and how it decays, depending on its mass $m_\phi$ and the mixing angle $\sin \theta$. We will discuss below the constraints based on the production processes. All the collider constraints that we have involved in our analysis are summarized in Table\,\ref{tab: collider constraints}.

\begin{table}[t!]
	\centering
	\begin{tabular}{ l | l l r }
	& Present & Future & Section \\
	\hline
	$\Upsilon$ decay
	& CLEO\,\cite{cleo_upsilon}, BABAR\,\cite{babar_upsilon0,babar_upsilon} & Belle\,II\,\cite{akimasa}
	& \ref{subsubsec: upsilon decay} \\
	$B$ decay
	& Belle\,\cite{belle, b_invisible_belle}, LHCb\,\cite{lhcb, Aaij:2015tna, Aaij:2016qsm},
	& Belle\,II\,\cite{Aushev:2010bq, b_invisible_belle II},
	& \ref{subsubsec: b decay} \\
	& BaBar\,\cite{Lees:2015rxq, babar, b_invisible_babar1, b_invisible_babar2} & LHCb\,\cite{Albrecht:2017odf} & \\
	$K$ decay
	& N48/2\,\cite{n48_2}, KTeV\,\cite{ktev1, ktev2}, E949\,\cite{E949},
	& NA62\,\cite{Martellotti:2015kna, Koval:2016hml} ,
	& \ref{subsubsec: k decay} \\
	& CHARM\,\cite{charm_exp, Bezrukov:2009yw}, KEK E391a\,\cite{Ahn:2009gb}
	& SHiP\,\cite{Alekhin:2015byh}, KOTO\,\cite{Beckford:2017gsf} & \\
	$H$ decay
	& LHC\,\cite{Khachatryan:2017mnf, Aaboud:2018fvk, Clarke:2015ala, Cheung:2014noa}
	& HL-LHC\,\cite{CMS:2013xfa, Bechtle:2014ewa}
	& \ref{subsubsec: higgs decay} \\
	Direct
	& LEP\,\cite{Acciarri:1996um} & -- -- --
	& \ref{subsubsec: direct production} \\
	\hline
	\end{tabular}
	\caption{\small \sl Collider constraints that we have imposed in our analysis. See also the caption of Table\,\ref{tab: apriori constraints}.}
	 \label{tab: collider constraints}
\end{table}

\subsubsection{Upsilon decay}
\label{subsubsec: upsilon decay}

Upsilons can produce the mediator through their decays. The most important decay channels are $\Upsilon(1S, 2S, 3S) \to \gamma\,\phi$ followed by the $\phi$ decay into a lepton pair such as $\mu^- \mu^+$ and $\tau^- \tau^+$. 
These upsilon decays allow us to detect the mediator through a narrow peak search in the di-lepton invariant mass spectrum at the mass region of $m_\phi < m_\Upsilon \simeq 10$\,GeV. These channels have been searched for by CLEO\,\cite{cleo_upsilon} and BaBar\,\cite{babar_upsilon0,babar_upsilon} collaborations. Since no signal has been detected so far, they put an upper limit on the following quantity:
\begin{equation}
	R_0 = \frac{ {\rm Br}(\Upsilon \to \gamma \phi)\,{\rm Br}(\phi \to \ell \ell) } { {\rm Br}(\Upsilon \to \mu^- \mu^+) } =
	\sin^2 \theta \frac{G_F m_b^2} {\sqrt{2} \pi \alpha}
	\left(1 - \frac{m_\phi^2}{m_\Upsilon^2} \right)\,{\rm Br}(\phi \to \ell \ell)\,{\cal F}_{\rm QCD}.
\end{equation}
$G_F$, $m_b$ and $\alpha$ are the Fermi constant, the bottom quark mass and the fine structure constant, respectively. The parameter ${\cal F}_{\rm QCD}$ takes a value between 0.5 to 1.5, 
originated from QCD bound state and relativistic corrections.
The branching fraction of the decay channel $\Upsilon \to \mu^- \mu^+$ have been measured precisely\,\cite{pdg}. Then, an upper limit on ${\rm Br}(\Upsilon \to \gamma \phi) \times {\rm Br}(\phi \to \ell \ell)$ is obtained at 90\% confidence level, depending on the mediator mass $m_\phi$\,\cite{cleo_upsilon, babar_upsilon0, babar_upsilon}. For instance, an upper limit on ${\rm Br}(\Upsilon \to \gamma \phi) \times {\rm Br}(\phi \to \mu^- \mu^+) \lesssim 3\times10^{-6}$ is obtained in the mass region of $m_\phi \lesssim 3.5$\,GeV, while ${\rm Br}(\Upsilon \to \gamma \phi) \times {\rm Br}(\phi \to \tau^- \tau^+) \lesssim 3 \times10^{-5}$ is obtained for 4\,GeV $\lesssim m_\phi \lesssim$ 8.5\,GeV. The limit becomes weaker when $m_\phi$ approaches to $m_\Upsilon$.

The mediator is assumed to promptly decay into a lepton pair in the above analysis. One might suspect that this method does not work because the mixing angle may become so small that the mediator is too long-lived. 
However, the method indeed works well but its reason is a bit complicated. 
We take the mediator mass to be $m_\phi = 220$\,MeV as an example, 
just above the $\mu^- \mu^+$ threshold giving the longest lifetime to make $\phi$ decay into $\mu^- \mu^+$ at each mixing angle.\footnote{Needless to say , the lifetime (thus, the decay length) becomes shorter if the mediator mass is heavier.} The decay width of the mediator is then estimated to be $\Gamma_\phi \simeq 10^{-9} \times \sin^2\theta$\,GeV. The constraint ${\rm Br}(\Upsilon\to \gamma \phi) \times {\rm Br}(\phi \to \mu^- \mu^+) \lesssim 3\times10^{-6}$ is 
then translated to $\sin \theta \lesssim 0.14$, so that the decay length of the mediator produced by the $\Upsilon$ decay is estimated to be $\gamma_\phi\,\tau_\phi\,c = {\cal O}(0.1)$\,mm 
with the speed of light $c$. Here, the so-called gamma factor is estimated to be $\gamma_\phi \simeq m_\Upsilon/(2 m_\phi) \simeq 25$. This decay length is much shorter than the present sensitivity of displaced vertex searches, which requires ${\cal O}(1)$\,cm\,\cite{Lees:2015rxq}.

The Belle\,II experiment will update the constraints in the near future with the integrated luminosity about fifty times more than that of the BaBar experiment, if no signal is detected there\,\cite{akimasa}. 
The constraints on the branching fractions will be about six times more severer than the present ones. 
As a result, for instance, the constraints will be ${\rm Br}(\Upsilon \to \gamma \phi) \times {\rm Br}(\phi \to \mu^- \mu^+) \lesssim 5\times 10^{-7}$ and ${\rm Br}(\Upsilon \to \gamma \phi) \times {\rm Br}(\phi \to \tau^- \tau^+) \lesssim 5 \times 10^{-6}$. Note that the decay length of the mediator is, at most, ${\cal O}(1)$\,mm even in the near future, so that its decay can be treated as a prompt one. 
We include all the constraints discussed here in our likelihood. 
It, however, turns out that the present constraints are not stronger than others. 
Those from $B (K)$ decays are more sensitive in the mass region of $m_\phi < m_B (m_K)$ with $m_B (m_K)$ being the $B (K)$ meson mass, while the LEP experiment gives a stronger constraint in $m_B < m_\phi < 9.2$\,GeV, as we will see in following discussions. On the other hand, the future constraints could play an important role to search for the mediator in a certain mass region of $\phi$.

\subsubsection{$B$ meson decay}
\label{subsubsec: b decay}

When the mediator $\phi$ is lighter than $B$ mesons ($m_B \simeq 5.3$\,GeV), it can be produced by their decays through the sub-process $b \to s \phi$, where it is induced mainly from the one-loop diagram composed of weak bosons and various quarks in the loop\,\cite{Krnjaic:2015mbs}. Among various $B$ meson decays, one of the efficient channels to search for the mediator at collider experiments is the charged $B$ meson decay, $B^{\pm} \to K^{\pm} \phi$. Its decay width is estimated to be\,\cite{kl_theory}:
\begin{eqnarray}
	\Gamma (B^\pm \to K^\pm \phi) =
	\frac{ |C_{sb}|^2\,F^2_K (m_\phi) } { 16 \pi m_B^3 }
	\left( \frac{ m_B^2 - m_K^2 } { m_b - m_s} \right)^2
	\sqrt{ (m_B^2 - m^2_K - m_\phi^2 )^2 - 4 m_K^2 m_\phi^2 },
	\label{eq: BtoKphi}
\end{eqnarray}
where the scalar form factor of $K^\pm$ is estimated to be $F_K (q) \simeq 0.33\,(1 - q^2/38\,{\rm GeV}^2)^{-1}$, while the coefficient of the FCNC effective interaction, $C_{sb}\,\bar{s}_L\,b_R\,\phi + h.c.$, is given by $|C_{sb}| \simeq |2 g^2 m_b m_t^2 V^*_{ts} V_{tb} \sin \theta|/(64\pi^2 m_W^2 v_H) \simeq 6.4\times 10^{-6} \sin \theta$ with $V_{tb}$ and $V_{ts}$ being the CKM matrix elements. Combined with the total decay width of the $B$ meson, $\Gamma_{B^+} \simeq 4.1\times 10^{-13}$\,GeV, the above formula allows us to compute the branching fraction of the decay channel $B^{\pm} \to K^{\pm} \phi$. It is also possible to consider other decay channels for the $\phi$ search such as $B^0 \to K^{*0} \phi$. Collider signals then depend on how the mediator decays and how long distance it travels before the decay. 
In this subsection, all experimental results of $B$ meson decay at present and expected null signal results in the near future 
are discussed.

\subsubsection*{$B$ meson decay with prompt $\phi$ decay}

When the decay length of the mediator is enough smaller than the detector size, its decay can be treated as a prompt one. In such a case, the decay channel $B^{\pm} \to K^{\pm} \phi \to K^{\pm} \mu^- \mu^+$ is frequently used to detect the mediator, as it gives a clean signal against backgrounds. At present, BaBar\,\cite{babar}, Belle\,\cite{belle} and LHCb\,\cite{lhcb} collaborations give upper limits on its branching fraction as $P_p\,{\rm Br}(B^{\pm} \to K^{\pm} \phi)\,{\rm Br}( \phi \to \mu^- \mu^+ ) \lesssim 3 \times 10^{-7}$, accepting that the theoretical prediction of the SM contribution is ${\rm Br} (B^\pm \to K^\pm \mu^- \mu^+)_{\rm SM} = (3.5 \pm 1.2) \times 10^{-7}$\,\cite{b_theory}. Here, the prefactor $P_p$ is the probability that the mediator decays inside the detector,
\begin{eqnarray}
	P_p \equiv \frac{1}{2} \int_0^\pi d\theta_\phi\,\sin \theta_\phi
	\left( 1 - \exp \left[ - \frac{l_{xy}}{\sin \theta_\phi} \frac{1}{\gamma \beta c \tau_\phi } \right] \right),
	\label{eq: definition of Pp}
\end{eqnarray}
where $\tau_\phi$ is the lifetime of the mediator, while $\theta_\phi$ is the angle between the direction of the mediator produced from the $B$ decay and that of the beampipe. The boost factor is estimated to be $\gamma \beta \simeq m_B/(2m_\phi)$, as the mediator is produced from the $B$ decay. The size of the detector is set to be $l_{xy} \simeq 25$\,cm referencing to those of Belle and BaBar detectors\,\cite{charm_eq}.

%
The theoretical error on the SM contribution overwhelms the experimental errors 
due to the uncertainty of the form factor in eq.\,(\ref{eq: BtoKphi}). 
Therefore, a significant improvement of the sensitivity is not expected even if more data is accumulated at e.g. the upcoming SuperKEKB experiment\,\cite{Aushev:2010bq}. 
On the other hand, if the theoretical error is much reduced, thanks to the development of lattice QCD simulations and/or 
the use of the characteristic feature coming from the narrow dilepton invariant mass peak from the mediator decay, 
the analysis can have a further improvement.
Since no convincing study on the issues has been performed yet, 
we do not involve the constraint from the prompt $B^{\pm} \to K^{\pm} \phi \to K^{\pm} \mu^- \mu^+$ decay to 
investigate the future prospects of the minimal WIMP model and involve it only for its present status.

\subsubsection*{$B$ meson decay with displaced $\phi$ decay}

When the decay length of the mediator $\phi$ becomes comparable to the detector size, displaced vertex is the most powerful channel to search for $\phi$. At present, BaBar\,\cite{Lees:2015rxq} and LHCb\,\cite{Aaij:2015tna, Aaij:2016qsm} collaborations give various stringent constraints. The BaBar collaboration has looked for the displaced vertex caused by decay channels $\phi \to e^- e^+$, $\mu^- \mu^+$, $\pi^- \pi^+$ and $K^- K^+$ by observing the invariant mass spectrum of the decay products, where the mediator is produced mainly from $B$ decays, $B \to X_s \phi$ with $X_s$ being a hadronic system with a strangeness. 
Since no excess over the SM prediction has been observed, 
they put a constraint on ${\rm Br}(B \to X_s \phi)\,{\rm Br}( \phi \to e^- e^+, \mu^- \mu^+, \pi^- \pi^+, K^- K^+)$, 
which is translated to that on the mixing angle, $\sin^2\theta \gtrsim 2 \times 10^{-8}$ at 90\% C.L., 
almost regardless of $m_\phi$ and $\tau_\phi$ in the region of 0.5\,GeV $\leq m_\phi \leq$ 1.5\,GeV 
and 1\,cm $\leq c \tau_\phi \leq$ 20\,cm. We adopt it in our likelihood analysis. 
More stringent limits on the mixing angle for regions $m_\phi \geq 1.5$\,GeV and $c \tau_\phi \geq 20$\,cm 
is obtained at LHCb (see below) and beam dump experiments, respectively.

On the other hand, the LHCb collaboration is recently searching for the mediator produced by both charged and neutral $B$-meson decays: $B^\pm \to K^\pm\,\phi \to K^{\pm}\,\mu^- \mu^+$ and $B^0 \to K^{*0}\,\phi \to K^{*0}\,\mu^- \mu^+$. Their latest report based on the negative result of the search using 3\,fb$^{-1}$ data at 7\,TeV and 8\,TeV running is found in Ref.\, \cite{Aaij:2016qsm}. We also involve this result in our analysis, where it gives a stringent constraint in the region of $m_\phi \geq 1.5$\,GeV.

In the near future, Belle\,II collaboration will update the constraint (given by the BaBar collaboration) with 100 times more data (50\,ab$^{-1}$)\,\cite{Aushev:2010bq}, if no signal is detected. On the other hand, the LHCb collaboration will accumulate data during the 13\,TeV running, where 300 times more $B$ mesons will be in data compared to the present one\,\cite{Albrecht:2017odf}. For the near future prospects in our analysis, we therefore adopt the BaBar and LHCb constraints mentioned above with their sensitivities on the mediator searches increased 10 and 17 times.

\subsubsection*{$B$ meson decay with very long-lived $\phi$ decay}

When the decay length of $\phi$ is much longer than the detector size, it is searched by utilizing the channel, $B^{\pm} \to K^{\pm} + {\rm missing}$, where the SM process, $B^\pm \to K^{\pm} \nu \bar{\nu}$, is a background against the signal. This channel is also used to search for the mediator decaying invisibly, $B^\pm \to K^\pm \phi \to K^\pm \chi \chi$, even if it is short-lived. At present, Belle\,\cite{b_invisible_belle} and BaBar\,\cite{b_invisible_babar1, b_invisible_babar2} collaborations put a constraint as $P_l\,{\rm Br}(B^{\pm} \to K^{\pm} \phi) + P_p\,{\rm Br}(B^{\pm} \to K^{\pm} \phi)\,{\rm Br}(\phi \to \chi \chi) \leq 1.6 \times 10^{-5}$. The factor $P_l$ is the probability that $\phi$ decays outside the detector,
\begin{eqnarray}
	P_l \equiv \frac{1}{2} \int_0^\pi d\theta_\phi\,\sin \theta_\phi
	\exp \left[ - \frac{l_{xy}}{\sin \theta_\phi} \frac{1}{\gamma \beta c \tau_\phi } \right],
	\label{eq: Pl}
\end{eqnarray}
where the values of $l_{xy}$ and $\gamma \beta$ are the same as those in eq.\,(\ref{eq: definition of Pp}). 
In future, the Belle\,II collaboration will update the sensitivity as $P_l\,{\rm Br}(B^{\pm} \to K^{\pm} \phi) + P_p\,{\rm Br}(B^{\pm} \to K^{\pm} \phi)\,{\rm Br}(\phi \to \chi \chi) \leq 5 \times 10^{-7}$\,\cite{b_invisible_belle II}. We include this constraint in our analysis for the near future prospects.

\subsubsection{Kaon decay}
\label{subsubsec: k decay}

The mediator is also produced from Kaon decays if it is lighter than Kaons ($m_K \simeq 0.5$\,GeV), which is induced from the sub-process $s \to d\,\phi$. The most efficient channels to search for $\phi$ are $K^\pm \to \pi^\pm + \phi$ and $K_L \to \pi^0 + \phi$, where their decay widths are given as follows\,\cite{Krnjaic:2015mbs}:
\begin{eqnarray}
	\Gamma(K^\pm \to \pi^\pm\,\phi) &=& \frac{ |C_{ds}|^2 } { 16 \pi m_{K^\pm}^3 }
	\left( \frac{ m_{K^\pm}^2 - m_{\pi^\pm}^2 } { m_s - m_d } \right)^2
	\sqrt{ (m_{K^+}^2 - m_{\pi^+}^2 - m_\phi^2 )^2 - 4 m_{\pi^+}^2 m_\phi^2 },
	\\
	\Gamma(K_L \to \pi^0 \phi) &=& \frac{ ({\rm Im}\,C_{ds})^2 } { 16 \pi m_{K_L}^3 }
	\left( \frac{ m_{K_L}^2 - m_{\pi^0}^2 } { m_s - m_d } \right)^2
	\sqrt { (m_{K_L}^2 - m_{\pi^0}^2 - m_\phi^2 )^2 - 4 m_{\pi^0}^2 m_\phi^2 }.
	\label{eq: K decays}
\end{eqnarray} 
Since scalar form factors for pions are close to unity\,\cite{Marciano:1996wy}, we neglect those in the above equations. On the other hand, the coefficient $C_{ds}$ comes from the FCNC effective interaction, $C_{sd}\,\bar{s}_L\,d_R\,\phi + h.c.$, where $C_{sd} \simeq (2 g^2_W m_s m_t^2 V^*_{ts} V_{td} \sin \theta)/(64 \pi^2 m_W^2 v_H) \simeq (1.2 + 0.5 i) \times 10^{-9} \sin \theta$ with 
the CKM matrix elements $V_{ts}$ and $V_{td}$. 
By given the total decay widths of the Kaons, $\Gamma_{K^\pm} = 5.3 \times 10^{-17}$\,GeV and $\Gamma_{K_L} = 1.286 \times 10^{-17}$\,GeV, we can compute their branching fractions. The Kaon decays are then followed by the $\phi$ decay, and the decay channels $\phi \to \mu^- \mu^-$, $e^- e^+$ are used, as it gives a clean signal and has less theoretical uncertainties than others.

\subsubsection*{Kaon decay with prompt $\phi$ decay}

When the mediator decays promptly, the most stringent limit on the branching fraction of the charged Kaon decay, $K^\pm \to \pi^\pm \phi \to \pi^\pm \mu^- \mu^+$, is put by the N48/2 collaboration\,\cite{n48_2}. Adopting the SM contribution, ${\rm Br}(K^+ \to \pi^+ \mu^- \mu^+)_{\rm SM} \simeq (8.7 \pm 2.8) \times 10^{-8}$, they give a limit as $P_p'\,{\rm Br}(K^+ \to \pi^+ \phi)\,{\rm Br}(\phi \to \mu^- \mu^+) \leq 4 \times 10^{-8}$ at 90\% C.L. with $P_p'$ being the probability that the mediator decays within the range of the longitudinal vertex resolution $\sigma_z \simeq 100$\,cm\,\cite{charm_eq}, namely $P_p' \equiv 1-{\rm exp}[-\sigma_z/(\gamma \beta c \tau_\phi)]$. Here, the boost factor is estimated to be $\gamma \beta \simeq 120$, because the mediator is produced by the Kaon decay with its momentum of 60\,GeV.

On the other hand, the KTeV collaboration\,\cite{ktev1, ktev2} put the other constraints using neutral Kaon decays as $P_p'\,{\rm Br}(K_L \to \pi^0 \phi)\,{\rm Br}(\phi \to e^- e^+,\,\mu^- \mu^+) \leq 2.8 \times 10^{-10},\, 3.8 \times 10^{-10}$ at 90\% C.L. with $\sigma_z$ and $\gamma \beta$ in $P_p'$ being replaced by those of the KTeV experiment, $\Delta l = 4$\,mm and $\gamma \beta \simeq 1$\,\cite{kl_theory}. Since SM predictions are ${\rm Br}(K_L \to \pi^0 \mu^- \mu^+) \sim {\rm Br}(K_L \to \pi^0 e^- e^+) \sim 3 \times 10^{-11}$, those have not been observed at the experiment. We include all the above constraints in our likelihood analysis for the present status of the minimal WIMP model.

In the near future, an improvement of the sensitivity on the search with the $K^\pm$ decay is not expected, for the systematic error at the scalar form factor already dominates. Moreover, there is no successor of the KTeV experiment, so that 
the constraints from the neutral Kaon decays are not expected to have any improvement in the near future.
We hence do not consider the constraints relevant to the above prompt Kaon decays in our likelihood analysis for the near future prospects of the minimal WIMP model.

\subsubsection*{Kaon decay with displaced $\phi$ decay}

When $\phi$ decays with the decay length of ${\cal O}(100)$\,m, it is efficiently searched at proton beam dump experiments. At present, the CHARM collaboration gives the stringent constraint on this search\,\cite{charm_exp}, where many Kaons and $B$ mesons are produced from the 400\,GeV proton beam which is dumped into the copper target. Then, the mediator is expected to be promptly produced from the channels, $K^\pm \to \pi^\pm \phi$, $K_L \to \pi^0 \phi$ and $B \to X_s \phi$. At the experiment, the detector is located at 480\,m away from the target with its size of 35\,m, and $\phi$ penetrating the wall is searched for with leptonic decay channels $\phi \to e^- e^+$ and $\mu^- \mu^+$.

Since zero signal event was observed as expected by background, 
the collaboration put a constraint on the number of signal events as $N_{\rm dec} \leq 2.3$ at 90\% C.L.\,\cite{Bezrukov:2009yw}. The total number of signal events at the experiment is estimated by\,\cite{Winkler:2018qyg}
\begin{eqnarray}
    N_{\rm dec} = N_{\rm p.o.t}
    \left[
        P^K_{\rm dec} \frac{\ell_H}{c \gamma_K \tau_K} n_K {\rm Br}(K \to \pi \phi) + P^B_{\rm dec} n_B {\rm Br}(B\to X_s \phi)
    \right],
\end{eqnarray}
where $N_{\rm p.o.t}$ is the number of protons on the target, and $n_{K(B)}$ is the number of $K(B)$ meson created per each incoming proton. Since Kaons are long-lived and absorbed in the target, the factor $\ell_H/(c \gamma_K \tau_K)$ is multiplied to its contribution with $\gamma_K$, $\tau_K$ and $\ell_H$ being the Lorentz factor, lifetime of the $K$ meson and the hadronic absorption length, respectively. The factors $P_{\rm dec}^K$ and $P_{\rm dec}^B$ are probabilities that the mediator decays inside the detection region,
\begin{eqnarray}
    P^{K(B)}_{\rm dec} = \eta^{K(B)}_{\rm geom} \eta^{K(B)}_{\rm rec}
    \left(
        -\exp\left[ -\frac{L_2}{\gamma \beta c \tau_\phi} \right] + \exp\left[ -\frac{L_1}{\gamma \beta c \tau_\phi} \right]
    \right),
	\label{eq: Pdec}
\end{eqnarray}
with $\eta^{K(B)}_{\rm geom}$ and $\eta^{K(B)}_{\rm rec}$ are the geometric and reconstruction efficiencies, respectively, assuming those are independent along the beam line. In the CHARM experiment, the values of the above parameters are $N_{\rm p.o.t} = 2.4 \times 10^{18}$, $n_K = 0.9$, $n_B = 3.2 \times 10^{-7}$, $\ell_H = 15.3$\,cm, $\eta^K_{\rm geom} \simeq 0.002$, $\eta^B_{\rm geom}\simeq 0.006$, $\eta^{K (B)}_{\rm rec} \simeq 0.5$, $\gamma_K \simeq 20\,{\rm GeV}/m_K$, $\gamma \beta m_\phi \simeq 10$\,GeV and $L_1 = L_2 - 35 = 480$\,m (the location of the detector), respectively. This constraint is included in our likelihood analysis to investigate the present status of the minimal WIMP model.

In the near future, the above constraint will be improved by the SHiP experiment at CERN SPS\,\cite{Alekhin:2015byh} if no signal is detected, where the 400\,GeV proton beam is dumped to the fixed target and the expected number of protons on the target is $N_{\rm p.o.t}=2 \times 10^{20}$. 
The SHiP detector is located at 69\,m away from the target with its size of 51\,m, so that the probability that the mediator decays inside the detection region is given by $P_{\rm dec}$ in eq.\,(\ref{eq: Pdec}) with $\eta^{K}_{\rm geom}\simeq 0.065$, $\eta^{B}_{\rm geom} \simeq 0.35$, $\gamma \beta m_\phi \simeq 25$\,GeV and $L_1 = L_2 - 51\,{\rm m} = 69$\,m. 
Reconstruction efficiencies below (above) two muon threshold are $\eta^{K,B}_{\rm rec}\simeq 0.4 (0.7)$. 
Then, the constraint will be updated as $N_{\rm dec} \leq 3$ at 95\% C.L., if no signal is detected there. 
Since the SHiP experiment can detect the decay channels of the mediator into $\pi\pi$ and $KK$, 
it will have more sensitivity for a heavy mediator ($\lesssim 4$\,GeV) than that of the CHARM experiment. We include this future-expected limit to study the future prospects of the minimal WIMP model.

\subsubsection*{Kaon decay with very-long lived $\phi$ decay}

When the lifetime of the mediator is very long, it is searched through the process, $K^\pm \to \pi^\pm + {\rm missing}$, where the SM process, $K^\pm \to \pi^\pm \nu \bar{\nu}$, becomes a background against the signal. In addition to the signal channel, $K^\pm \to \pi^\pm\,({\rm long-lived}\,\phi)$, another channel, $K^\pm \to \pi^\pm \phi \to \pi^\pm \chi \chi$, also contributes to the signal when the mediator is twice heavier than the WIMP, as in the case of the $B$ meson decay described at the last paragraph of section\,\ref{subsubsec: b decay}. At present, the E949 collaboration put a stringent constraint on the branching fraction $P_l\,{\rm Br}(K^+ \to \pi^+ \phi) + (1 - P_l)\,{\rm Br}(K^+ \to \pi^+ \phi)\,{\rm Br}(\phi \to \chi \chi)$, where $P_l$ is defined in eq.\,(\ref{eq: Pl}) with the size of the detector and the boost factor being replaced by $l_{xy} \simeq 145$\,cm and $\gamma \beta \simeq 1$, respectively. The constraint on the branching fraction at 90\% C.L. is found in Fig.\,18 of Ref.\,\cite{E949}.

On the other hand, such a very long-lived mediator is also searched by using the neutral Kaon decay, $K_L　\to \pi^0 + {\rm missing}$. Then, the SM process, $K_L \to \pi^0 \nu \bar{\nu}$, becomes a background against the signal. At present, the KEK E391a experiment put a constraint on the branching fraction as $P_l\,{\rm Br}(K_L \to \pi^0 \phi) + (1 - P_l)\,{\rm Br}(K_L \to \pi^0 \phi)\,{\rm Br}(\phi \to \chi \chi) \leq 2.6\times 10^{-8}$ with $l_{xy}$ and $\gamma \beta m_\phi$ used in $P_l$ being replaced by $l_{xy} \simeq 1$\,m and $\gamma \beta m_\phi \simeq 1$\,GeV, respectively\,\cite{Ahn:2009gb}. We involve the above two constraints from the charged and neutral Kaon decay experiments in our likelihood analysis to investigate the present status of the minimal WIMP model.
 
In the near future, the constraint from the charged Kaon decay will be improved by the NA62 experiment as $P_l'\,{\rm Br}(K^+ \to \pi^+ \phi) + (1 - P_l')\,{\rm Br}(K^+ \to \pi^+ \phi)\,{\rm Br}(\phi \to \chi \chi) \lesssim 10^{-11}$ if no signal is detected, where $P_l' = \exp[-l_z/(\gamma \beta c \tau_\phi)]$ with $l_z = 65$\,cm and $\gamma \beta m_\phi = 37.5$\,GeV\,\cite{Martellotti:2015kna, Koval:2016hml}. On the other hand, the constraint from the natural Kaon decay will be improved as $P_l\,{\rm Br}(K_L \to \pi^0 \phi) + (1 - P_l)\,{\rm Br}(K_L \to \pi^0 \phi)\,{\rm Br}(\phi \to \chi \chi) \lesssim 1.46 \times 10^{-9}$ at the KOTO experiment with $P_l$ being the same as that for the KEK E391a experiment\,\cite{Beckford:2017gsf}, which is close to the so-called Grossman-Nir bound. Since the branching fractions of the decay channels, ${\rm Br}(K^+ \to \pi^+ \phi)$ and ${\rm Br}(K_L \to \pi^0 \phi)$, are at the same order, the constraint from the charged Kaon decay is stronger than the neutral one in the most of the parameter region. Despite this fact, we involve both the constraints in our likelihood analysis for the sake of comprehensiveness to investigate the future prospects of the minimal WIMP model.

\subsubsection{Higgs decay}
\label{subsubsec: higgs decay}

The mediator can also be produced from the Higgs decay, $h \to \phi \phi$, and this process is indeed being investigated by measuring the Higgs boson property carefully at the LHC experiment. The latest result of the measurement is consistent with the SM prediction, so that we have various constraints on the decay process depending on the decay length of the mediator.

\subsubsection*{Higgs decay with prompt $\phi$ decay}

When the mediator promptly decays into two leptons, it is searched for through the Higgs decay channel, $h \to 4\ell$. 
At present, the ATLAS collaboration put a constraint on the branching fraction of the four muon process, $(P_p')^2\,{\rm Br}(h \to \phi \phi) \times {\rm Br}(\phi \to \mu \mu)^2$, where $P_p'$ is the probability that the decay is considered to be a prompt one at the experiment and defined as $P_p' = 1 - \exp[-\sigma/(\gamma \beta c \tau_\phi)]$ with $\sigma = 1$\,mm and $\gamma \beta = m_h/(2 m_\phi)$, respectively. The limit on the branching fraction at 90\% C.L. is found in Ref.\,\cite{Aaboud:2018fvk}. On the other hand, the CMS collaboration put constraints on similar branching fractions with the same $\sigma$ and $\gamma \beta$ by considering various leptonic channels, $h \to 4\mu$, $2\mu 2\tau$ and $4\tau$. Constraint on the branching fraction of each channel is obtained from Ref.\,\cite{Khachatryan:2017mnf} using the same prefactor $P_p'$. We involve these constraints in our analysis for the present status of the minimal WIMP model.\footnote{Both ATLAS and CMS collaborations assumed in their analyses that the Higgs boson decays into two pseudo-scalars, which means that different branching fractions of Higgs and mediator decays are used. We have corrected those to apply their constraints to our case where the Higgs boson decays into two scalars.}

In the near future, both the collaborations will update their constraints at the high-luminosity upgrade of the LHC experiment (HL-LHC)\,\cite{CMS:2013xfa}, if no signal is detected. Since we can expect about a hundred times more data at the HL-LHC than the current one, the sensitivity of the search will be improved by one order of magnitude if the statistical error dominates. For the future prospects of the model, we hence involve the same constraints on the above branching fractions in our analysis with making the limits ten times severer.

\subsubsection*{Higgs decay with displaced $\phi$ decay}

When the mediator decays with the decay length of 0.1--10\,m, it is searched for by the displaced vertex analysis at the LHC experiment. The mediator is produced mainly from the Higgs decay, $pp \to h + X \to \phi \phi + X$, when the mixing angle, $\sin \theta$, is small. Then, two displaced vertices are formed by the mediator decay if it is long-lived, and those are detected via the decays $\phi \to \mu\mu$, $ee$ and $\pi\pi$. Here, the decay products are collimated as the mediator is boosted. At present, the ATLAS collaboration put a constraint on the branching fraction of the Higgs decay with displaced vertices using various decay channels of the mediator\,\cite{Aad:2014yea}. These results are summarized in Ref.\,\cite{Clarke:2015ala}, where an upper limit on the branching fraction, ${\rm Br}(h \to \phi \phi)$, is given at 90\% C.L. as a function of the mediator mass and the mixing angle (lifetime). Roughly speaking, the region of ${\rm Br}(h \to \phi \phi) \gtrsim 30\%$ is excluded when 0.3\,GeV $\lesssim m_\phi \lesssim 60$\,GeV and 0.1\,m $\lesssim c \tau_\phi \lesssim$ 10\,m. We involve the constraint in Ref.\,\cite{Clarke:2015ala} in our analysis for the present status of the minimal WIMP model. 
Note that this constraint is not applied in the $m_\phi \geq 2 m_\chi$ region, 
as the $\phi \to \chi \chi$ decay was not considered in Ref.\,\cite{Clarke:2015ala}.

Since the systematic error already dominates the statistical one to put the constraint on the branching fraction ${\rm Br}(h \to \phi \phi)$ (with displaced vertices) at the present LHC experiment, it is difficult to expect a significant improvement of the sensitivity on this search at the HL-LHC experiment. We therefore do not consider the constraint from the search for the Higgs decay with displaced vertices to study the future prospects of the minimal WIMP model.

\subsubsection*{Higgs decay with very long-lived $\phi$ decay}

When the mediator is very long-lived, the decay process, $h \to \phi \phi$, contributes to the invisible decay width of the Higgs boson. In addition, the other decay channels exist, which are $h \to \chi \chi$ and $h \to \phi \phi \to 4\chi$, and always contribute to the invisible decay width of the Higgs boson without respect to the decay length of the mediator. As a result, the new physics contribution to the branching fraction of the invisible Higgs decay is given as follows:
\begin{eqnarray}
	{\rm Br}(h \to {\rm inv.})_{\rm BSM} =
	P_{\ell_{30}}^2\,{\rm Br}(h \to \phi \phi) +
	{\rm Br}(h \to \chi \chi) +
	(1 - P_{\ell_{30}}^2)\,{\rm Br}(h \to \phi\phi \to 4 \chi),
\end{eqnarray}
where $P_{\ell_{30}} = {\rm exp}[-\ell_{30}/(\gamma \beta c \tau_\phi)]$ and $1 - P_{\ell 30}$ are the probabilities that the mediator decays outside and inside the detector, respectively, with the size of detector and the boost factor being $\ell_{30} = 30$\,m and $\gamma \beta \simeq m_h/(2m_\phi)$. At present, the constraint on the branching fraction is given as ${\rm Br}(h \to {\rm inv.})_{\rm BSM} \leq 0.19$ at 90\% C.L.\cite{Cheung:2014noa} from Higgs precision measurements, and we include this in our analysis for the present status of the minimal WIMP model.

The constraint on this branching fraction will be improved at the HL-LHC experiment with 3\,ab$^{-1}$ data. 
If no signal is detected there\,\cite{Bechtle:2014ewa}, we can set 
${\rm Br}(h \to {\rm inv.})_{\rm BSM} \leq 0.05$ at 90\% C.L. 
We include it in our analysis for the future prospects of the minimal WIMP model.

\subsubsection{Direct production}
\label{subsubsec: direct production}

When the mediator is heavier than the $B$ meson and the mixing angle is not very suppressed, the mediator is also efficiently searched for by its direct production at high-energy colliders. The mediator is produced by the Bremsstrahlung process, $f \bar{f} \to V^* \to V \phi$, with $f$ and $V$ being a SM fermion and a weak gauge boson, respectively, and the mediator always decays promptly in the parameter region of our interest. At present, the LEP experiment put the most severe limit on the mixing angle ($\sin \theta$) as a function of the mediator mass ($m_\phi$). To be more precise, the L3 collaboration have used the hadronic decay channel of the mediator associated with $Z$ boson decays, $Z \to \nu \bar{\nu}$, $e^- e^+$ and $\mu^- \mu^+$, where its result is found in Ref.\,\cite{Acciarri:1996um}. Similarly, the ALEPH collaboration have also used the hadronic decay channel of the mediator, but only associated with the decay $Z \to \nu \bar{\nu}$. Its result is found in Ref.\,\cite{Buskulic:1993gi}, which is slightly weaker than that of the L3 collaboration. On the other hand, the OPAL collaboration have utilized the inclusive $Z$ production with its leptonic decays, $Z \to e^- e^+$ and $\mu^- \mu^+$, where its result is found in Ref.\,\cite{Abbiendi:2002qp}. It is, however, applicable for a very light mediator, say as light as $10^{-6}$\,GeV, and it dose not play an important role in the analysis.

The LHC experiment is also searching for the mediator produced by the above direct production process\,\cite{CMS:2016mtd}, though it is less sensitive than the LEP experiment. Moreover, the obtained limit on the mixing angle is already dominated by systematic errors, it seems difficult to expect a significant improvement even at the HL-LHC experiment. We therefore involve the limit obtained by the L3 collaboration in our likelihood analysis to investigate the present status of the minimal WIMP model, and do not consider the expected constraint from the direct mediator production for the near future prospects of the WIMP model.

\section{Results}
\label{sec: results}

We are now in a position to present the result of our likelihood analysis. We first explain the framework of our analysis in some details, as it plays a crucial role in our study. Then, we will present various results obtained in our analysis and discuss their implications to investigate the present status and near-future prospects of the minimal WIMP model.

\subsection{Simulation framework}
\label{subsec: simulation framework}

\begin{table}[t!]
    \centering
    \begin{tabular}{l|ll}
        Likelihood type & Present & Future \\
        \hline
        Step & Preselection criteria, LHCb, & -- -- -- \\
        & Kinematical equilibrium, BBN & \\
        Poisson & CHARM, XENON1T, CRESST, & SHiP \\
        & Darkside-50 & \\
        Half-Gaussian & CLEO, BABAR, Belle, & SuperCDMS-SNOLAB, LZ, \\
        & LHCb, N48/2, KTeV, & NEWS-SNOLAB, Belle\,II, LHCb \\
        & E949, KEK E391a, LHC, LEP & NA62, KOTO, HL-LHC \\
        Gaussian & Relic abundance, Plank($\Delta N_{\rm eff}$) & CMB-S4($\Delta N_{\rm eff}$) \\
        \hline
    \end{tabular}
    \caption{\small Summary of likelihood distributions used in our analysis.}
    \label{tab: likelihood}
\end{table}

There are many released limits and data used in this work. Because of lacking discovery of new physics, the majority of them are based on the $90\,\%$ upper or lower confidence limit on a one-dimensional physical observable.\footnote{The terminology $90\,\%$\,C.L. for $\delta\chi^2=2.71$ in one-tail Gaussian likelihood distribution is sometimes used.} Nevertheless, for some other constraints such as relic density, because of its clear discovery, it is nature to describe such a likelihood probability by a two-tail Gaussian with a narrow peak. Except above two types of limits, theoretical constraint is used to veto the parameter space whose answer is often physical or not physical. 
As summarized in Table\,\ref{tab: likelihood}, we have used four types of likelihood function in our analysis: 
(i) Gaussian likelihood, 
(ii) Half Gaussian likelihood with the central value being fixed to be zero, 
(iii) Poisson likelihood for counting experiments, and 
(iv) Step function likelihood. 
In the following, we will present the usage of these four likelihood functions.

The most powerful advantage of the global analysis is to increase the statistics by combining different data sets which allows us to remove more parameter spaces from different corners. However, in order to conservatively exclude parameter space by adding the statistics, a proper likelihood function is needed and its tail at the exclusion region is particularly important. Once the central value and error bar are given, the Gaussian likelihood is
\begin{eqnarray}
\label{eq:gaulike}
	{\cal L}_{\rm Gau.} \propto \exp \left[ -\frac{\chi^2}{2} \right],
	\qquad {\rm where} \qquad
	\chi^2 = \left[ \frac{ {\rm Prediction} - {\rm Center~~value} } { {\rm Error~~bar} } \right]^2.
\label{eq: gaulike}
\end{eqnarray}
On the other hand, for counting experiments, if expected event number and observed event number are provided, 
Poisson distribution is usually adopted for its likelihood function, 
\begin{eqnarray}
\label{eq:poslike}
	{\cal L}_{\rm Pos.} \propto 
	\frac{e^{-\left( s+b\right)} (s+b)^o}{o!},
\end{eqnarray}
where $s$ and $b$ are expected signal and background events, while $o$ is observed events.

It is usually having null signal detection in new physics search. In such a search, a lower limit or upper limit of an observable is reported. Such a limit with some information given by experimental collaboration, one can of course reconstruct and verify some likelihood used in the experimental collaboration. Because of expansive background simulation and likelihood computation time, it will not be realistic to repeat whole procedure to reconstruct such a precise likelihood as used in the experimental collaboration. Therefore, we model the likelihood based on two reasonable assumptions: null signal detection and Gaussian distribution with the alignment of $90\,\%$\,C.L. The likelihood of such a distribution called half Gaussian is almost identical as eq.\,(\ref{eq: gaulike}) but with two differences; The center value is set to be zero because of null signal detection and the error-bar-squared can be obtained by the $90\,\%$\,C.L. limit divided by $2.71$ to align $90\,\%$ Gaussian one tail limit. Note that such a half Gaussian distribution shall be more conservative than precise one because new physics signal is expected to have an excess than background in the future.

Some experimental upper/lower limits are built based on multi-dimensional variables which are hard to reconstruct their likelihoods. For example, the limits of the LHC Higgs displaced vertex search in Sec.\,\ref{subsubsec: higgs decay} are based on three dimensional variables, $m_\phi$, $\sin\theta$ and the branching fraction of $h \to \phi\phi$. For a sake of simplicity, one may just use a hard cut for such an experimental constraint
to answer whether the parameter space is allowed or excluded. 
Sometimes, one could perform a scan with the constraints described in case (ii) by using a step function, but it is particularly avoided in our analysis because it will lose the advantage of combined analysis as discussed in the previous paragraph.

Finally, when performing two-dimensional contour figures in this paper, we use the method "Profiled Likelihood", namely the minimum Chi-squared method. Like frequentist, such a method allows us to get rid of unwanted/not interesting parameters by taking the maximum likelihood along the direction of other unwanted/not interesting dimensions.

Regarding to the sampling, as seen in Appendices\,\ref{app: initial parameter region} and \ref{app: equilibrium condition}, the likelihood function is flat within vast 1sigma allowed region but dramatically changing at the 2sigma tails. Therefore, in our analysis, we adopt an unusual strategy to scan our parameter space. We first scan the parameter space with the range in eq.\,(\ref{eq: the parameters}) and the conditions in Table\,\ref{tab: apriori constraints}. Note that two dark matter parameters $m_\chi$ and $c_s$ are not scanned in this step. After collecting ${\cal O}(10^6)$ allowed points, we use it to present the parameter space constrained by the preselection criteria in Appendix\,\ref{app: initial parameter region}. In the second step, based on the former collected data set, we varied $m_\chi$ and $c_s$ for each points to fit the relic abundance and kinematical equilibrium conditions. However, the parameter $\lambda_\Phi$ does not affect the dark matter phenomenology at all. In this step, we evaluated ${\cal O}(10^8)$ points but only $\sim 5\times10^6$ points fulfill the relic abundance and kinematical equilibrium conditions. 
In the third step, based on the ${\cal O}(10^8)$ points collected from the former scan, we evaluate all the likelihoods with the constraints given in Tables\,\ref{tab: cosmological and astrophysical constraints} and \ref{tab: collider constraints}. 
Finally, we use the knowledge we learned from previous scans as the new prior distribution to do several more sophisticated scans. 
We employed two scan tools (\texttt{emcee}\,\cite{ForemanMackey:2012ig} and \texttt{MultiNest}\,\cite{Feroz:2008xx}) to undertake the sampling. 
With total ${\cal O}(10^8)$ likelihood evaluation done by the tools \texttt{emcee} and \texttt{MultiNest}, 
we found the coverage is good enough. We combine all the scans from the former steps to draw our figures.

\subsection{Present status}
\label{subsec: status}

Results of our analysis for the present status of the minimal WIMP model projected on the $(m_\chi, m_\phi)$- and $(m_\phi, |\sin \theta|)$-planes are shown in the left and right panels of Fig.\,\ref{fig: present status}, respectively, while those projected on all planes of input parameters are found in appendix\,\ref{app: present status}.

\begin{itemize}

\item[\bf (a)]
In the left panel, the upper limit on the mediator mass $m_\phi$ comes from the relic abundance condition. As mentioned above, most of parameter regions satisfy the relation $m_\phi \lesssim m_\chi$, while the exception is the $s$-channel resonance-enhanced region with $m_\phi \sim 2 m_\chi$. The latter case appears only for $m_\phi \gtrsim {\cal O}(1)$\,GeV to be consistent with collider constraints.

\item[\bf (b)]
The lower limit on the WIMP mass is obtained by the combination of collider constraints and kinematic equilibrium condition. When it is lighter than ${\cal O}(10)$\,MeV, the mixing angle is constrained to be below $10^{-3}$ due to Kaon experiments. The mixing angle is, however, required to be larger than $10^{-3}$ to satisfy the kinematical equilibrium condition, as shown in the plot on the $(m_\chi, |\sin \theta|)$-plane in Fig.\,\ref{fig: After AP RE KE constraints}, which leads to the limit, $m_\chi \gtrsim 10$\,MeV.

\item[\bf (c)]
On the other hand, the lower limit on $m_\phi$ in the range of $m_\chi \lesssim 500$\,MeV comes from the $\Delta N_{\rm eff}$ constraint discussed in section \ref{subsubsec: neff}, while the limit in the range of $m_\chi \gtrsim 3$\,GeV is from the direct dark matter detection constraint, for the spin-independent scattering cross section between the dark matter and a nucleon is proportional to $m_\phi^{-4}$. The lower limit on $m_\phi$ in the range of  500\,MeV $\lesssim m_\chi \lesssim$ 3\,GeV is obtained in a complicated way: it is from the combination of the direct detection constraint and the kinematical equilibrium condition. When the dark matter mass is below a few GeV, the freeze-out temperature becomes lower than the QCD phase transition, so that a mixing angle is required not to be very suppressed in order to maintain the kinematical equilibrium.\footnote{The process $\phi f \to \phi f$ via the $h-\phi-\phi$ coupling does not work (with `$f$' being a SM fermion), for only light SM fermions exist in the universe and their couplings to $h$ are suppressed by small Yukawa couplings.} However, such an unsuppressed mixing angle (as well as a small $m_\phi$) leads to a large spin-independent scattering cross section between the dark matter and a nucleon and ruled out by the direct dark matter detection.

\item[\bf (d)]
In the right panel, the lower limit on the mixing angle is from the BBN constraint discussed in section\,\ref{subsubsec: bbn}, $\tau_\phi \leq 1$\,s in the range of $m_\phi\geq 2m_\pi$ and $\tau_\phi \leq 10^5$\,s in the range of $m_\phi < 2m_\pi$. The spike structure at $m_\phi \sim 1$\,GeV comes from the uncertainty on the decay width of the mediator discussed in section\,\ref{subsec: phi decay}, while another one at $m_\phi \sim 200$\,MeV is from the quick change of the decay width due to the threshold of the $\phi \to \mu^- \mu^+$ channel.

\item[\bf (e)]
On the other hand, the upper limit on the angle comes mainly from collider constraints. For $m_\phi \lesssim 500$\,MeV, the limit is set by Kaon experiments such as CHARM. For 500\,MeV $\lesssim m_\phi \lesssim$ 5\,GeV, B meson experiments set the limit. The long island at $m_\phi \simeq 1$\,GeV and $10^{-3} \lesssim |\sin \theta| \lesssim 10^{-1}$ is again due to the theoretical uncertainty of the $\phi$-decay width. The reason why this island region is isolated is that the displaced vertex search of $B \to X_s \phi$ at the BaBar experiment ruled out the parameter region of 0.5\,GeV $\lesssim m_\phi \lesssim$ 1.5\,GeV and $5 \times 10^{-4} \lesssim |\sin \theta| \lesssim 10^{-3}$. The step-like structure at $m_\phi \sim$ 2--4\,GeV is because $\phi \to g g, c \bar{c}$, $\tau^- \tau ^+$ opens and the branching fraction of the observable channel, e.g. $\phi \to \mu^- \mu^+$, is suppressed.

\item[\bf (f)]
When the mediator becomes heavier than 5\,GeV, the limit on the mixing angle is set by the direct dark matter detection at underground experiments. This is because the dark matter is required to be heavier than the mediator in most of the parameter region, for the relic abundance condition is satisfied by the $\chi \chi \to \phi \phi$ annihilation process, and the direct dark matter detection put a sever limit when $m_\chi \gtrsim 5$\,GeV. Only the exception is found in extended region at $m_\phi \sim$ several 10\,GeV and $|\sin \theta| \sim 10^{-2}$, where the relic abundance condition is satisfied by the $s$-channel resonance-enhanced annihilation $\chi \chi \to \phi \to f \bar{f}$.

\item[\bf (g)]
The SN1987A constraint\,\cite{Krnjaic:2015mbs} is over depicted on the $(m_\phi, |\sin \theta|)$-plane as a darker transparent color region at $5 \cdot 10^{-7} \lesssim |\sin\theta| \lesssim 3 \cdot 10^{-5}$ and $m_\phi \lesssim \mathcal{O}(100)$\,MeV. 
It is seen that the SN1987A constraint is potentially important, 
for it restricts the parameter region that is not restricted by other constraints. 
The exclusion region by the SN1987A constraint can be understood as follows: 
The mediator interaction is too weak to affect the SN cooling when $|\sin\theta|\lesssim 5\cdot 10^{-7}$, 
while the mediator decays or trapped inside the SN when $|\sin\theta|\gtrsim 3\cdot 10^{-5}$ and does not contribute the cooling. 
When $m_\phi \gtrsim T_c \simeq 30$\,MeV with $T_c$ being the critical core temperature of SN1987A, 
the mediator cannot be thermally, thus efficiently, produced.
\end{itemize}

\begin{figure}[t!]
	\centering
	\includegraphics[height=2.9in, angle=0]{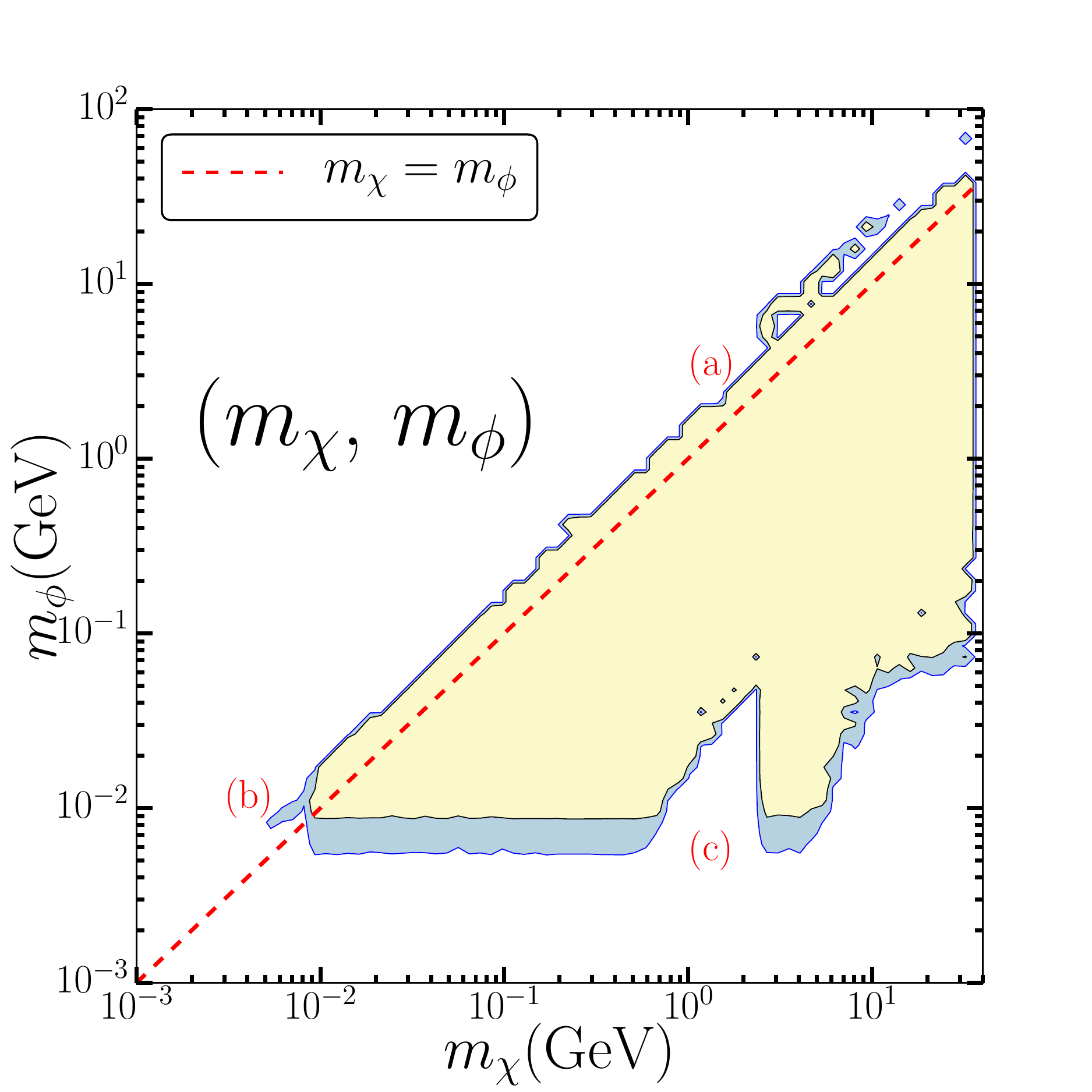}
	\qquad
	\includegraphics[height=2.9in, angle=0]{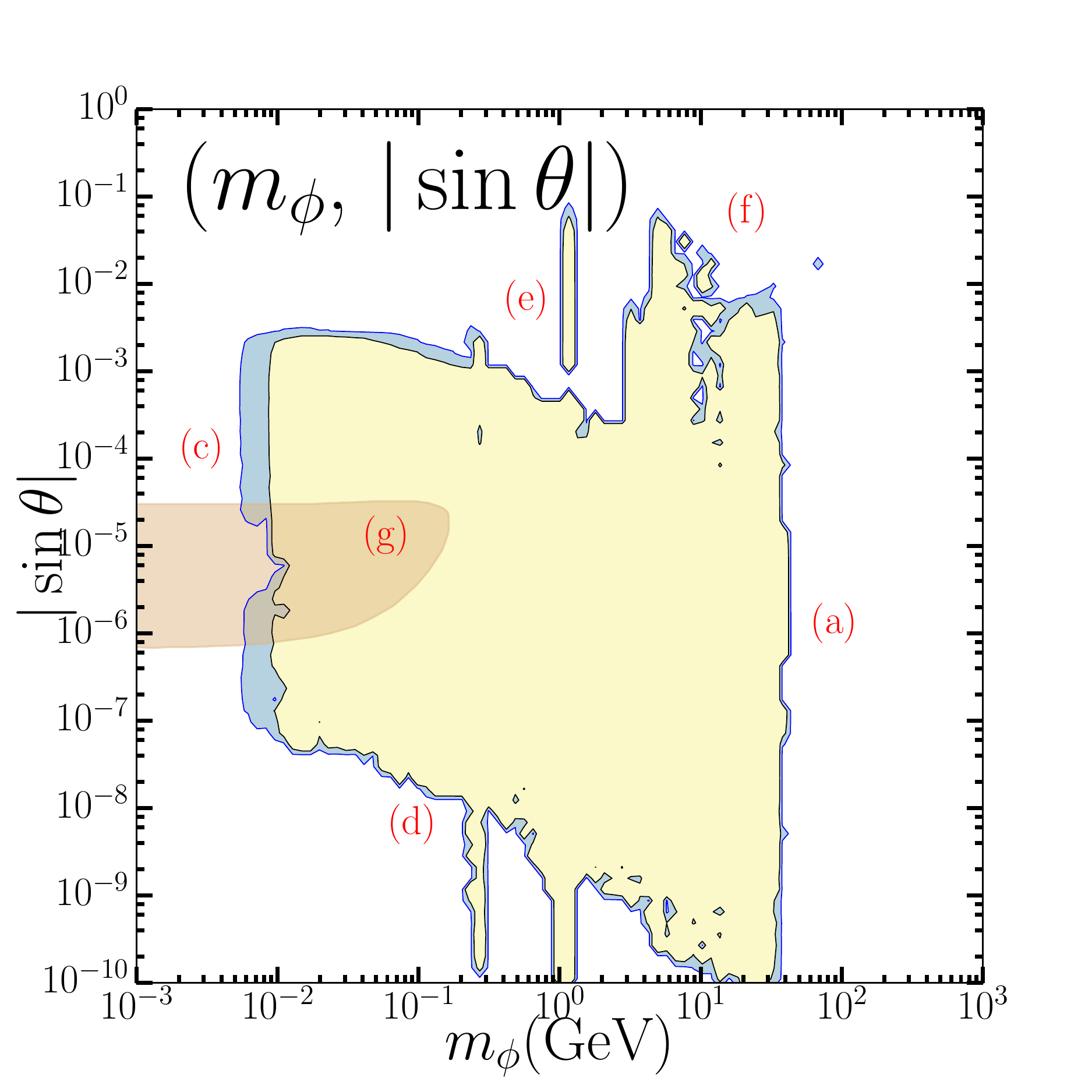}
	\caption{\small \sl Results of our analysis for the present status of the minimal WIMP model projected on the $(m_\chi,m_\phi)$- and $(m_\phi,|\sin\theta|)$-planes at 68\% C.L. (yellow)  and 95\% C.L. (blue). Please see figures in appendix\,\ref{app: present status} for those who are interested in results projected on  all planes of input parameters. The alphabet on each edge of the contour corresponds to each paragraph of the main text.}
	\label{fig: present status}
\end{figure}

\subsection{Future prospects}
\label{subsec: prospects}

\begin{figure}[t!]
	\centering
	\includegraphics[height=2.9in, angle=0]{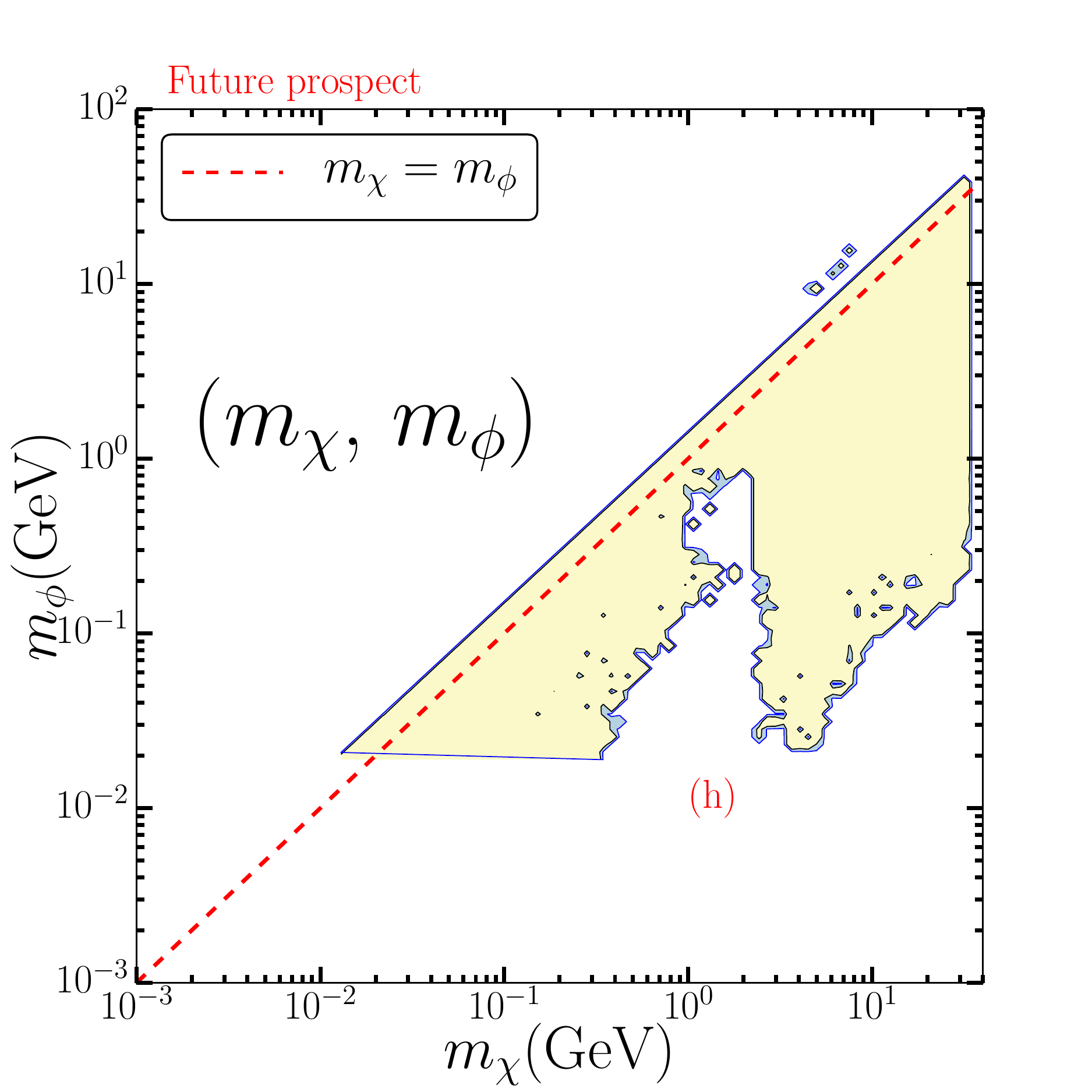}
	\qquad
	\includegraphics[height=2.9in, angle=0]{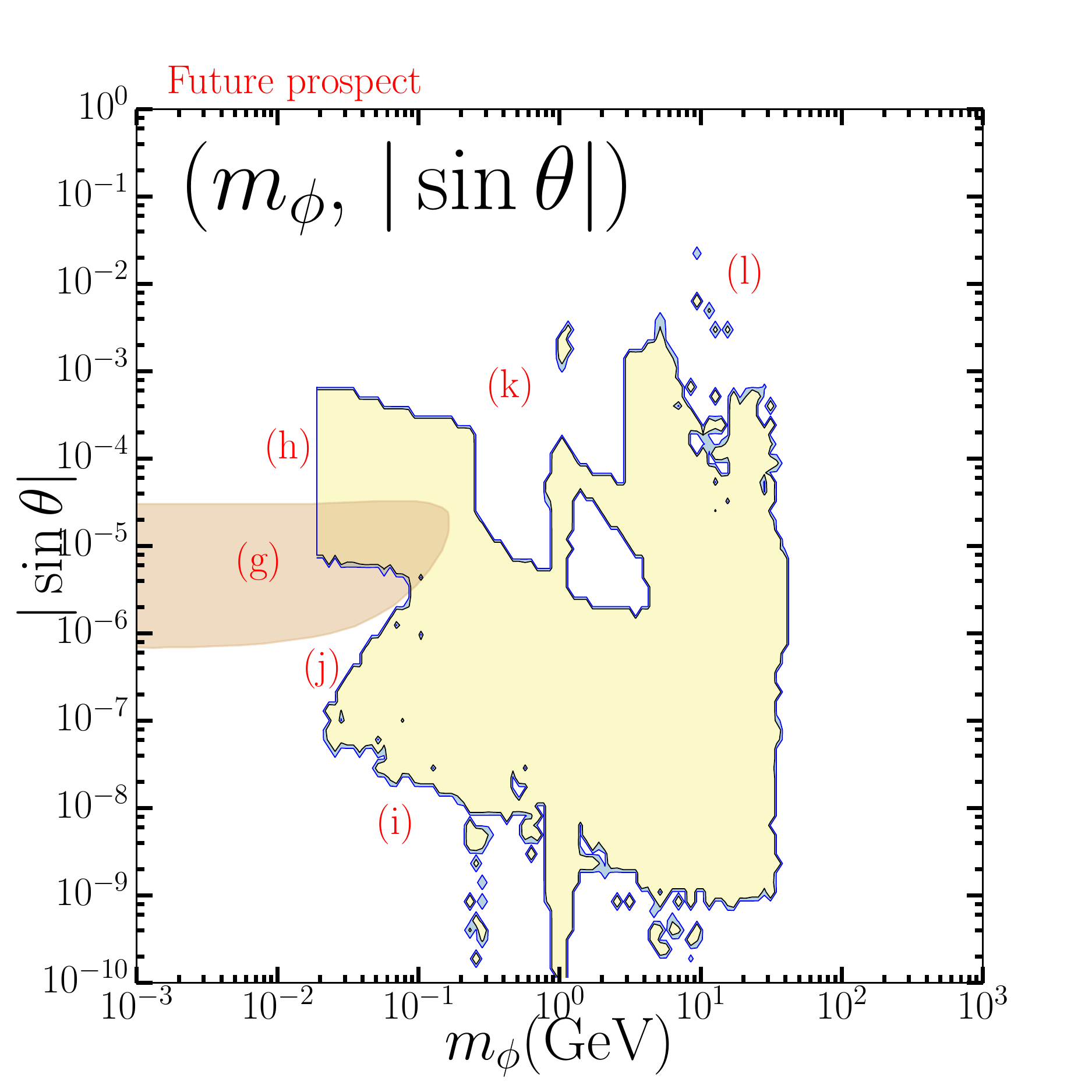}
	\caption{\small \sl Results of our analysis for the future prospects of the minimal WIMP model projected on the $(m_\chi, m_\phi)$- and $(m_\chi, |\sin \theta|)$-planes at 68\% C.L. (yellow) and 95\% C.L. (blue). Please see figures in appendix\,\ref{app: future prospects} for those who are interested in results projected on all planes of input parameters. The alphabet on each edge of the contour corresponds to each paragraph of the main text.}
	\label{fig: future prospects}
\end{figure}

Results of our analysis for the future prospects of the minimal WIMP model projected on the $(m_\chi, m_\phi)$- and $(m_\chi, |\sin \theta|)$-planes are shown in the left and right panels of Fig.\,\ref{fig: future prospects}, respectively, where we have assumed that dark matter signals as well as mediator signals are not detected at any near future experiments/observations discussed in the previous section. Our results projected on all planes of input parameters are again found in appendix\,\ref{app: future prospects}.

\begin{itemize}
\item[{\bf (h)}]
In the left panel of Fig.\,\ref{fig: future prospects}, comparing with the one in Fig.\,\ref{fig: present status}, the lower limit on $m_\phi$ becomes severer as $m_\phi \gtrsim 20$\,MeV in the range of $m_\chi \lesssim 300$\,MeV due to the future-expected constraint from the $\Delta N_{\rm eff}$ measurement. Together with the relic abundance condition, it also gives a lower limit on the WIMP mass. On the other hand, the lower limit on the mediator mass in the range of $m_\chi \gtrsim 300$\,MeV becomes stronger than those in Fig.\,\ref{fig: present status} because the significant upgrade will be expected at the direct dark matter detection in the near future. The void region, which is located at $m_\chi \sim 1$\,GeV and $m_\phi \sim $ a few hundred MeV, is because of the constraint from the SHiP experiment. 
In addition, the mixing angle in this region is required to be enough large from the kinematical equilibrium condition when $m_\chi \sim 1$\,GeV, while such a mixing angle will be ruled out if no signal is detected at the experiment.

\item[{\bf (i)}]
In the right panel of Fig.\,\ref{fig: future prospects}, the lower limit on the mixing angle $|\sin \theta|$ in the range of $m_\phi \lesssim 10$\,GeV is not very much different from the previous one in Fig.\,\ref{fig: present status}. On the other hand, when $m_\phi \gtrsim 10$\,GeV, the region of $|\sin \theta| \lesssim {\cal O}(10^{-9})$, where it was survived in Fig.\,\ref{fig: present status}, is removed in Fig.\,\ref{fig: future prospects}. This is because the search for the Higgs invisible decay at the HL-LHC experiment rules out the region if no signal of the decay is detected there.

\item[{\bf (j)}]
The excluded region at $m_\phi \lesssim 100$\,MeV and $|\sin \theta| \sim 10^{-6}$ is from the direct dark matter detection at future underground experiments. This is because, as discussed in appendix\,\ref{app: supplemental figures}, the dark matter mass must be larger than 200\,MeV to satisfy the kinematical equilibrium condition in this parameter region, and a small mediator mass leads to a large scattering cross section between the dark matter and a nucleon despite of a suppressed mixing angle. The sensitive future direct dark matter detection thus start proving this region.

\item[{\bf (k)}]
On the other hand, the upper limit on the mixing angle is also very much improved. The SHiP experiment will change the landscape significantly at the region of 0.3\,GeV $\lesssim m_\phi \lesssim$ 4\,GeV and $2 \times 10^{-6} \lesssim |\sin \theta| \lesssim 10^{-4}$, if no mediator signal is detected. Moreover, LHCb and Belle\,II experiments could also put severe constraints on $|\sin\theta|$ in the range of 0.5\,GeV $\lesssim m_\phi \lesssim$ 5\,GeV, while the direct dark matter detection at future underground experiments could put a severer constraint on the mixing angle $|\sin \theta|$ in the range of $m_\phi \gtrsim 5$\,GeV.

\item[{\bf (l)}]
Finally, let us comment on the resonant annihilation region, where the relation $m_\phi \sim 2 m_\chi$ holds. As seen in the right panel of Fig.\,\ref{fig: future prospects}, this region could still survive in the near future at $m_\phi \sim 2m_\chi \sim 10$\,GeV and $|\sin \theta| \sim 10^{-2}$, though it obviously shrunk compared to the region in Fig.\,\ref{fig: present status}. This is because that the sensitivity of the direct dark matter detection is significantly improved when the dark matter mass is greater than several GeV.
 
\end{itemize}

\subsection{Implication of the results}
\label{subsec: implication}

\begin{figure}[t!]
	\centering
	\includegraphics[height=2.8in, angle=0]{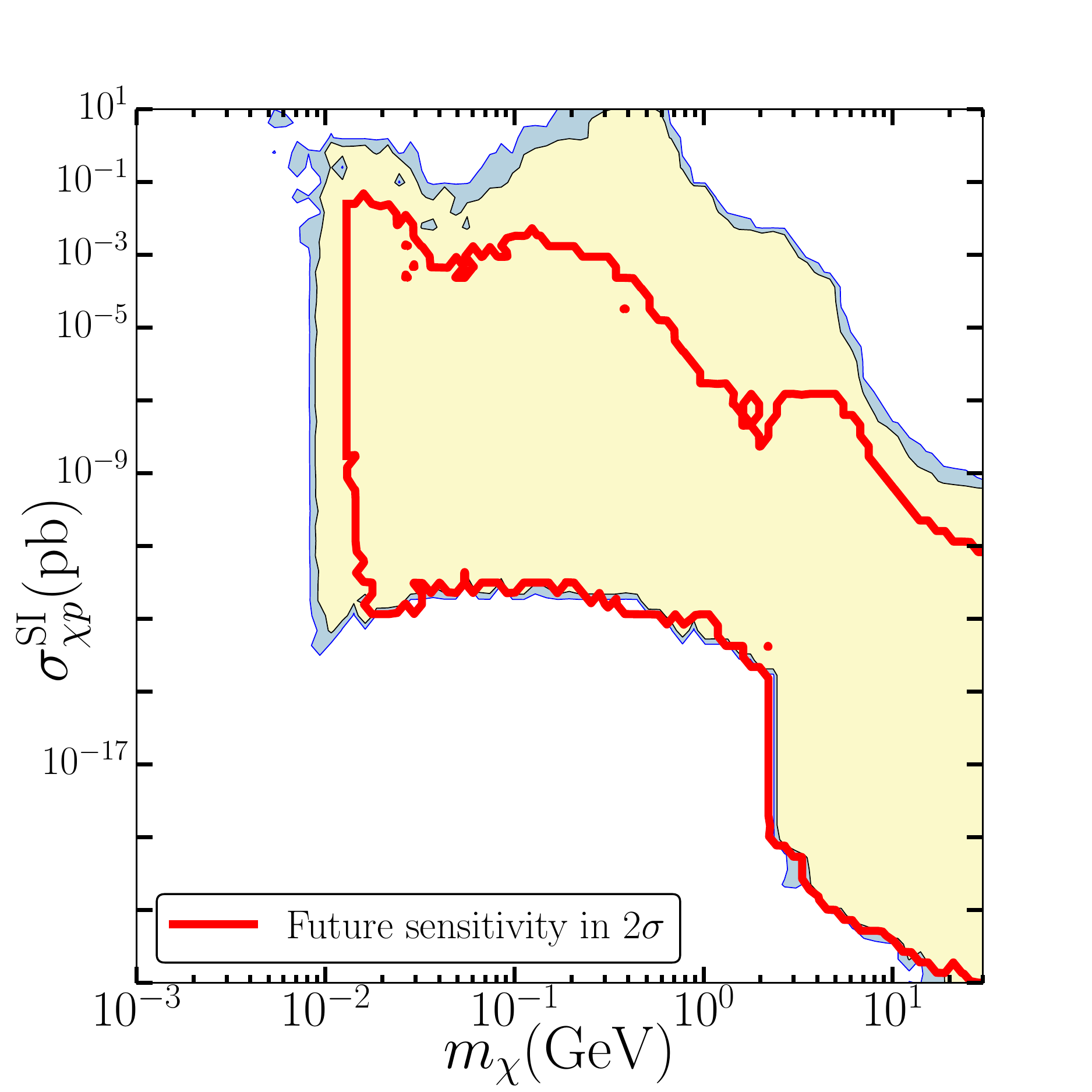}
	\qquad
	\includegraphics[height=2.8in, angle=0]{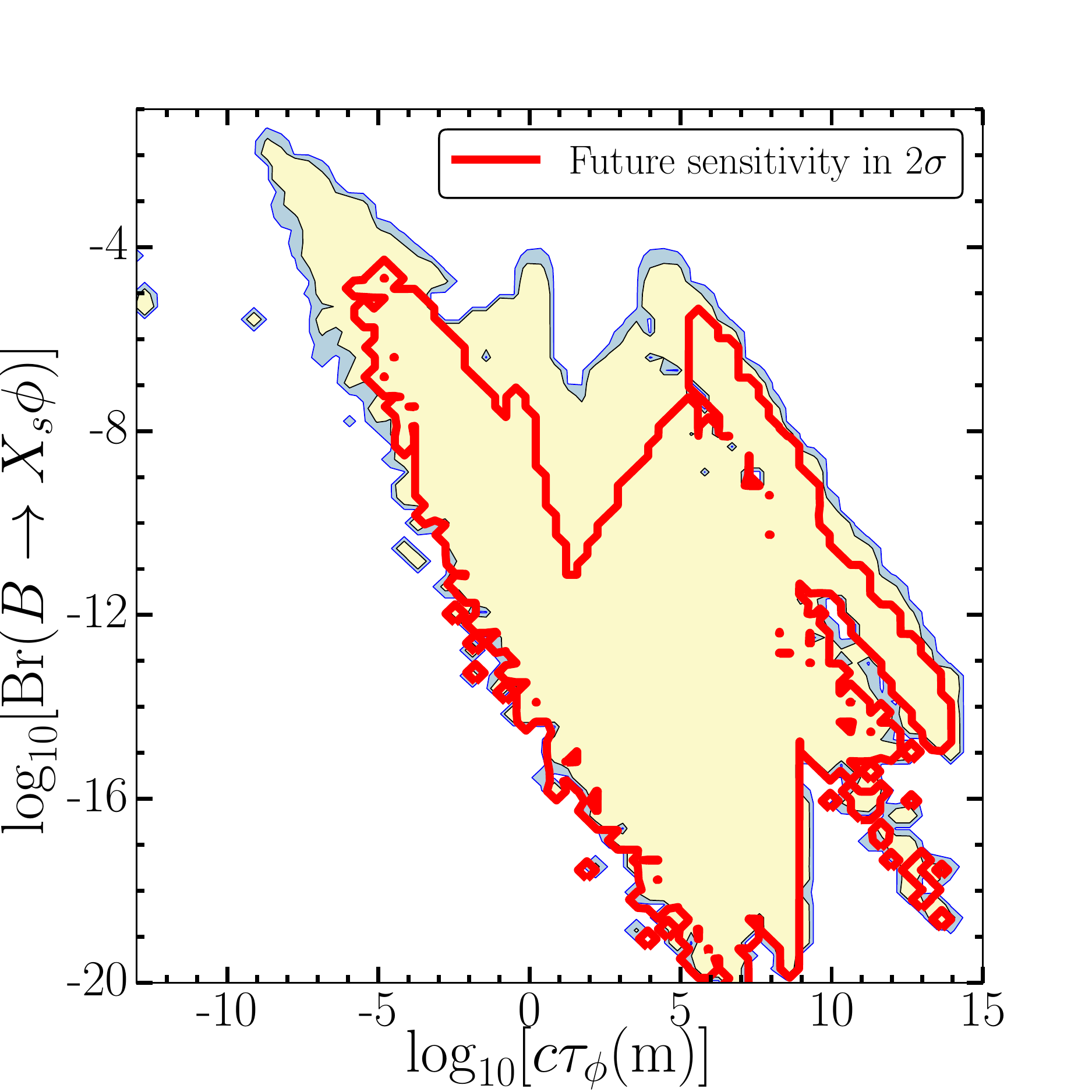}
	\\
	\includegraphics[height=2.8in, angle=0]{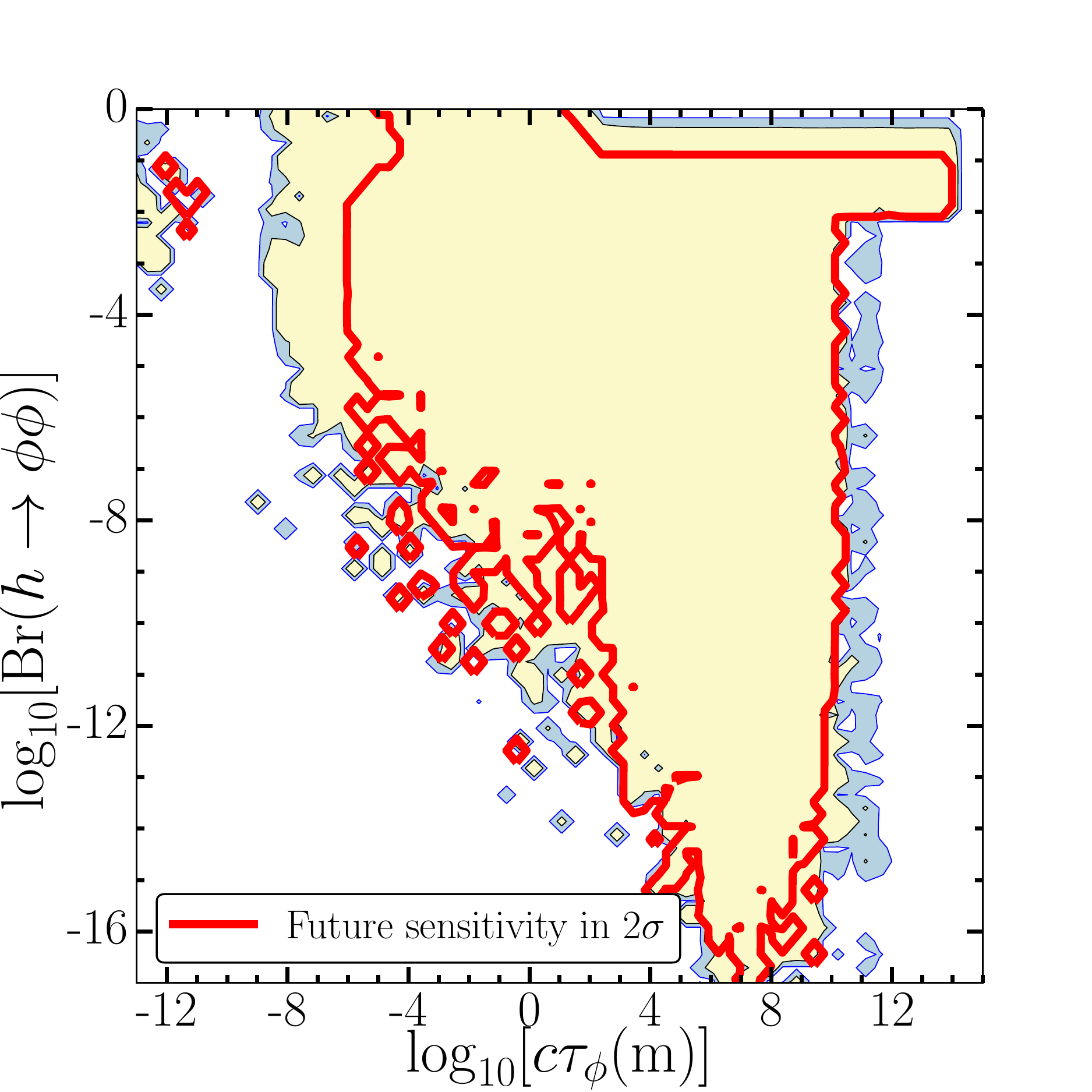}
	\qquad
	\includegraphics[height=2.8in, angle=0]{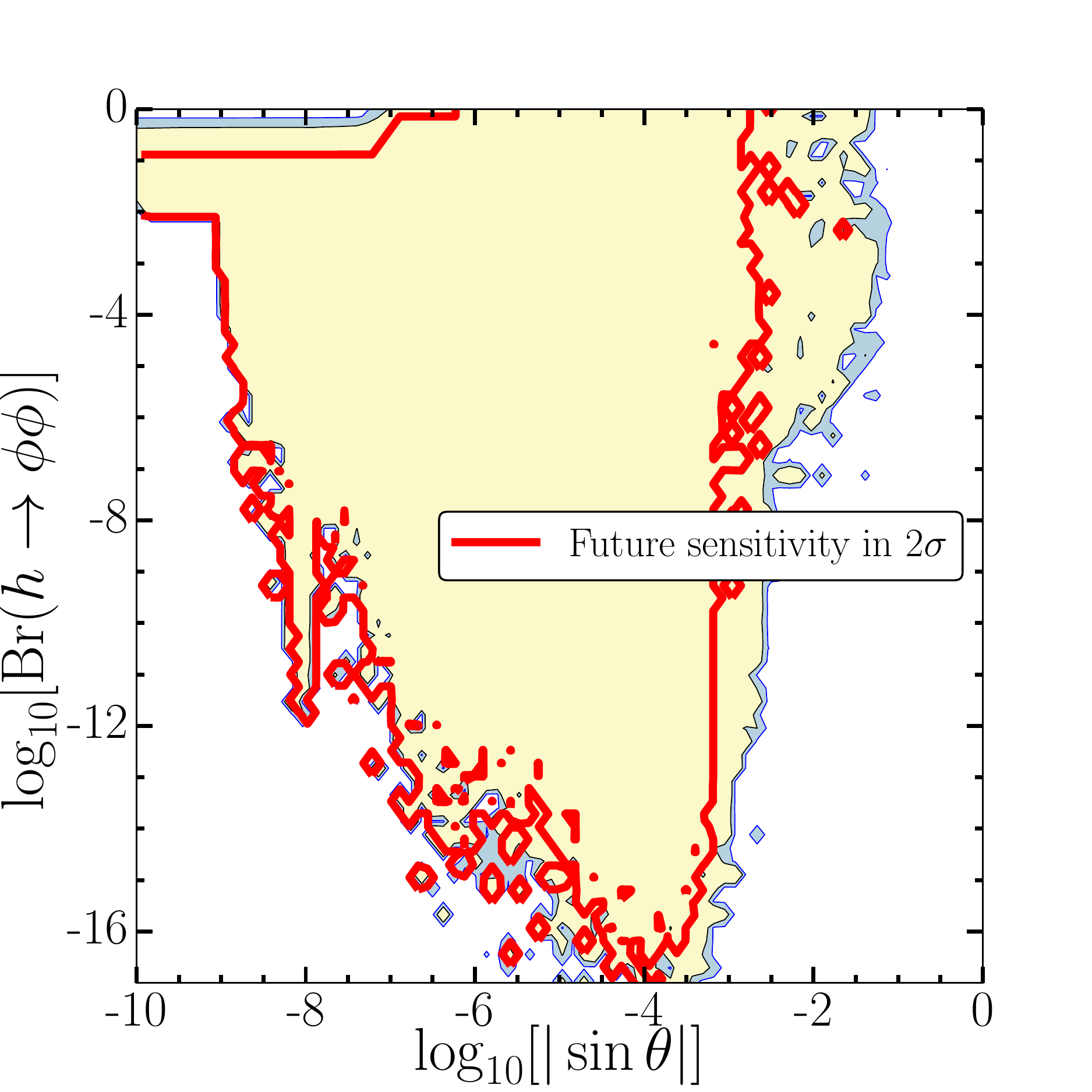}
	\caption{\small \sl Results of our analysis for the present status of the minimal WIMP model projected on the $(m_\chi,m_\phi)$- and $(m_\phi,|\sin\theta|)$-planes at 68\% C.L. (yellow)  and 95\% C.L. (blue). Please see figures in appendix\,\ref{app: present status} for those who are interested in results projected on  all planes of input parameters.}
	\label{fig: implication}
\end{figure}

Here, we consider implication of the results discussed in previous two sub-sections. In Fig.\,\ref{fig: implication}, some results projected on various observables are shown, where the region consistent with all present experimental results discussed so far is depicted using the same color code as those in Figs.\,\ref{fig: present status} and \ref{fig: future prospects}. Moreover, the region which will survive (at 95\,\% C. L.) even if no dark matter and mediator signals are detected in the near future is also over depicted as a red line. We discuss below implication of the results in each panel of Fig.\,\ref{fig: implication} in some detail.

In the top left panel, the results are projected on the $(m_\chi, \sigma_{\rm SI})$-plane, where $\sigma_{\rm SI}$ is the spin-independent scattering cross section between the WIMP and a nucleon, as defined in section\,\ref{subsubsec: direct detection}. As expected, the direct dark matter detection is effective to directly search for the WIMP as far as $\sigma_{\rm SI}$ is large enough.\footnote{The improvement of the lower limit on $m_\chi$ in the near future is because of the provisional $\Delta N_{\rm eff}$ constraint.} On the other hand, it can be also seen that the wide region with a very small value of $\sigma_{\rm SI}$ remains survived even in future, especially when $m_\chi$ is larger than ${\cal O}(1)$\,GeV. This originates in the survived parameter region with a suppressed mixing angle ($\sin \theta \sim 0)$, and thus other experiments are required to test it.

In the top right panel, we present the favored regions for $c \tau_\phi$ vs. ${\rm Br}(B \to X_s \phi)$ with $\tau_\phi$ and ${\rm Br}(B \to X_s \phi)$ being the lifetime of the mediator and the decay branching fraction of the $B$ meson into a mediator and a hadronic system with a strangeness\,\cite{Clarke:2015ala}, respectively. It is found that the region with $c \tau_\phi \simeq (10^{-3}$ -- $10^3)$\,m will be well searched for thanks to the flavor experiments. Belle\,II and SHiP experiments will play significant roles for the search in regions of $10^{-3}$\,m $\lesssim c \tau_\phi \lesssim 1$\,m and 1\,m $\lesssim c \tau_\phi \lesssim 10^3$\,m, respectively. On the other hand, the direct dark matter detection plays an important role to search for the region of $c \tau_\phi \lesssim 10^{-3}$, for both the WIMP and the mediator are required to be heavy enough to have such a short mediator lifetime. Finally, we found that the direct dark matter detection and BBN constraints will play some roles to search for the region of $c \tau_\phi \gtrsim 10^5$\,m, though the wide parameter region survives even in the future due to the small mixing angle.

In the bottom left panel, we present the favored regions for $c \tau_\phi$ vs. ${\rm Br}(h \to \phi \phi)$, where ${\rm Br}(h \to \phi \phi)$ is the  branching fraction of the exotic Higgs decay into two mediators. As discussed above, the direct dark matter detection will play a crucial role in the near future, in particular to search for the region of $c \tau_\phi \lesssim 10^{-5}$\,m. On the other hand, other experiments will also play important roles to search for the region of $c \tau_\phi \gtrsim 10^{-5}$\,m. For instance, LHC experiment will search for the region by observing the exotic Higgs decay when its branching fraction is large enough, while cosmological observations ($\Delta N_{\rm eff}$ measurement, etc.) and flavor experiments (Belle\,II and SHiP experiments, etc.) will search for the same region but when the branching fraction of the exotic Higgs decay is very suppressed.

It is worth emphasizing here that the constraint from the exotic Higgs decay plays a complemental role to others as shown in the bottom right panel. Magnitude of WIMP and mediator signals is proportional to the mixing angle squared in most of experiments and observations, while that of the exotic Higgs decay does not rely on the mixing angle. This is because the mediator is required to have an interaction to SM particles with enough magnitude to satisfy the kinematical equilibrium condition, and it is achieved by the $\phi-\phi-h$ interaction (originating in the $\Phi^2 |H|^2$ operator) when the mixing angle (originating in the $\Phi |H|^2$ operator) is suppressed. As a result, the precise measurement of the nature of the Higgs boson will be mandatory in the future to test the minimal WIMP model, and it could be done by future lepton colliders such as ILC\,\cite{Fujii:2017vwa}, CEPC\,\cite{CEPC-SPPCStudyGroup:2015csa} and FCC-ee\,\cite{dEnterria:2016fpc}.

\section{Summary}
\label{sec: summary}

We have studied a minimal (renormalizable) model with a light fermionic WIMP and a light scalar mediator whose decay width is computed with uncertainties from non-perturbative QCD effects properly taken into account. In order to investigate the present status and future prospects of the light WIMP and the light scalar mediator, we have performed a comprehensive likelihood analysis involving all robust constraints obtained so far.
In addition, we also discuss those future sensitivities which will be obtained in the near future (if no WIMP signals are detected) from particle physics experiments as well as cosmological and astrophysical observations. We have carefully involved a kinematical equilibrium condition assuming that the (chemical) freeze-out of the light WIMP occurred when it was in kinematically equilibrium with the thermal bath composed of SM particles. We have paid particular attention to a possible case that the light WIMP can be in the kinematical equilibrium through existent mediator particles at the freeze-out epoch even if the WIMP does not have an interaction directly to SM particles with enough magnitude.

A very wide parameter region is still surviving at present in the region of 10\,MeV $\lesssim m_\phi \lesssim m_\chi$, and it is found that many kinds of experiments and observations are required to test the WIMP in the near future. Direct dark matter detection experiments will play a crucial role in particular to search for the WIMP with the mass greater than ${\cal O}(100)$\,MeV, while flavor experiments will play a significant role to search for the mediator in the wide region of its mass. Moreover, precise cosmological observations such as the $\Delta N_{\rm eff}$ measurement are mandatory to search for a very light mediator whose mass is of the order of 10\,MeV.

On the other hand, a wide parameter region will remain survived in the near future even if no WIMP/mediator signals are detected in the experiments. 
In principle, the experiments rely on the mixing angle (originating in the $\Phi |H|^2$ operator) 
to detect WIMP/mediator signals, while all cosmological conditions (relic abundance and kinematical equilibrium conditions) can be satisfied even if the mixing angle is very suppressed. This is because the annihilation process within the dark sector, the $\chi \chi \to \phi \phi$ process, 
satisfies the relic abundance, while the kinematical equilibrium condition is satisfied through the other interaction between the mediator and SM particles, namely the $\phi-\phi-h$ interaction (originating in the $\Phi^2 |H|^2$ operator). 
As a result, a future experiment which is sensitive to this interaction is mandatory to test the remaining parameter region, 
and it can be achieved by measuring the nature of the Higgs boson precisely. 
In particular, future lepton colliders such as ILC\,\cite{Fujii:2017vwa}, CEPC\,\cite{CEPC-SPPCStudyGroup:2015csa} and FCC-ee\,\cite{dEnterria:2016fpc} will provide an ideal environment for this measurement.

\vspace{0.5cm}
\noindent
{\bf Acknowledgments}
\vspace{0.1cm}\\
\noindent
We would like to thank M. Ibe for fruitful discussion about cosmological constraints on the light mediator. This work is supported by the Grant-in-Aid for Scientific research from the Ministry of Education, Science, Sports, and Culture (MEXT), Japan and from the Japan Society for the Promotion of Science (JSPS), Nos. 16H02176, 17H02878 and 26104009 (for S.~Matsumoto), as well as by the World Premier International Research Center Initiative (WPI), MEXT, Japan. 
Y.S. Tsai is funded in part by Chinese Academy of Sciences Taiwan Young Talent Programme, Grant No. 2018TW2JA0005.
\newpage
\appendix

\section{Preselection criteria}
\label{app: initial parameter region}

\begin{figure}[t!]
	\includegraphics[height=1.32in,angle=0]{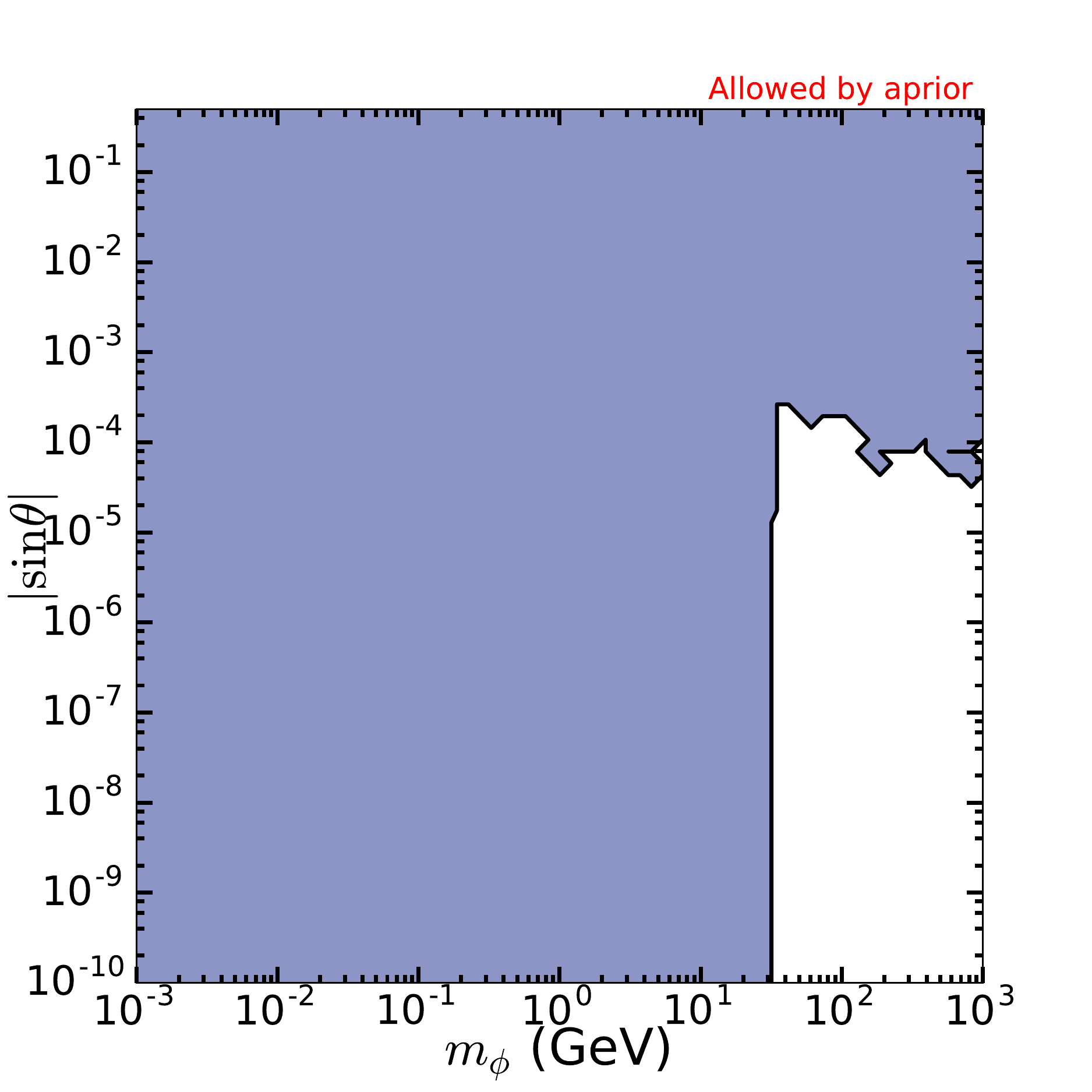}
	\newline
	\includegraphics[height=1.32in,angle=0]{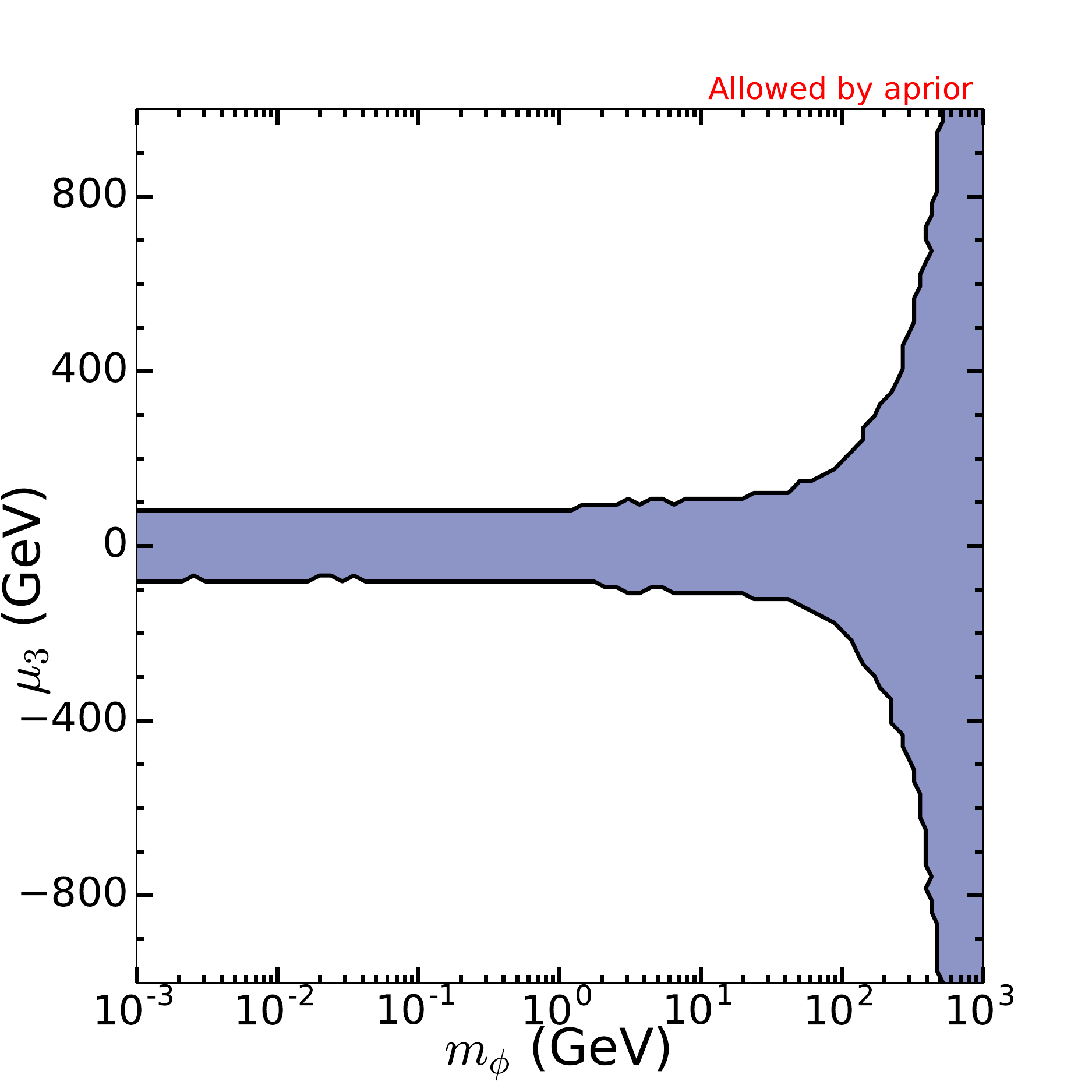}
	\includegraphics[height=1.32in,angle=0]{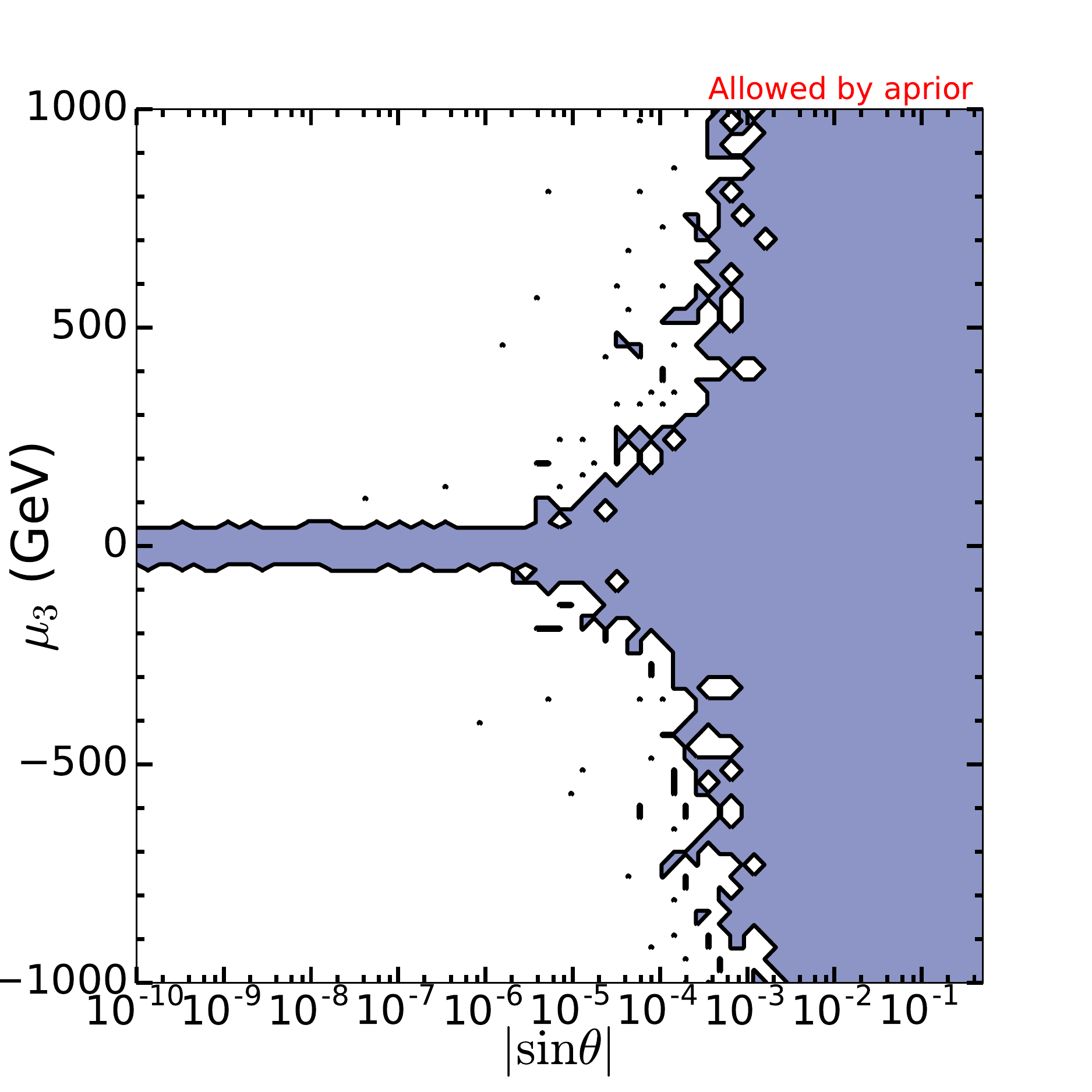}
	\newline
	\includegraphics[height=1.32in,angle=0]{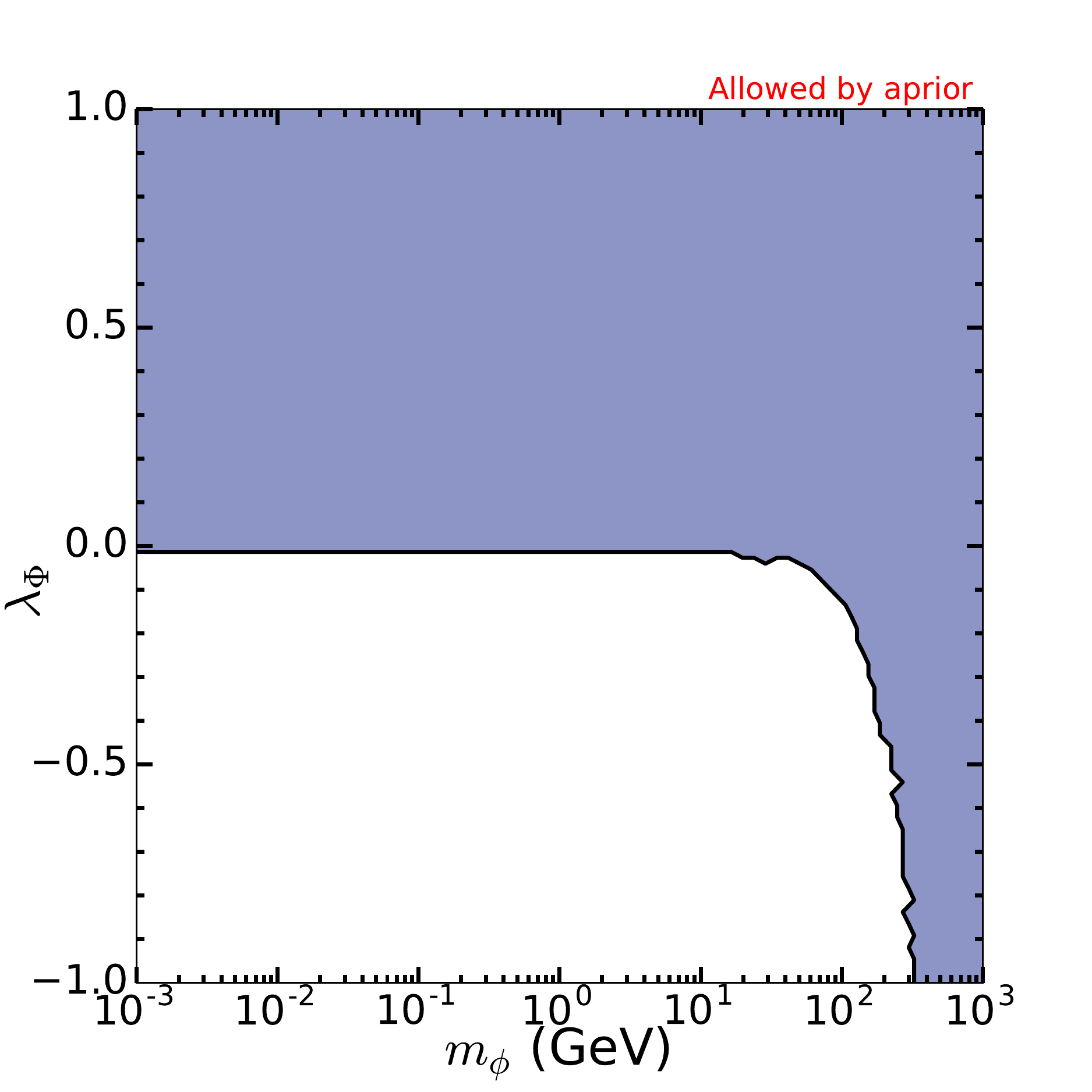}
	\includegraphics[height=1.32in,angle=0]{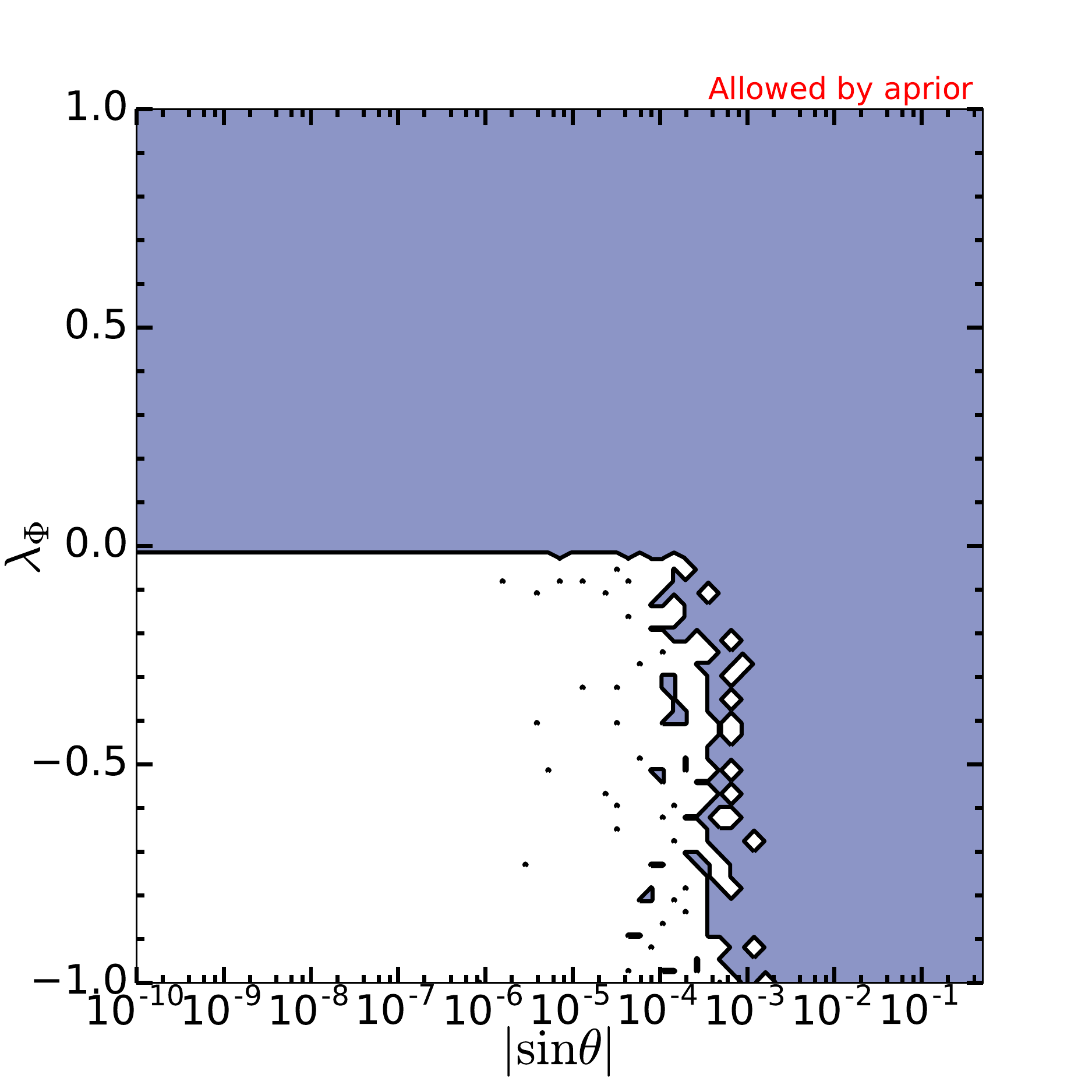}
	\includegraphics[height=1.32in,angle=0]{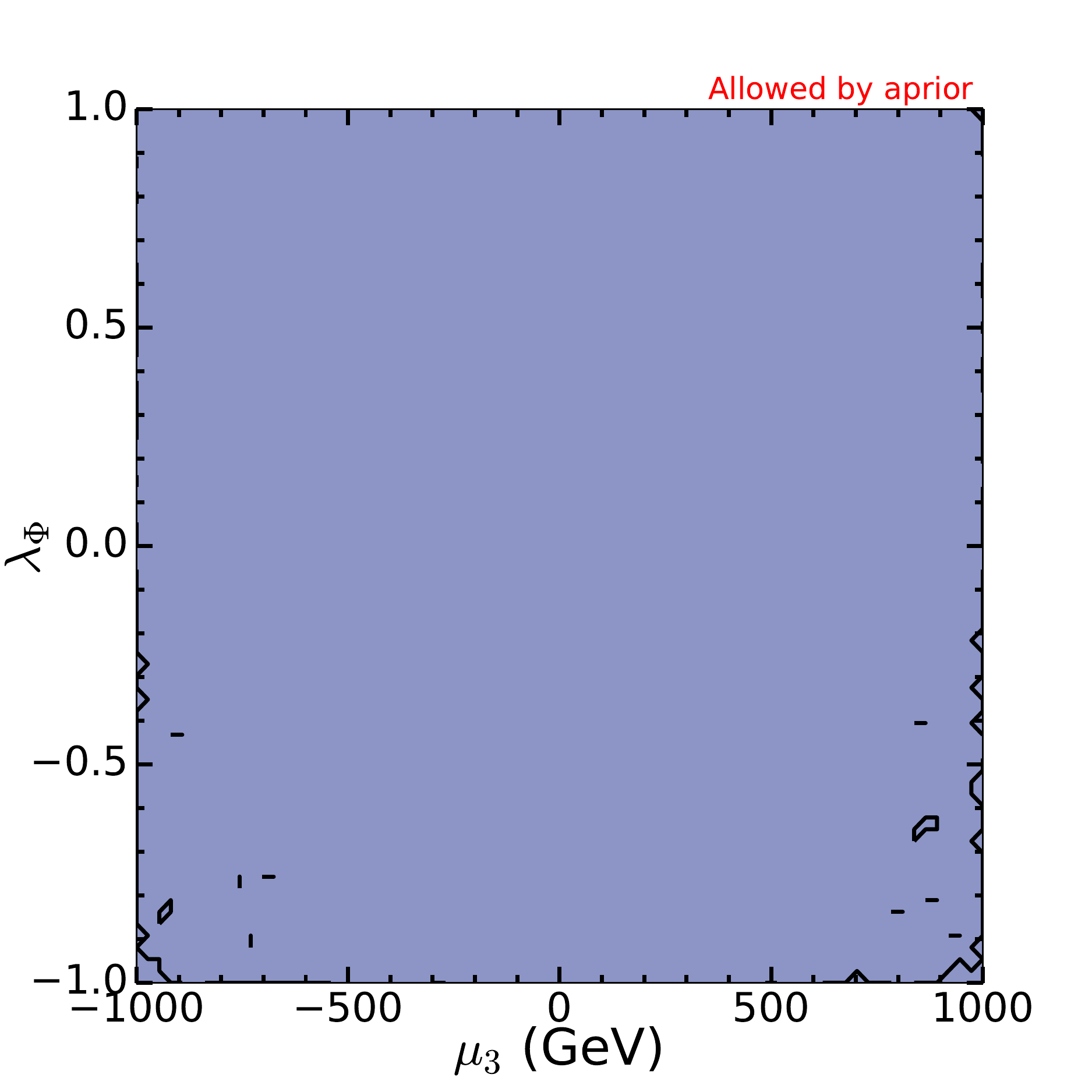}
	\newline
	\includegraphics[height=1.32in,angle=0]{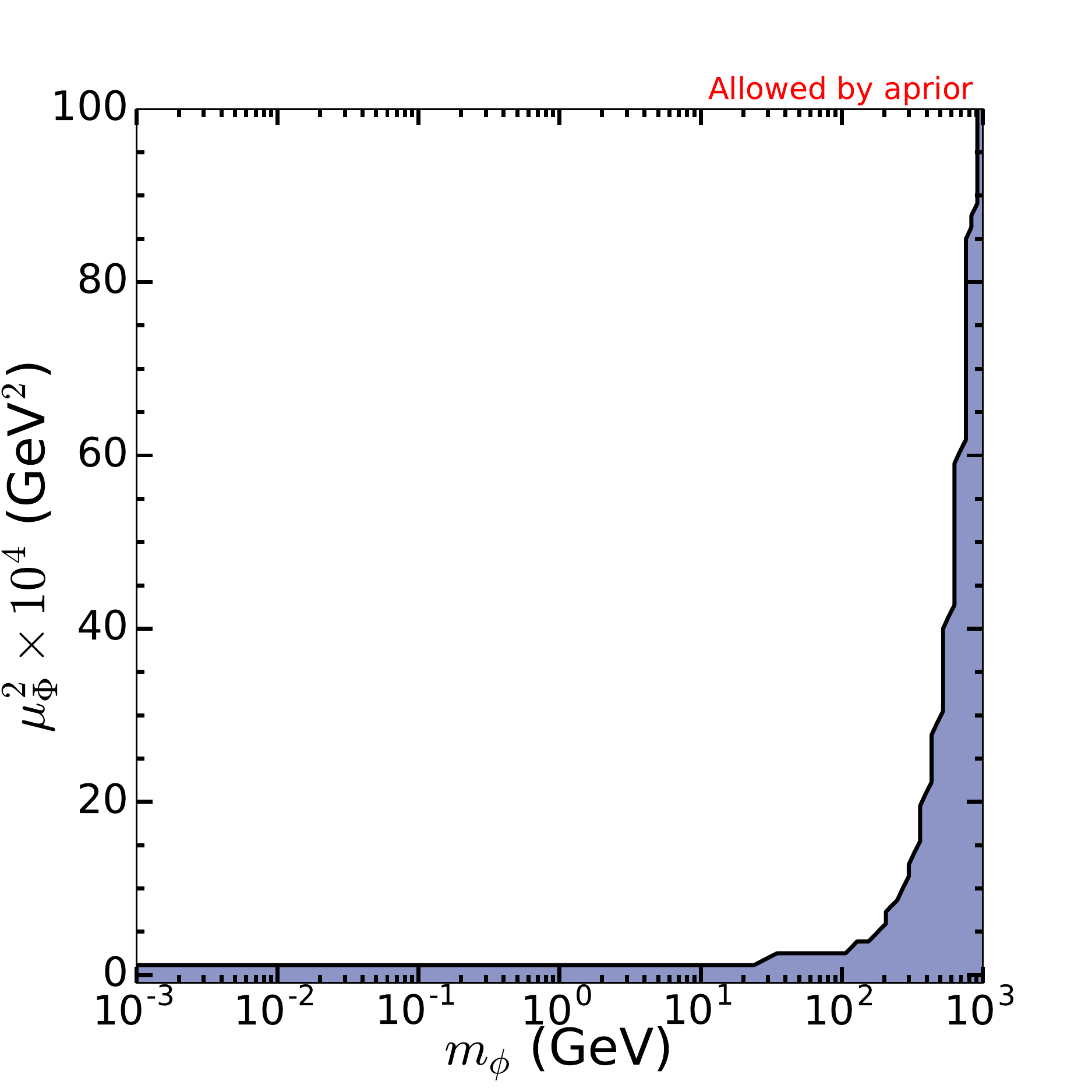}
	\includegraphics[height=1.32in,angle=0]{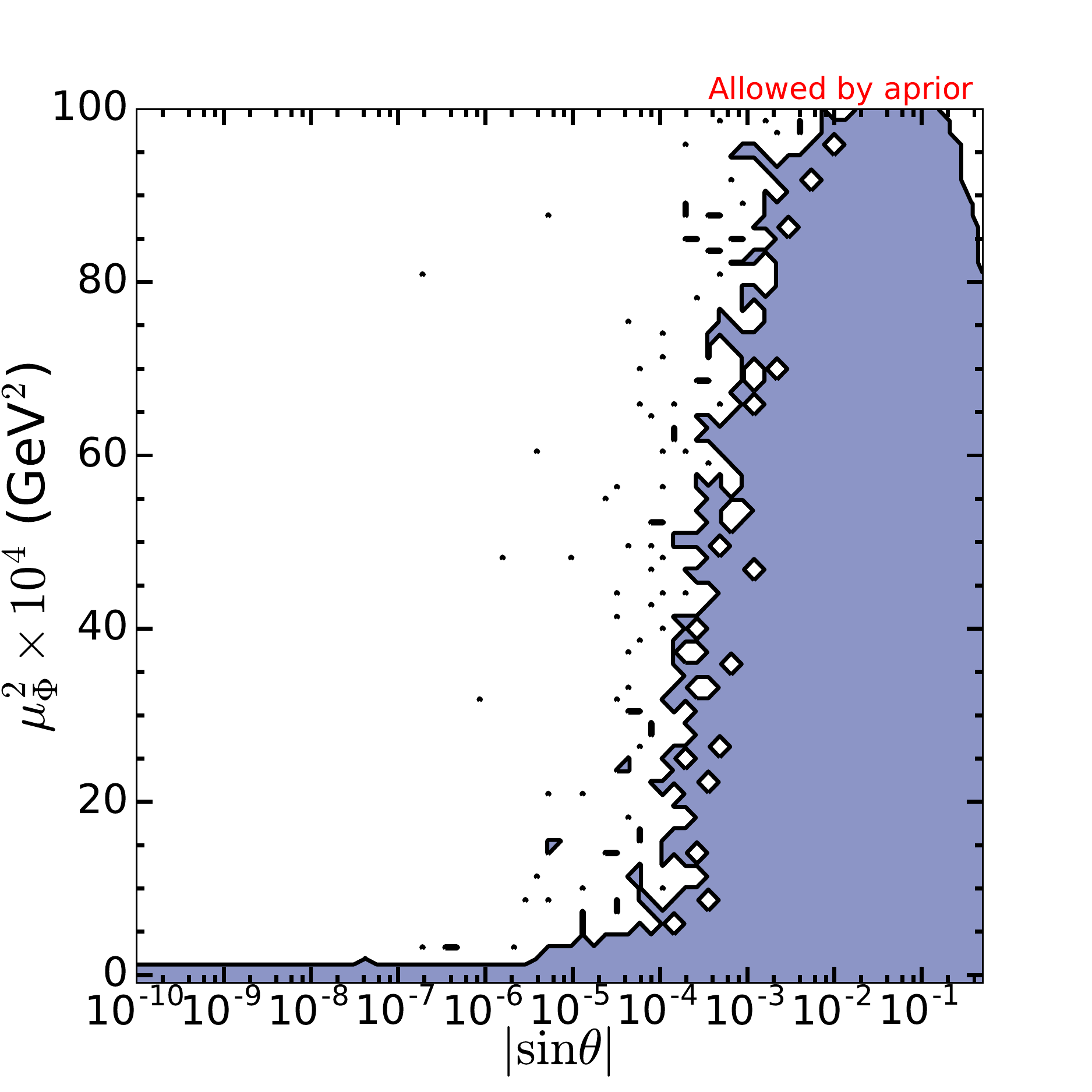}
	\includegraphics[height=1.32in,angle=0]{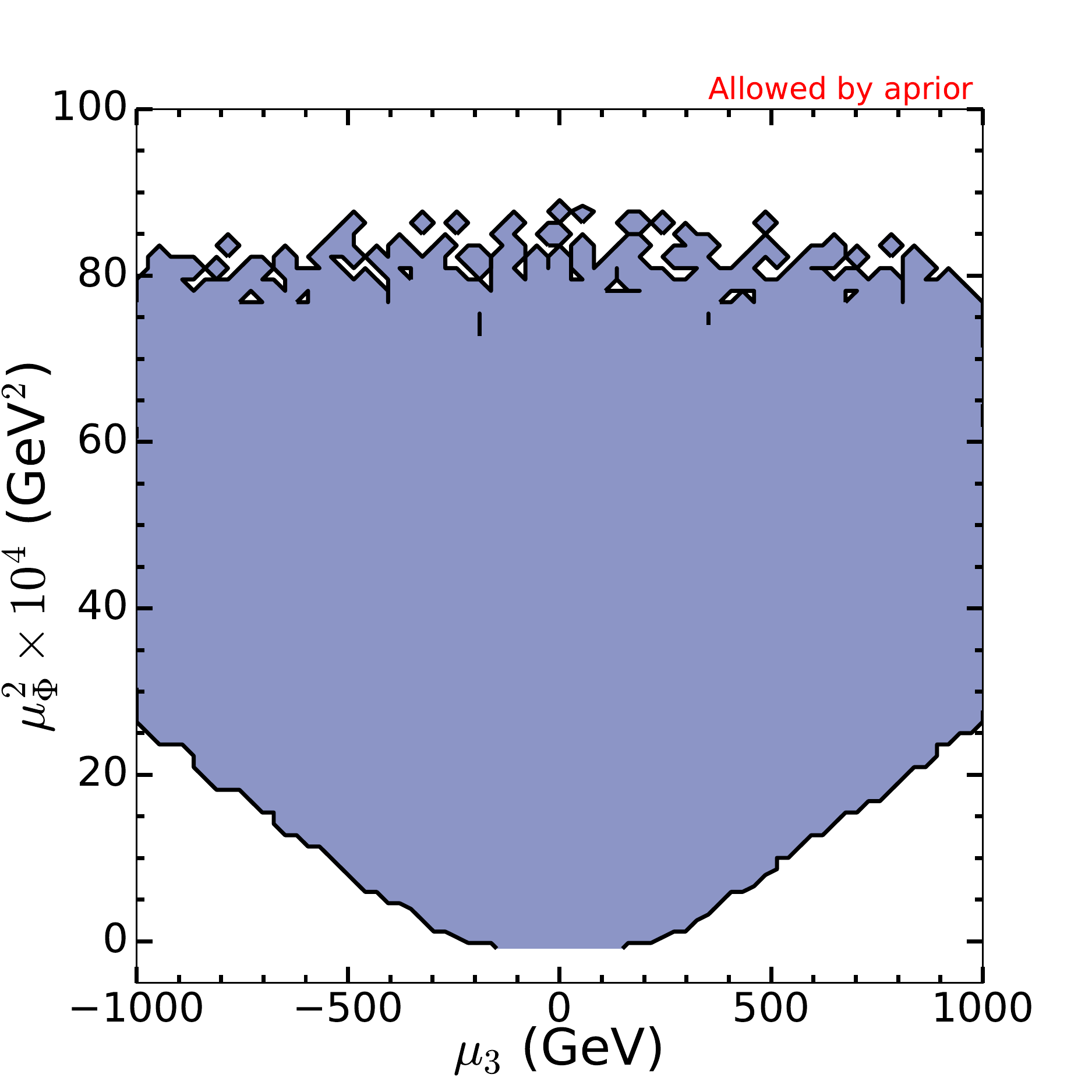}
	\includegraphics[height=1.32in,angle=0]{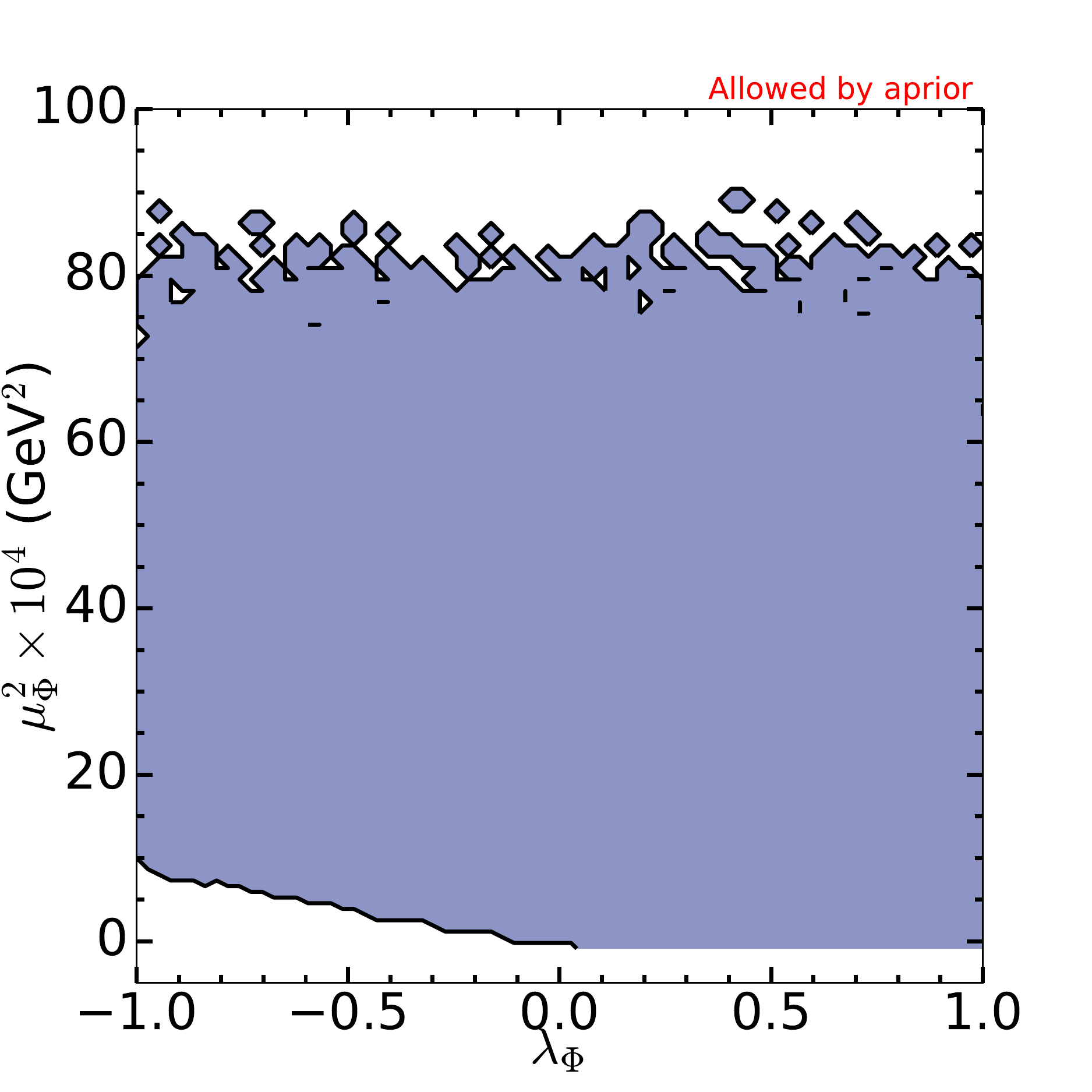}
	\caption{\small \sl Model parameter region survived after applying preselection criteria in section\,\ref{subsec: apriori constraints}.}
	\label{fig: After apriori constraints}
\end{figure}

As mentioned in section\,\ref{subsec: apriori constraints}, we apply preselection criteria on model parameters as $c_p = 0$, $|\theta| \leq \pi/6$ and $V(\eta, \xi) \geq V(v_\Phi, v_H)$ in the range of $|\xi| \leq 1$\,TeV and $|\eta| \leq 1$\,TeV, where $\xi$ and $\eta$ are defined as $\Phi = \eta$ and $H = (0, \xi/\sqrt{2})^T$, respectively. The region of the parameters survived after applying the criteria is shown in Fig.\,\ref{fig: After apriori constraints}. The survived parameter region is further investigated by the subsequent likelihood analysis, as discussed in section\,\ref{sec: constraints}.

Since the mediator mass is eventually constrained to be less than, or at least the same order of $m_\chi$ to satisfy the relic abundance condition, as seen in section\,\ref{subapp: RA and KE conditions}, let us focus on the region of $m_\phi \lesssim {\cal O}(10)$\,GeV and discuss how the preselection criteria work to restrict the model parameter space of the model. First, it can be seen in the result on the $(m_\phi, \mu_\Phi^2)$-plane that the parameter $\mu_\Phi^2$ is constrained to be $|\mu_\Phi^2| \lesssim (100\,{\rm GeV})^2$, because otherwise $m_\phi$ would be much heavier than ${\cal O}(10)$\,GeV. Next, the quartic coupling of the field $\Phi$ is constrained to be $\lambda_\Phi \gtrsim 0$ as seen in the result on the $(m_\phi, \lambda_\phi)$-plane to make our vacuum stable, because the contribution from the quadratic term $\mu_\Phi^2 \Phi^2$ is small. Finally, the parameter of the cubic term is constrained to be $|\mu_3| \lesssim 100$\,GeV, as seen in the result on the $(m_\phi, \mu_3)$-plane, to suppress the negative contribution to the potential in the region of $\Phi = \eta \lesssim 0$.

\section{Kinematical equilibrium condition}
\label{app: equilibrium condition}

In this appendix, we first write down all the reaction rates $\Gamma_{\chi {\rm SM}}$ $\Gamma_{\chi \phi}$ and $\Gamma_{\phi {\rm SM}}$ in a concrete form. Then, we present the model parameter region survived after applying the kinematical equilibrium condition as well as the preselection criteria and the relic abundance condition.

\subsection{The reaction rates}

We first consider the reaction rate between the WIMP and SM particles $\Gamma_{\chi {\rm SM}}$. Considering the fact that the WIMP is always non-relativistic while SM particles are relativistic during the freeze-our process, the reaction rate $\Gamma_{\chi {\rm SM}}$ is described by the following formula:
\begin{eqnarray}
	\Gamma_{\chi {\rm SM}} = \frac{2T_f}{m_{\chi}}
	\sum_i \langle \sigma_{\chi {\rm SM}_i} v \rangle_{T_f} \, n_{{\rm SM}_i}(T_f),
	\label{eq: Gamma WIMP SM}
\end{eqnarray}
where $n_{{\rm SM}_i}(T_f)$ is the number density of the SM particle `$i$' at the freeze-out temperature. Since relativistic SM particles dominantly contribute to the rate, its approximate form is given by $g_i \zeta(3) T_f^3/\pi^2$ if it is bosonic while $3 g_i \zeta(3) T_f^3/(4 \pi^2)$ 
if it is fermionic.
Here, $g_i$ is the color and spin degrees of freedom of the particle `$i$'.\footnote{The exact formula of the number density, $\int d^3k\,g_i/[\exp(-E_k/T) \mp 1]$, is used in our numerical analysis.} The prefactor $2 T_f/m_\chi$ comes from the fact that about $m_\chi/(2 T_f)$ times collisions are required to maintain the kinematical equilibrium, because the typical momentum transfer in a single collision is estimated to be $\Delta q^2/q^2 \sim 2 T_f/m_\chi \ll 1$ in a stochastic process\,\cite{Boehm:2013jpa, Gondolo:2012vh, Binder:2016pnr}. The thermal-averaged scattering cross section $\langle \sigma_{\chi {\rm SM}_i} v \rangle_{T_f}$ is calculated according to the following formula\,\cite{Gondolo:1990dk}:
\begin{eqnarray}
	\left\langle \sigma_{\chi {\rm SM}_i} v \right\rangle_{T_f}
	= \frac{ 2\pi^2 T_f \int_{(m_\chi + m_{{\rm SM}_i})^2}^{\infty} ds\,
	s^{3/2} \sigma_{\chi {\rm SM}_i}
	\left(1 - \frac{2(m^2_\chi + m^2_{{\rm SM}_i})}{s} + \frac{(m^2_\chi - m^2_{{\rm SM}_i})^2}{s^2} \right) K_1(s^{1/2}/T_f) }
	{ \left[ 4 \pi T_f m^2_\chi K_2(m_\chi/T_f) \right] \left[ 4\pi T_f m^2_{{\rm SM}_i} K_2(m_{{\rm SM}_i}/T_f) \right] },
	\label{eq: thermal average}
\end{eqnarray}
where $K_n(z)$ is the modified Bessel function of the 2nd kind. When the freeze-out temperature is less than the QCD scale, namely $T_f \leq \Lambda_{\rm QCD} \simeq 155$\,MeV\,\cite{Bhattacharya:2014ara}, the scattering process between the WIMP and an electron or a muon by the exchange of $\phi$ in the $t$-channel dominantly contributes to the reaction rate. On the other hand, when $T_f \geq \Lambda_{\rm QCD}$, scattering processes with various SM particles contribute to the rate. Those are scattering processes of the WIMP with a muon, a tau lepton as well as strange, charm and bottom quarks.

We next consider the reaction rate between the WIMP and the mediator, $\Gamma_{\chi \phi}$. Since the mediator can be either relativistic or non-relativistic at the freeze-out temperature depending on its mass $m_\phi$, the rate is given by the formula which is similar to that in eq.\,(\ref{eq: Gamma WIMP SM}):
\begin{eqnarray}
	\Gamma_{\chi \phi} = F\left( \frac{m_\chi}{T_f}, \frac{m_\phi}{T_f} \right) \langle \sigma_{\chi \phi} v \rangle n_\phi(T_f),
\end{eqnarray}
where $n_\phi(T_f)$ is the number density of the mediator at the freeze-out temperature. The prefactor $F(x_1,x_2)$ is the extension of that in eq.\,(\ref{eq: Gamma WIMP SM}), and it can be estimated based on the following discussion. First, when $x_1 \gg 1$ \& $x_2 \ll 1$, or $x_1 \ll 1$ \& $x_2 \gg 1$, the prefactor should be the same as the one in eq.\,(\ref{eq: Gamma WIMP SM}), namely $F(x_1, x_2) = 2/x_1$ or $2/x_2$. Next, when both arguments are small enough, the prefactor $F(x_1, x_2)$ should be ${\cal O}(1)$, for the momentum transfer $\Delta q^2/q^2$ between relativistic particles is so large that a single collision is (almost) enough to maintain the kinematical equilibrium. Detailed calculation of the momentum transfer between relativistic particles expects that the prefactor is $F(x_1, x_2) \simeq 1/2$. Finally, when the both arguments are (almost) the same and large enough, the prefactor should be ${\cal O}(1)$ again, for the momentum transfer is expected to be efficient between two non-relativistic particles whose masses are (almost) the same. Taking the above discussion into account, we use the following formula for estimating the prefactor $F(x_1, x_2)$:
\begin{eqnarray}
	F(x_1, x_2) &=&
	\left\lbrace
	\begin{array}{lcl} 
		1/2 & {\rm for} & x_1 \leq 4~~\&~~x_2 \leq 4\\
		2/x_1 & {\rm for} & x_1 \geq 4~~\&~~x_2 \leq 4\\
		2/x_2 & {\rm for} & x_1 \leq 4~~\&~~x_2 \geq 4\\
		{\rm NR}(x_1,x_2) & {\rm for} & x_1 \geq 4~~\&~~x_2 \geq 4,
	\end{array}
	\right. \nonumber \\
	{\rm NR}(x_1,x_2) &=&\,
	\left\lbrace
	\begin{array}{lcl}
		x_2/(2 x_1) & ~~\,{\rm for} & x_1 \geq x_2\\
		x_1/(2 x_2) & ~~\,{\rm for} & x_2 \geq x_1.
	\end{array}
	\right.
	\label{eq: prefactor}
\end{eqnarray}
The scattering process between the WIMP and the mediator $\phi$ comes from diagrams exchanging the WIMP in the $s$- and $u$-channels as well as a diagram exchanging the mediator in the $t$-channel. Its thermal-averaged scattering cross section, $\langle \sigma_{\chi \phi} v \rangle$, is calculated using the same formula as that in eq.\,(\ref{eq: thermal average}) with $m_{{\rm SM}_i}$ being replaced by the mediator mass $m_\phi$.

We finally consider the reaction rate between the mediator $\phi$ and SM particles. Three different processes contribute to the rate: the (inverse) decay process $\Gamma_{\phi \leftrightarrow {\rm SMs}}$, the scattering process $\Gamma_{\phi {\rm SM} \leftrightarrow \phi {\rm SM}}$ and the absorption (emmision) process $\Gamma_{\phi {\rm SM} \leftrightarrow {\rm SMs}}$. The contribution from {\bf the (inverse) decay process} is given by the thermal-average of the total decay width of the mediator $\Gamma_\phi$ that has been discussed in section\,\ref{subsec: phi decay}. Its explicit form is given as follows:
\begin{equation}
	\Gamma_{\phi \leftrightarrow {\rm SMs}} =
	\left \langle \frac{1}{\gamma} \right \rangle \Gamma_{\phi} =
	\frac{ \int_1^\infty d\gamma \, \gamma^{-1} \, [\gamma (\gamma^2 - 1)^{1/2}\,e^{-\gamma m_\phi/T}] }
	{ \int_1^\infty d\gamma \, [\gamma (\gamma^2 - 1)^{1/2}\,e^{-\gamma m_\phi/T}] } \Gamma_\phi,
\end{equation}
where $\gamma$ is the so-called the Lorentz gamma factor, and the distribution in the square brackets is the Maxwell-Juttner distribution. The contribution from {\bf the scattering process} is given by a similar formula discussed in eq.\,(\ref{eq: Gamma WIMP SM}). Since the mediator can be non-relativistic or relativistic at the freeze-out temperature $T_f$, while relativistic SM particles dominantly contribute to the reaction rate, the explicit form of the contribution is given as follows:
\begin{eqnarray}
	\Gamma_{{\phi {\rm SM} \leftrightarrow \phi {\rm SM}}} = \sum_i 
	F\left( \frac{m_\phi}{T_f}, \frac{m_{{\rm SM}_i}}{T_f} \right)
	\langle \sigma_{\phi {\rm SM}_i} v \rangle_{T_f} \, n_{{\rm SM}_i}(T_f),
	\label{eq: Gamma phi SM}
\end{eqnarray}
When $T_f \leq \Lambda_{\rm QCD}$, the scattering process between $\phi$ and an electron or a muon dominantly contributes to the reaction rate. On the other hand, when $T_f \geq \Lambda_{\rm QCD}$, scattering processes of $\phi$ with a muon, a tau lepton as well as strange, charm and bottom quarks contribute to the rate. Corresponding thermal-averaged scattering cross section are calculated using the same formula as that in eq.\,(\ref{eq: thermal average}) with $m_\chi$ being replaced by $m_\phi$. The contribution from {\bf the absorption (emmision) process} is given by the same formula as that in eq.\,(\ref{eq: Gamma phi SM}) again:
\begin{eqnarray}
	\Gamma_{\phi {\rm SM} \leftrightarrow {\rm SM}s} = \sum_i 
	F\left( \frac{m_\phi}{T_f}, \frac{m_{{\rm SM}_i}}{T_f} \right)
	\langle \sigma'_{\phi {\rm SM}_i} v \rangle_{T_f} \, n_{{\rm SM}_i}(T_f),
\end{eqnarray}
where the scattering cross section $\sigma_{\phi {\rm SM}_i}$ in eq.\,(\ref{eq: Gamma phi SM}) is replaced by the cross section of the absorption (emmision) process, $\sigma'_{\phi {\rm SM}_i}$. After the QCD phase transition, $T_f \leq \Lambda_{\rm QCD}$, the following three processes contribute to this reaction rate: $\phi \gamma \to f \bar{f}$, $\phi f \to \gamma f$, $\phi \bar{f} \to \gamma \bar{f}$, where $f$ is an electron or a muon, while $\bar{f}$ is its anti-particle. On the other hand, when $T_f \geq \Lambda_{\rm QCD}$, in addition to the above processes, other three processes, $\phi g \to q \bar{q}$, $\phi q \to g q$ and $\phi \bar{q} \to g \bar{q}$ also contribute to the rate with $q$ ($\bar{q}$) being a quark (anti-quark). Their corresponding thermal-averaged cross sections are calculated by the same formula as that in eq.\,(\ref{eq: Gamma phi SM}). As a result, the reaction rate between the mediator $\phi$ and SM particles is given by the sum of all processes mentioned above: $\Gamma_{\phi{\rm SM}} = \Gamma_{\phi \leftrightarrow {\rm SMs}} + \Gamma_{{\phi {\rm SM} \leftrightarrow \phi {\rm SM}}} + \Gamma_{\phi {\rm SM} \leftrightarrow {\rm SM}s}$.

\begin{figure}[t!]
	\centering
	\includegraphics[height=2.2in,angle=0]{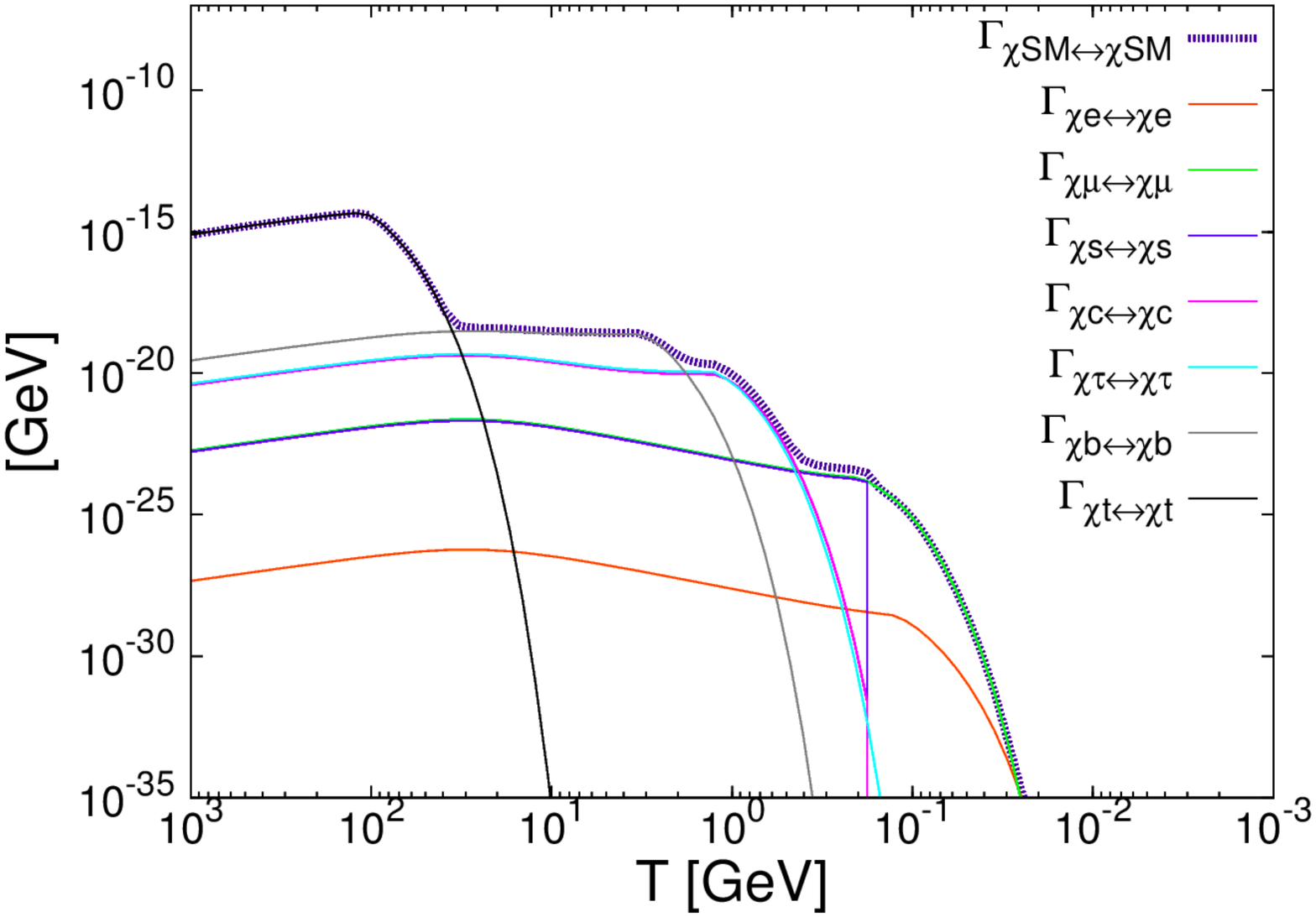}
	\includegraphics[height=2.2in,angle=0]{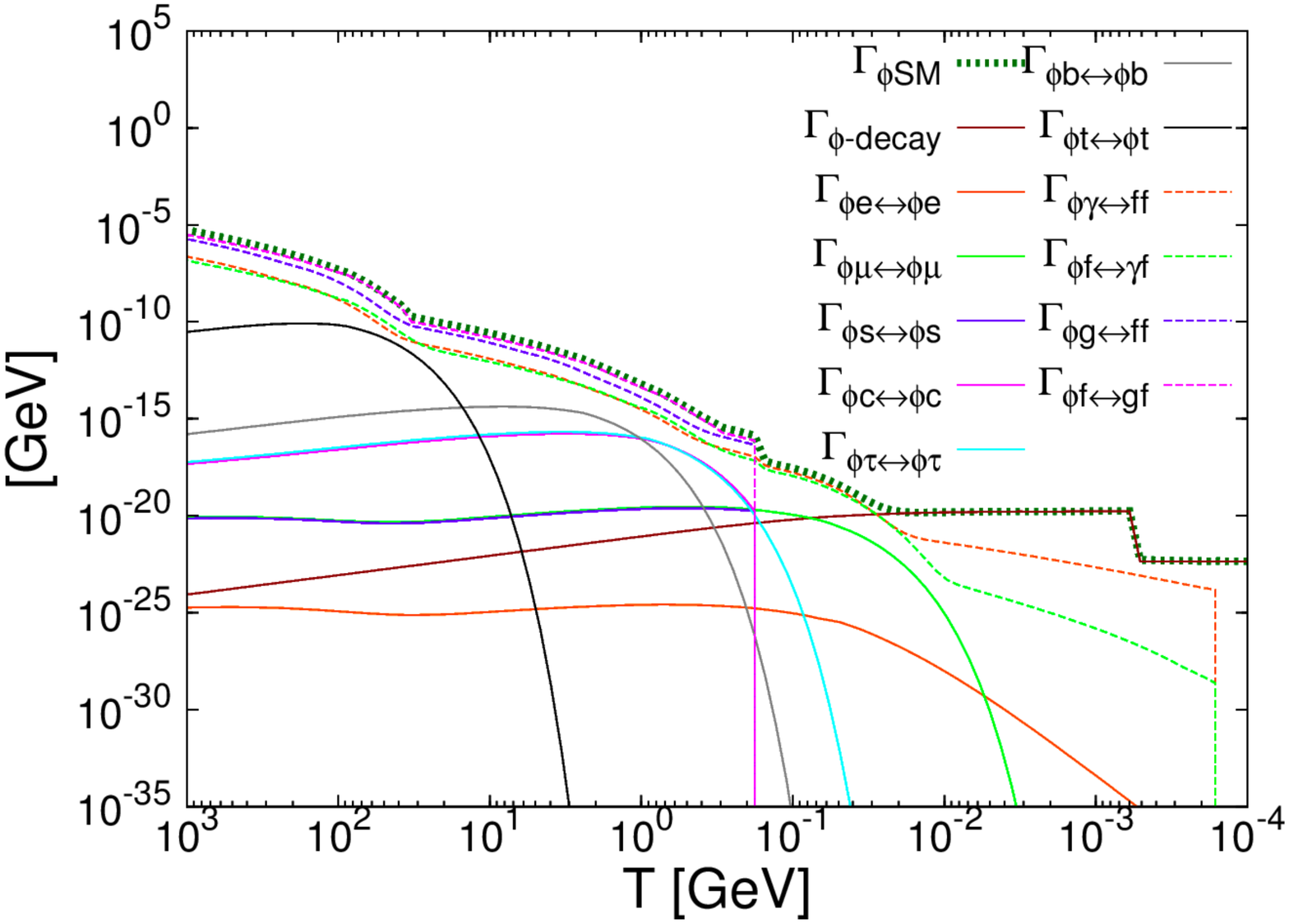}
	\caption{\small \sl Processes contributing to the rates $\Gamma_{\chi {\rm SM}}$ (left panel) and $\Gamma_{\phi {\rm SM}}$ (right panel) around the freeze-out temperature $T_f$. Model parameters are set to be the same as those adopted in Fig.\,\ref{fig: rates}.}
	\label{fig: rates2}
\end{figure}

As a demonstration, we show how each concrete process contributes to the reaction rates $\Gamma_{\chi {\rm SM}}$ and $\Gamma_{\phi {\rm SM}}$ in the left and the right panels of Fig.\,\ref{fig: rates2}, respectively, at around the freeze-out temperature $T_f$. Model parameters are set to be the same as those adopted in Fig.\,\ref{fig: rates}, namely ($m_\chi$, $c_s$, $m_\phi$, $\sin \theta$, $\mu_3$) are fixed to be (200\,MeV, 0.022, 100\,MeV, 10$^{-3}$, 10\,MeV).

\subsection{Parameter region after the equilibrium condition applyied}
\label{subapp: RA and KE conditions}

\begin{figure}[t!]
	\includegraphics[height=1.52in,angle=0]{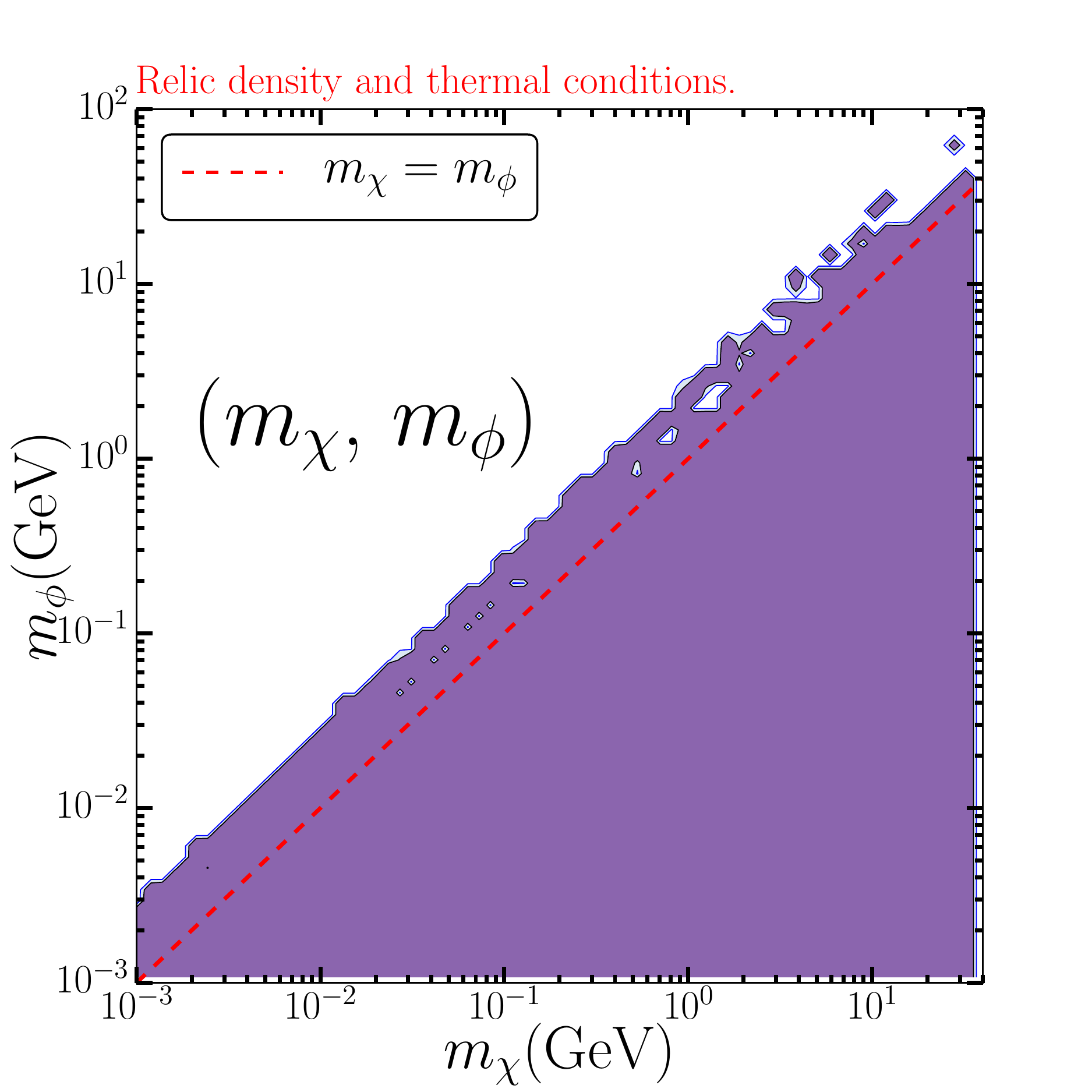}
	\newline
	\includegraphics[height=1.52in,angle=0]{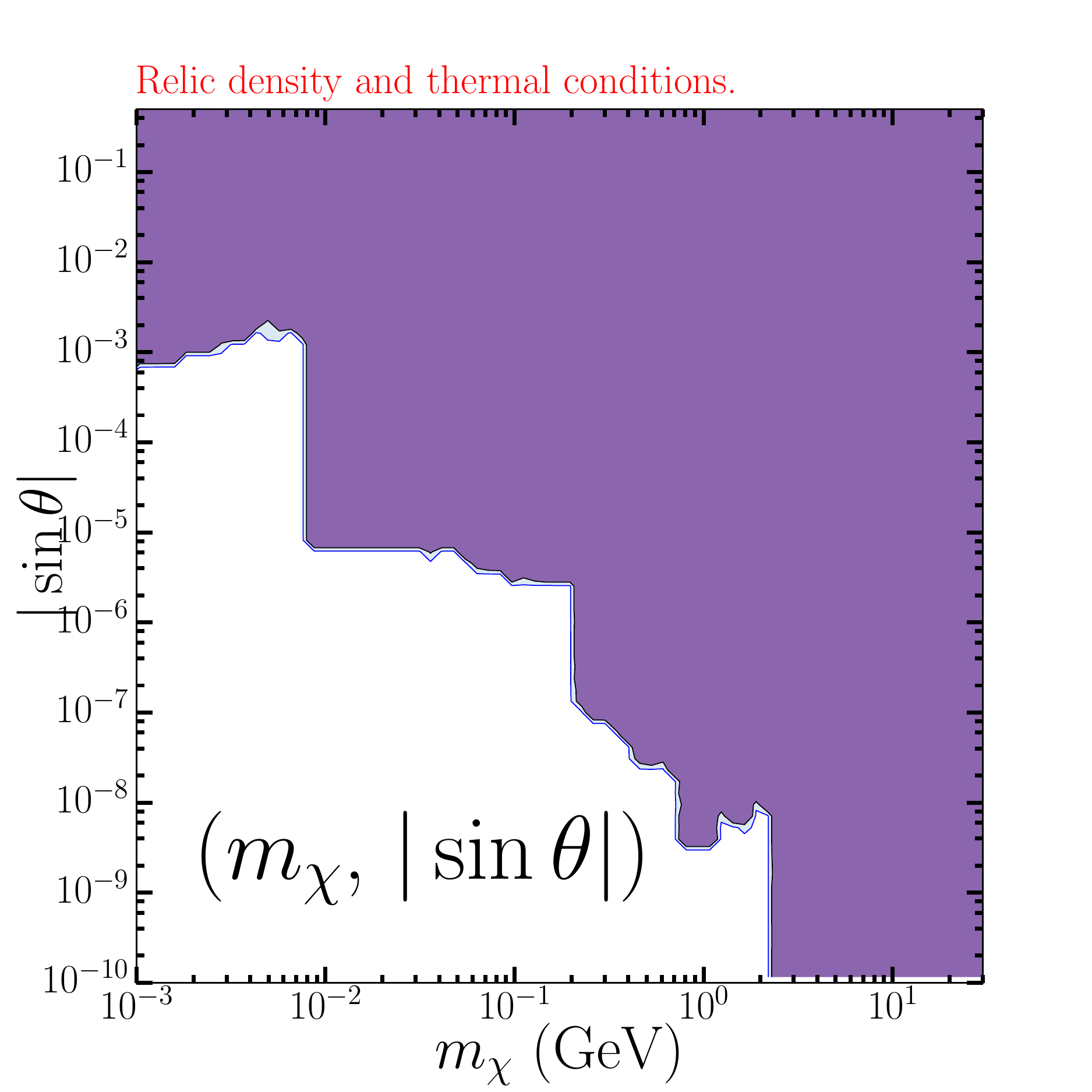}
	\includegraphics[height=1.52in,angle=0]{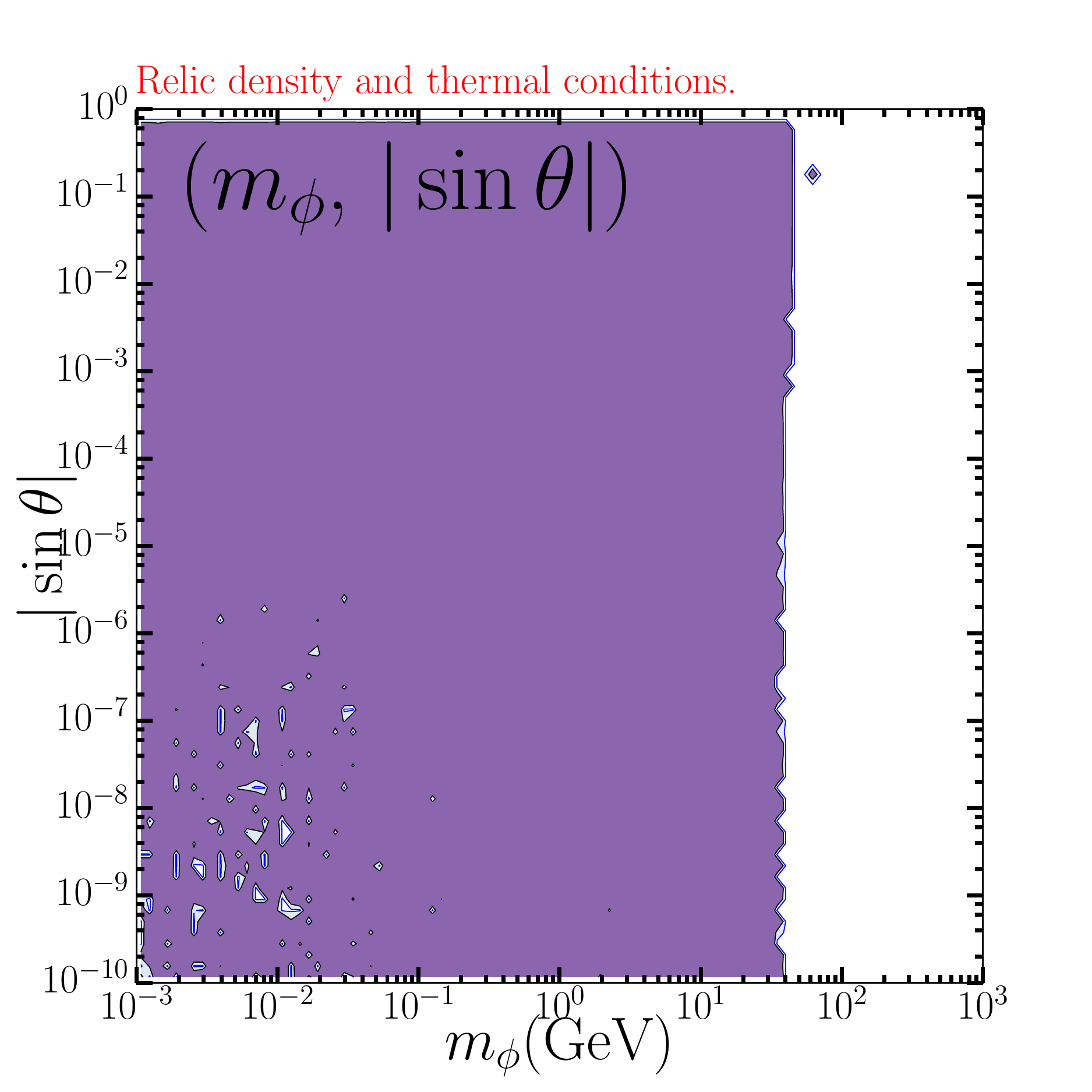}
	\newline
	\includegraphics[height=1.52in,angle=0]{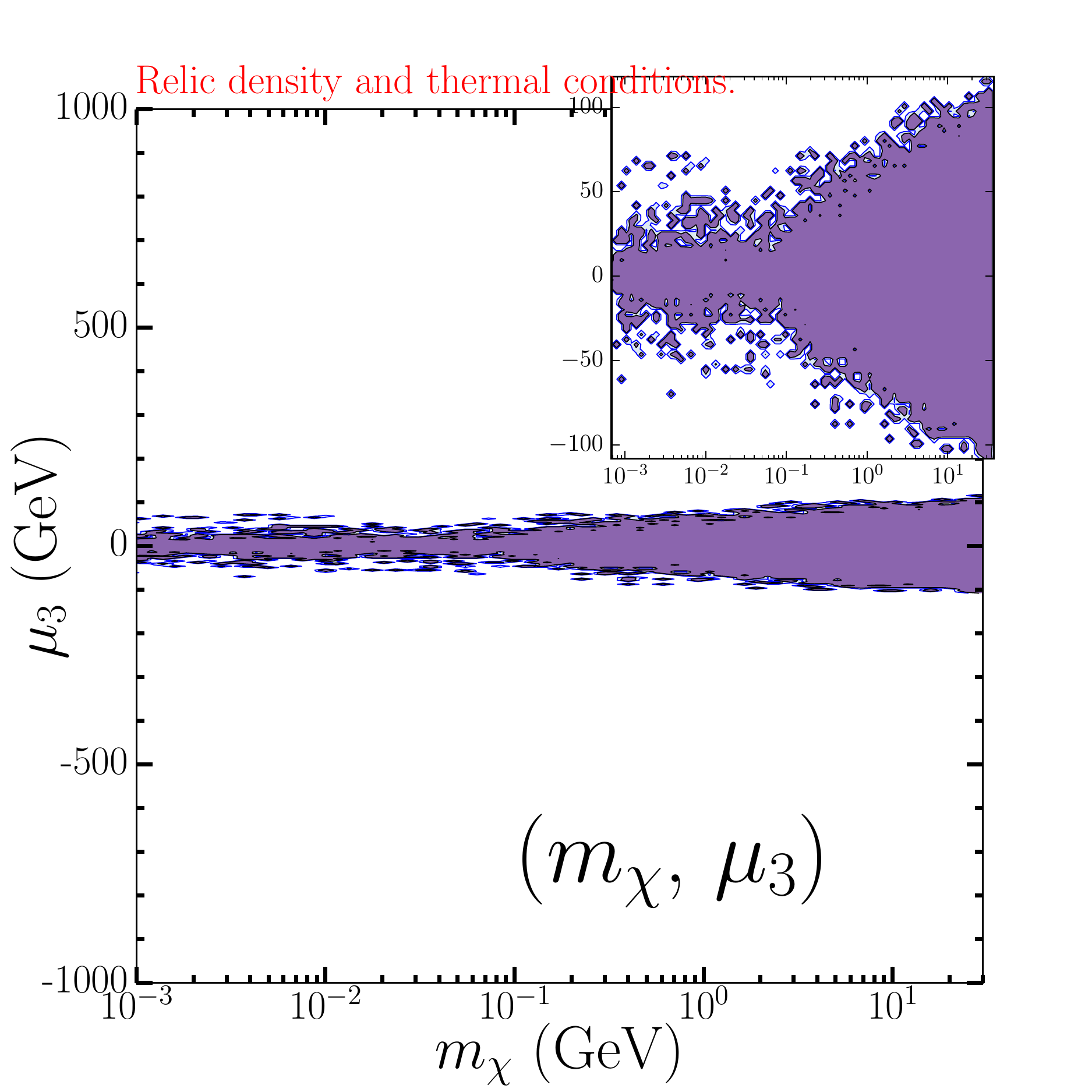}
	\includegraphics[height=1.52in,angle=0]{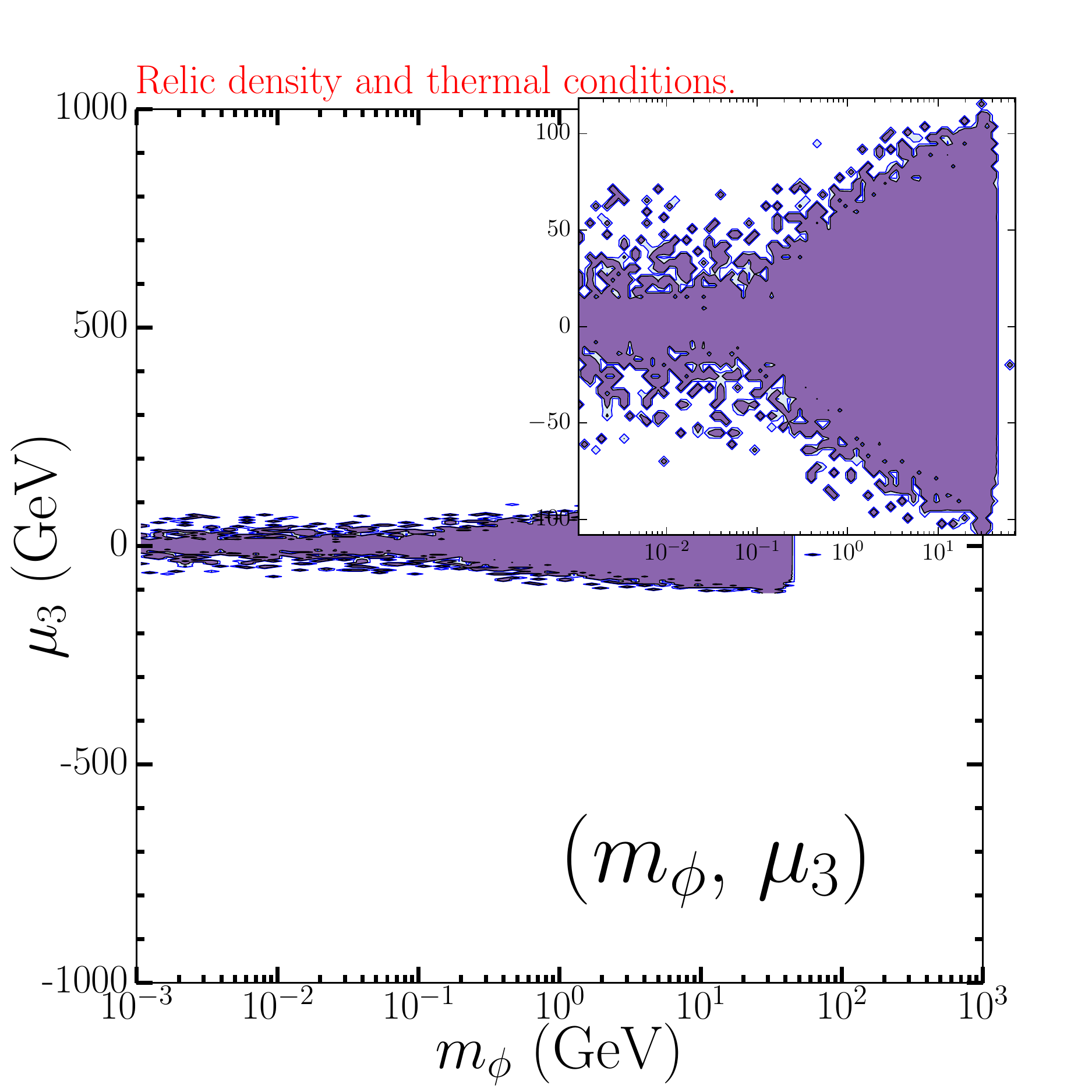}
	\includegraphics[height=1.52in,angle=0]{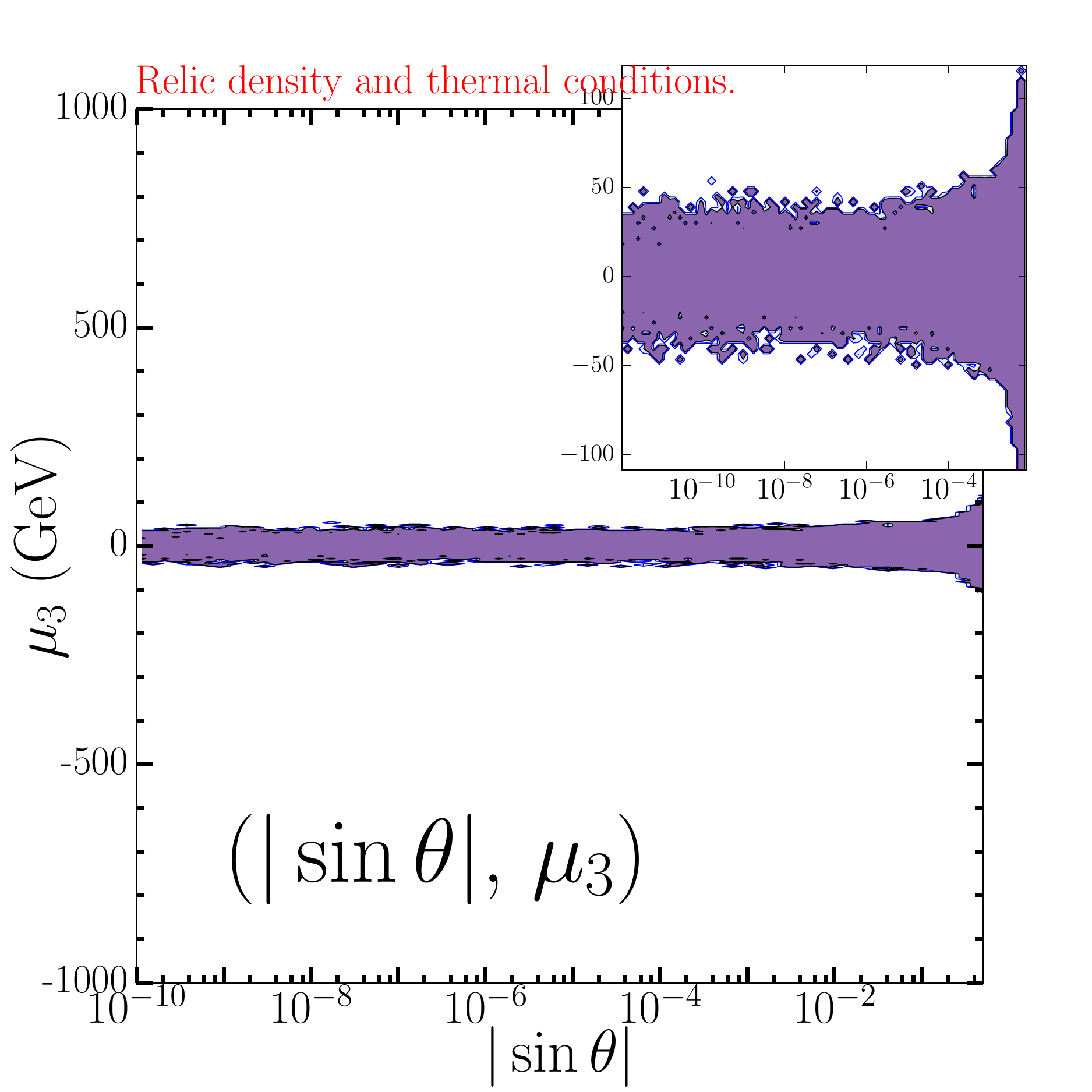}
	\newline
	\includegraphics[height=1.52in,angle=0]{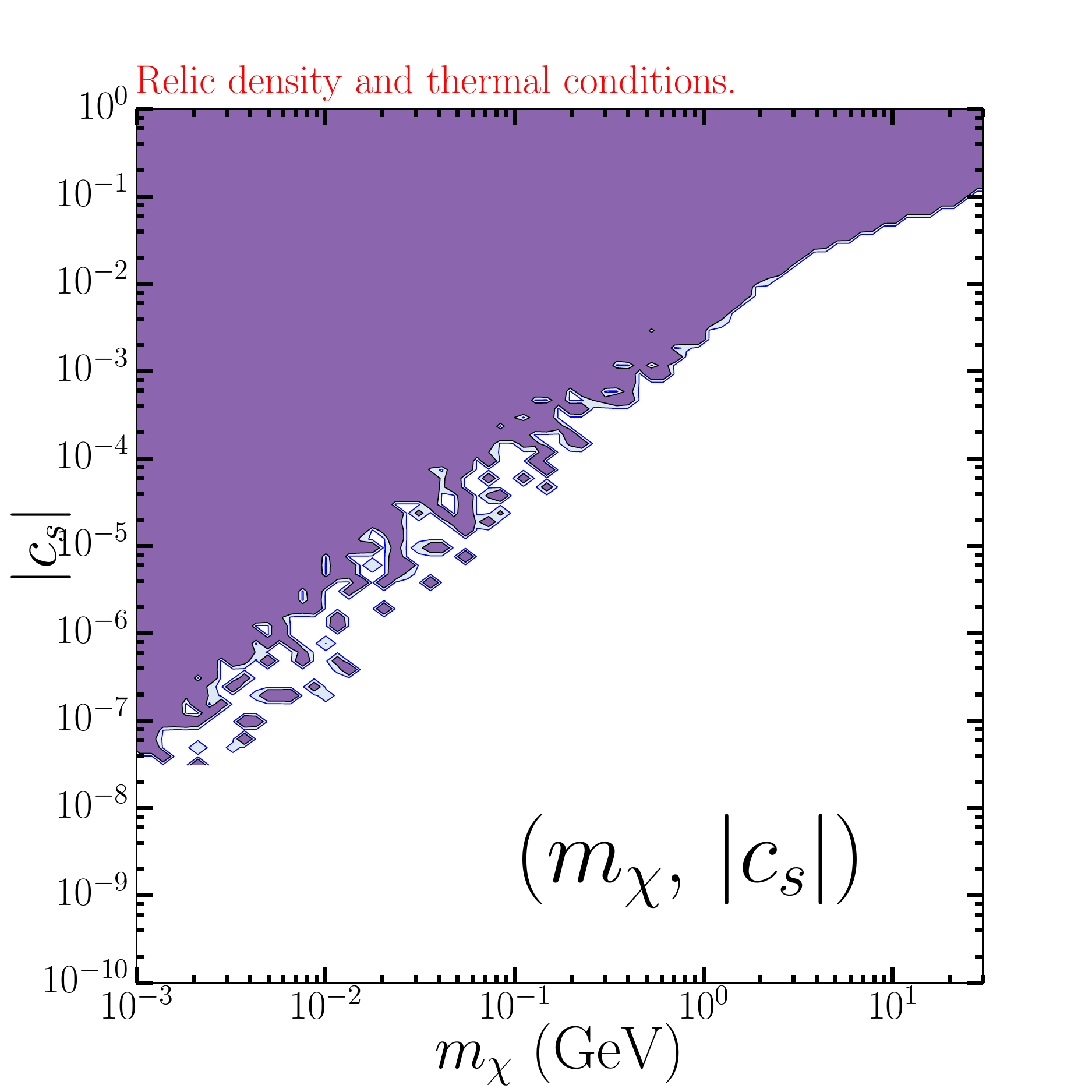}
	\includegraphics[height=1.52in,angle=0]{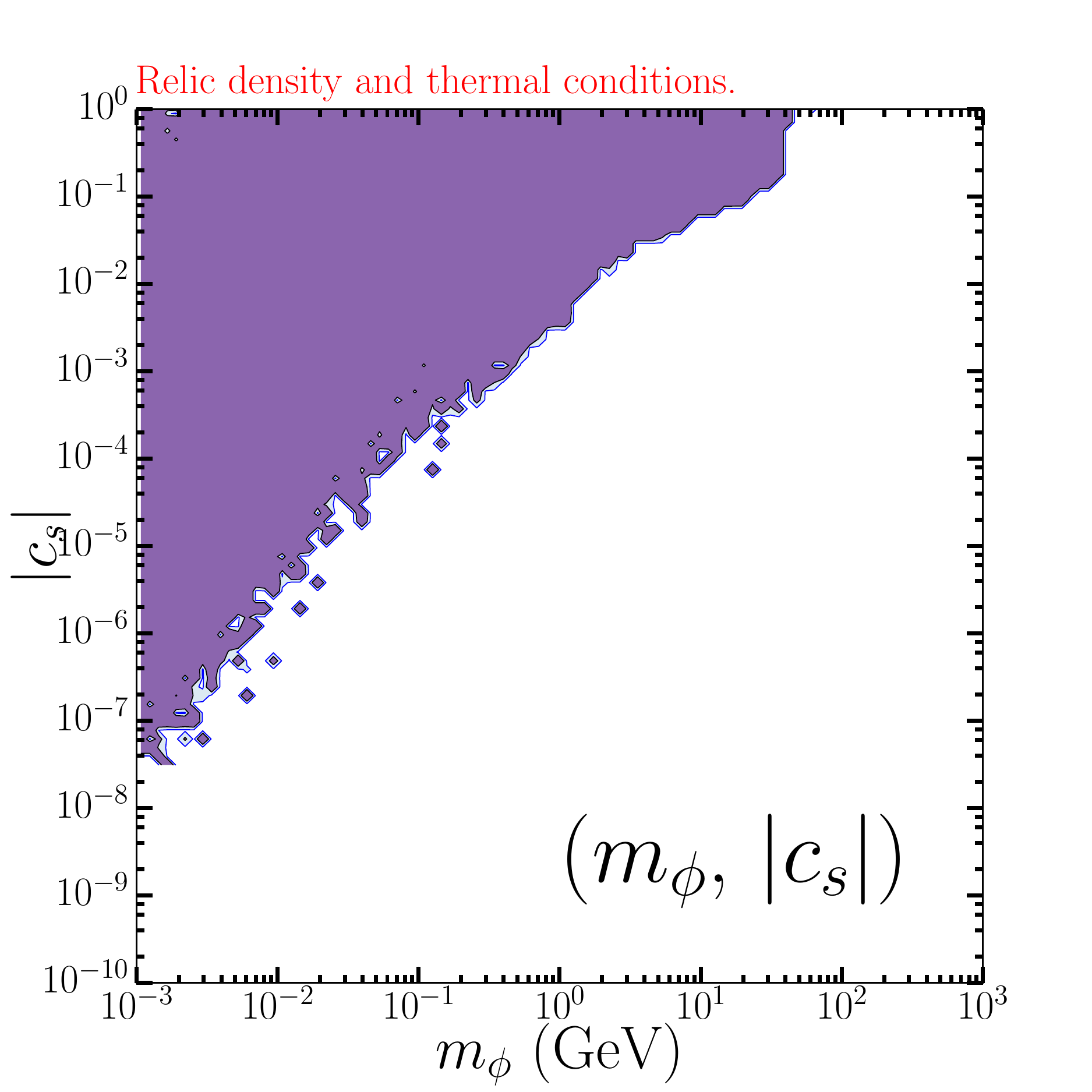}
	\includegraphics[height=1.52in,angle=0]{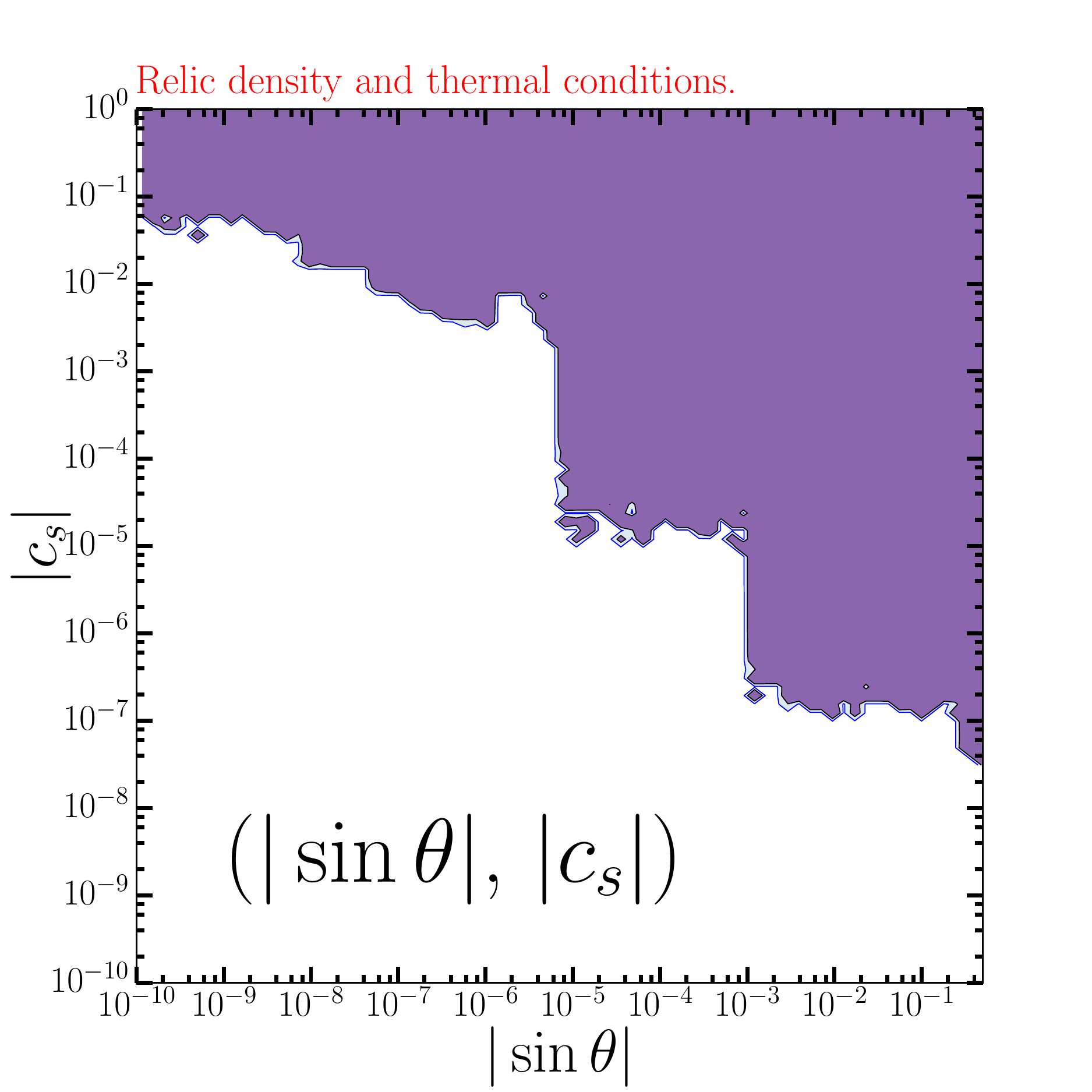}
	\includegraphics[height=1.52in,angle=0]{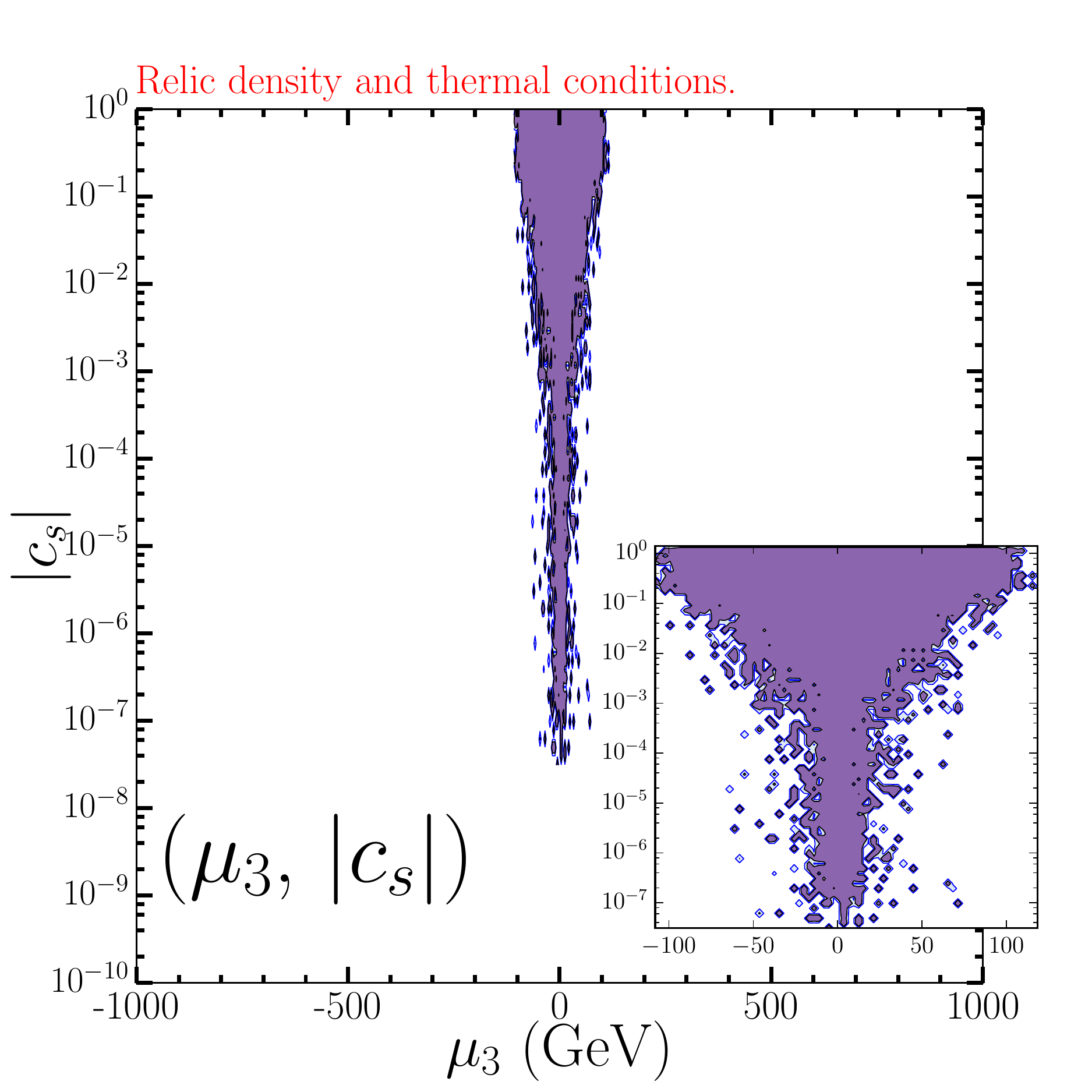}
	\caption{\small \sl Model parameter region survived after applying relic abundance and kinematical equilibrium conditions discussed in section\,\ref{subsec: cosmological constraints} as well as preselection criteria in the previous appendix.}
	\label{fig: After AP RE KE constraints}
\end{figure}

Here, we present the model parameter region survived after applying the kinematical equilibrium condition as well as the preselection criteria and the relic abundance condition. As we mentioned in section\,\ref{subsubsec: parameters}, model parameters ($m_{\chi}$, $c_s$, $m_\phi$, $\theta$, $\mu_3$) among seven independent parameters ($m_{\chi}$, $c_s$, $m_\phi$, $\theta$, $\mu_3$, $\mu_\Phi^2$, $\lambda_\Phi$) are relevant to the following discussion, so that we show the result in Fig.\,\ref{fig: After AP RE KE constraints} for the five parameters. In other words, parameters $\mu_\Phi^2$ and $\lambda_\Phi$ become nuisance parameters in the following phenomenological studies in this paper.

First, from the result on the $(m_\chi, m_\phi)$-plane, the mediator mass shall be at most around the WIMP mass to satisfy the relic abundance condition. It leads to the fact that the mediator is lighter than ${\cal O}(10)$\,GeV as long as we are discussing the light WIMP scenario. The upper limit on the mediator mass is also seen from the results on other planes spanned by $m_\phi$.

Next, as seen from the result on the $(m_\chi, |\sin \theta|)$-plane, the lower limit on $|\sin \theta|$ exists when $m_\chi \lesssim 2$\,GeV, which is required by the kinematical equilibrium condition. On the other hand, when $m_\chi \gtrsim 2$\,GeV, the mixing angle is not bounded from below. This is because $\phi$ can be in a kinematical equilibrium with SM particles not by the mixing angle but through interactions with an unsuppressed coupling $\lambda_{\Phi H}$ and not-so-suppressed Yukawa couplings.\footnote{The kinematical equilibrium is maintained by the $\phi f \to \phi f$ process with the Higgs boson being exchanged in the t-channel, where $f$ is a SM fermion. This process is, however, suppressed when the freeze-out temperature is less than the scale of the QCD phase transition, so that we have $m_\chi \sim 20 T_f \gtrsim 20 \Lambda_{\rm QCD} \sim 2$\,GeV.}

Third, the result on the planes spanned by the tri-linear coupling parameter $\mu_3$ shows that the parameter is indeed restricted to be around ${\cal O}(100)$\,GeV, as addressed in appendix\,\ref{app: initial parameter region}.

Fourth, the coupling constant between the dark matter and the mediator $|c_s|$ is bounded from below due to the relic abundance condition, as can be seen from the result on the $(m_\phi, |c_s|)$-plane. Here, one might worry about that the relic abundance of the WIMP becomes too small in the region with small $m_\chi$ and large $|c_s|$, however this region is satisfied by the case where $m_\chi$ is slightly lighter than 2$m_\phi$ but enough larger than $m_\phi$, namely the relic abundance condition is satisfied by the $\chi \chi \to \phi \to f \bar{f}$ process instead of $\chi \chi \to \phi \phi$.

Fifth, the reason why the result on the $(m_\phi, |c_s|)$-plane is similar to that on the $(m_\chi, |c_s|)$-plane is simply because the relation $m_\phi \leq {\cal O}(m_\chi)$ holds in the whole parameter region.

Finally, on the $(|\sin \theta|, |c_s|)$-plane, the reason why the region with small $|\sin \theta|$ and small $|c_s|$ is excluded is as follows. When $|\sin \theta|$ is small, $m_\chi$ must be heavy enough to satisfy the kinematical equilibrium condition, as seen from the result on the $(m_\chi, |\sin \theta|)$-plane. On the other hand, $m_\chi$ must be light enough when $c_s$ is small to satisfy the relic abundance condition, as seen in the result on the $(m_\chi, |c_s|)$-plane. As a result, small $|\sin \theta|$ and small $|c_s|$ are not simultaneously realized due to the contradiction between the two conditions.

\section{Supplemental figures}
\label{app: supplemental figures}

\subsection{Present status}
\label{app: present status}

\begin{figure}[t!]
	\includegraphics[height=1.52in,angle=0]{current_mx_mphi_like}
	\newline
	\includegraphics[height=1.52in,angle=0]{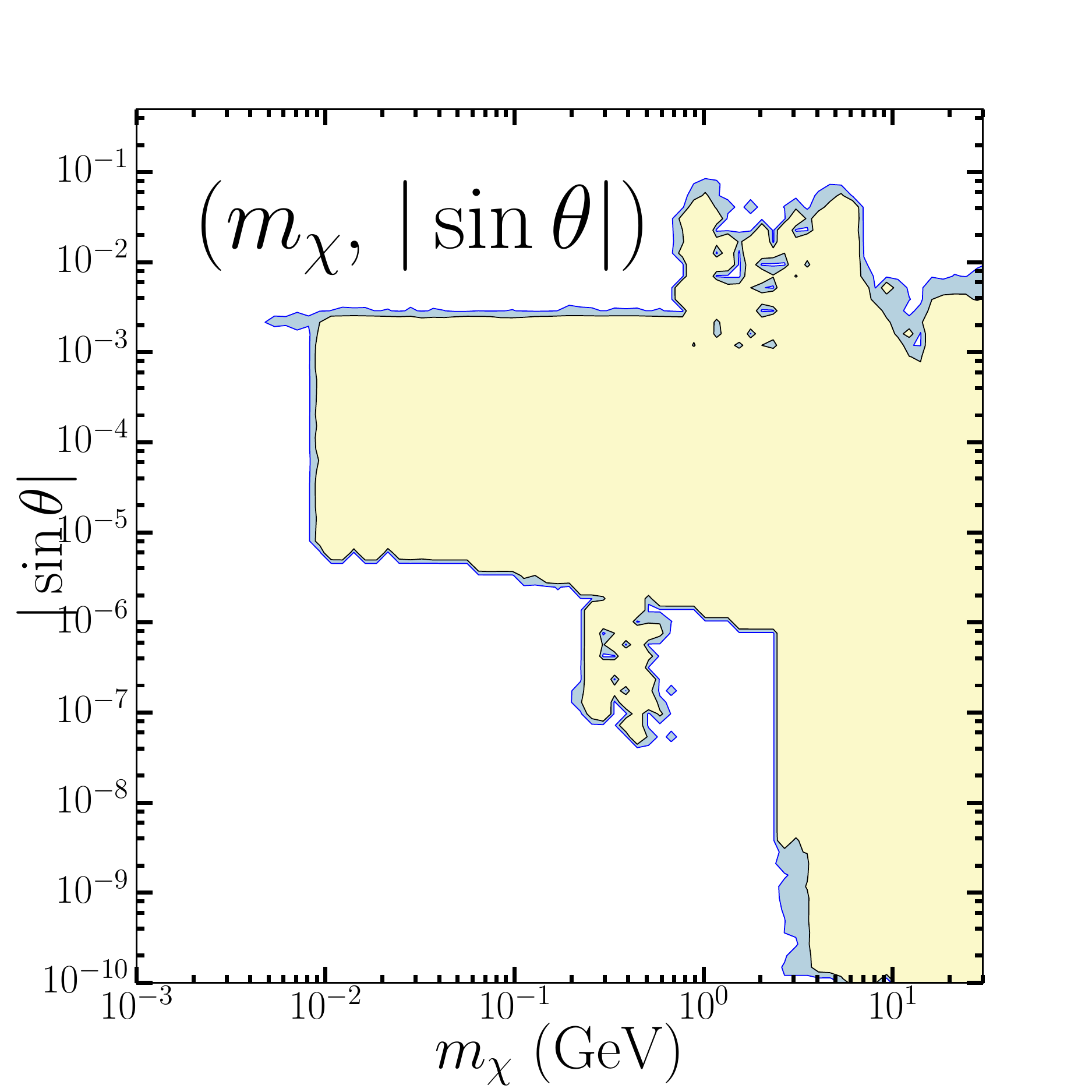}
	\includegraphics[height=1.52in,angle=0]{current_mphi_logsth_like}
	\newline
	\includegraphics[height=1.52in,angle=0]{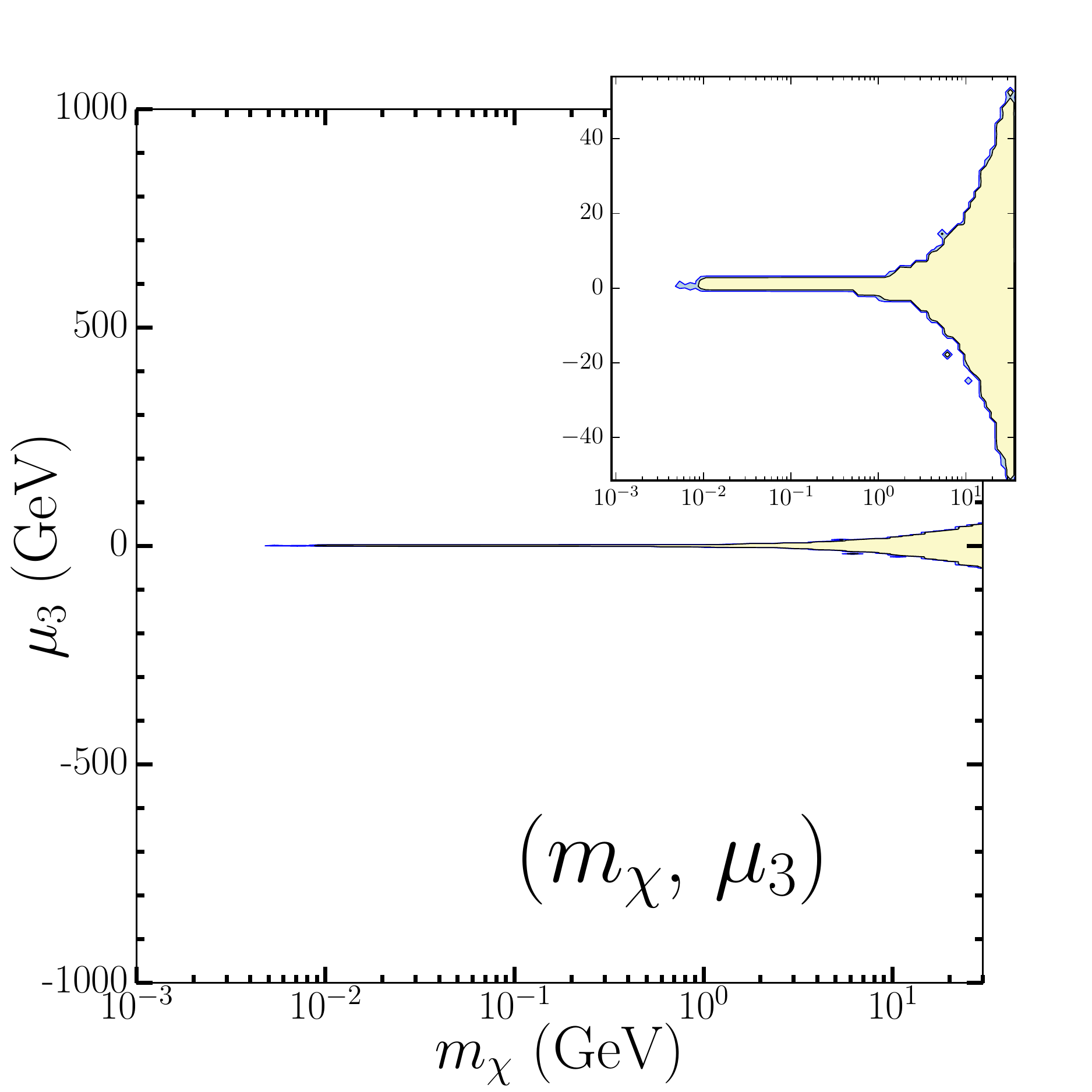}
	\includegraphics[height=1.52in,angle=0]{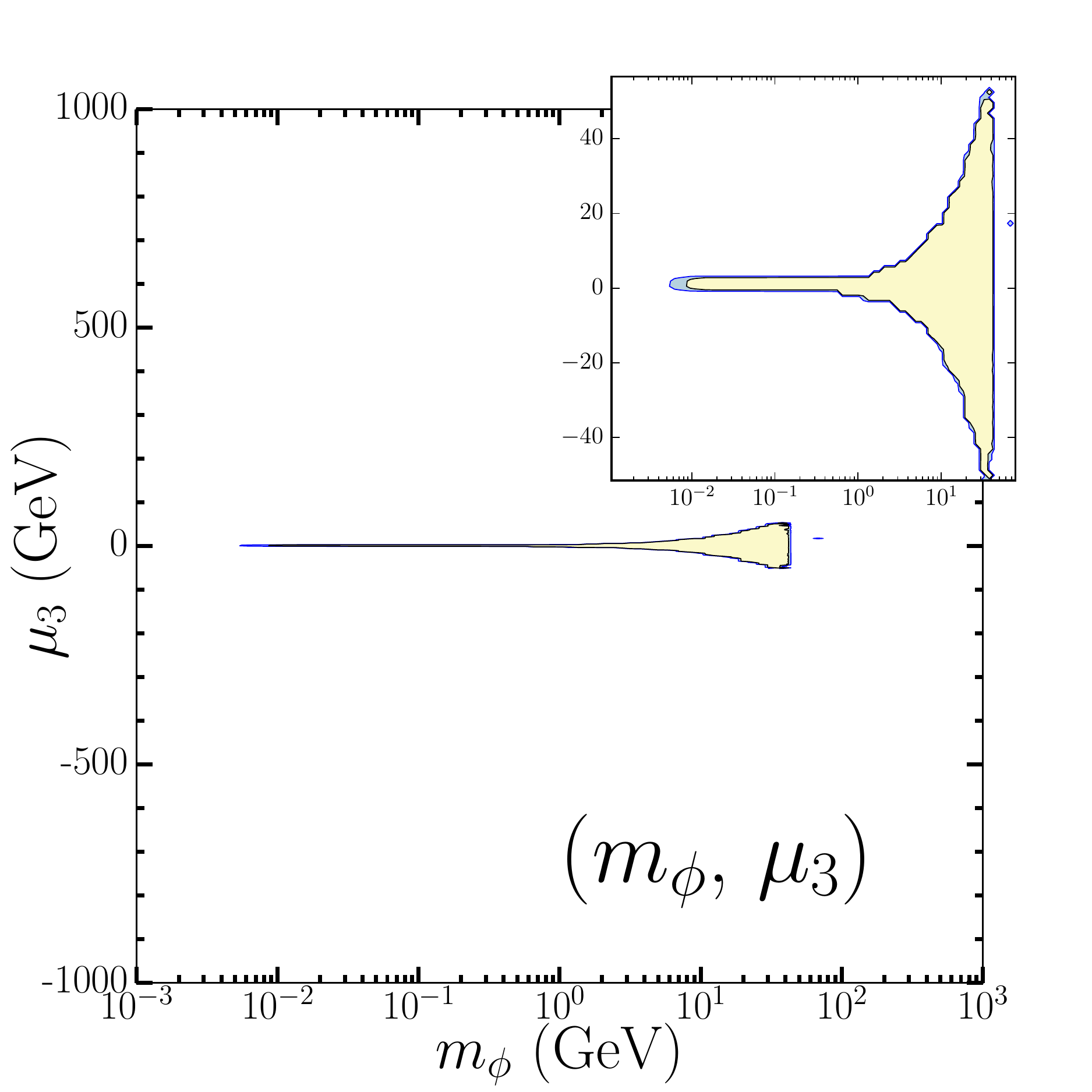}
	\includegraphics[height=1.52in,angle=0]{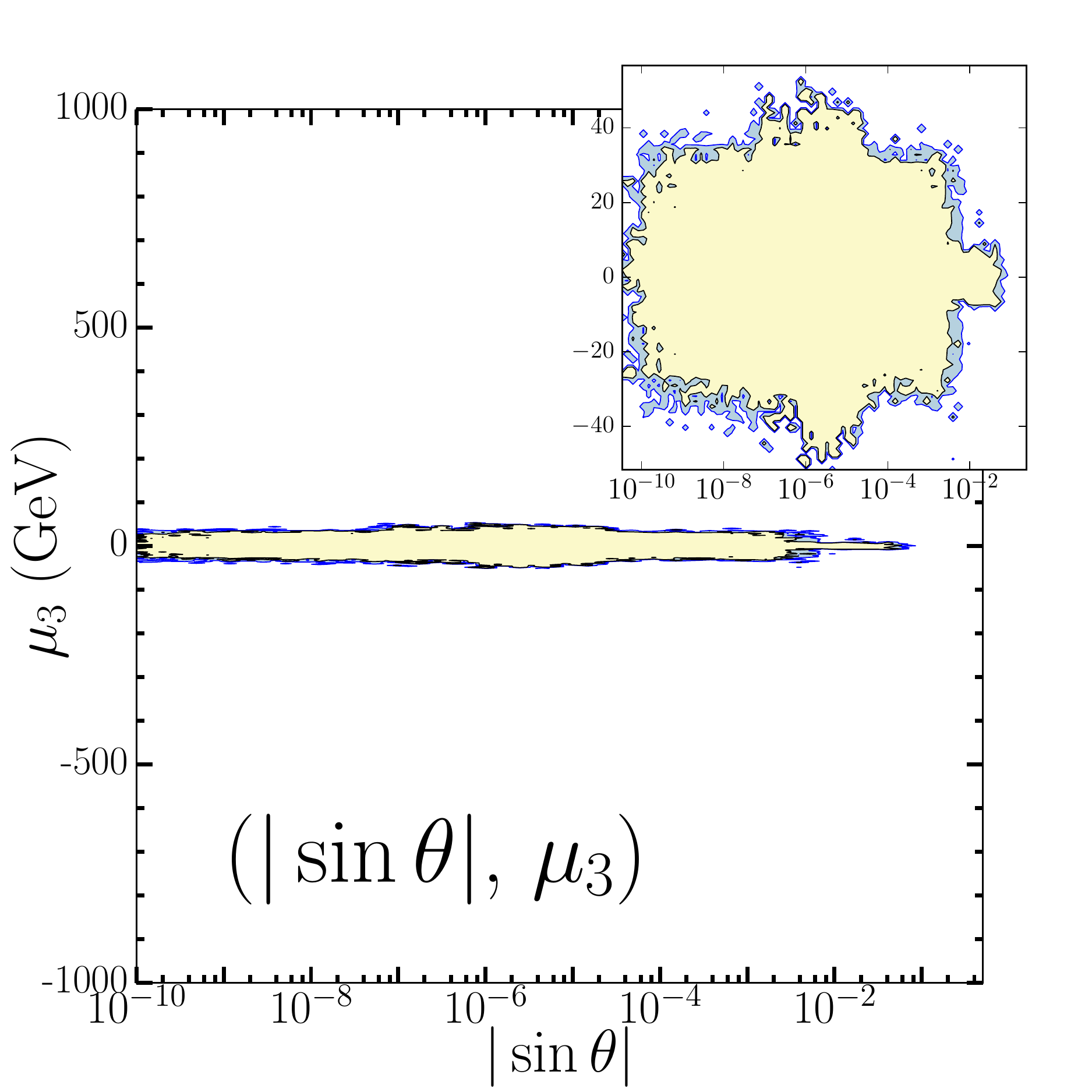}
	\newline
	\includegraphics[height=1.52in,angle=0]{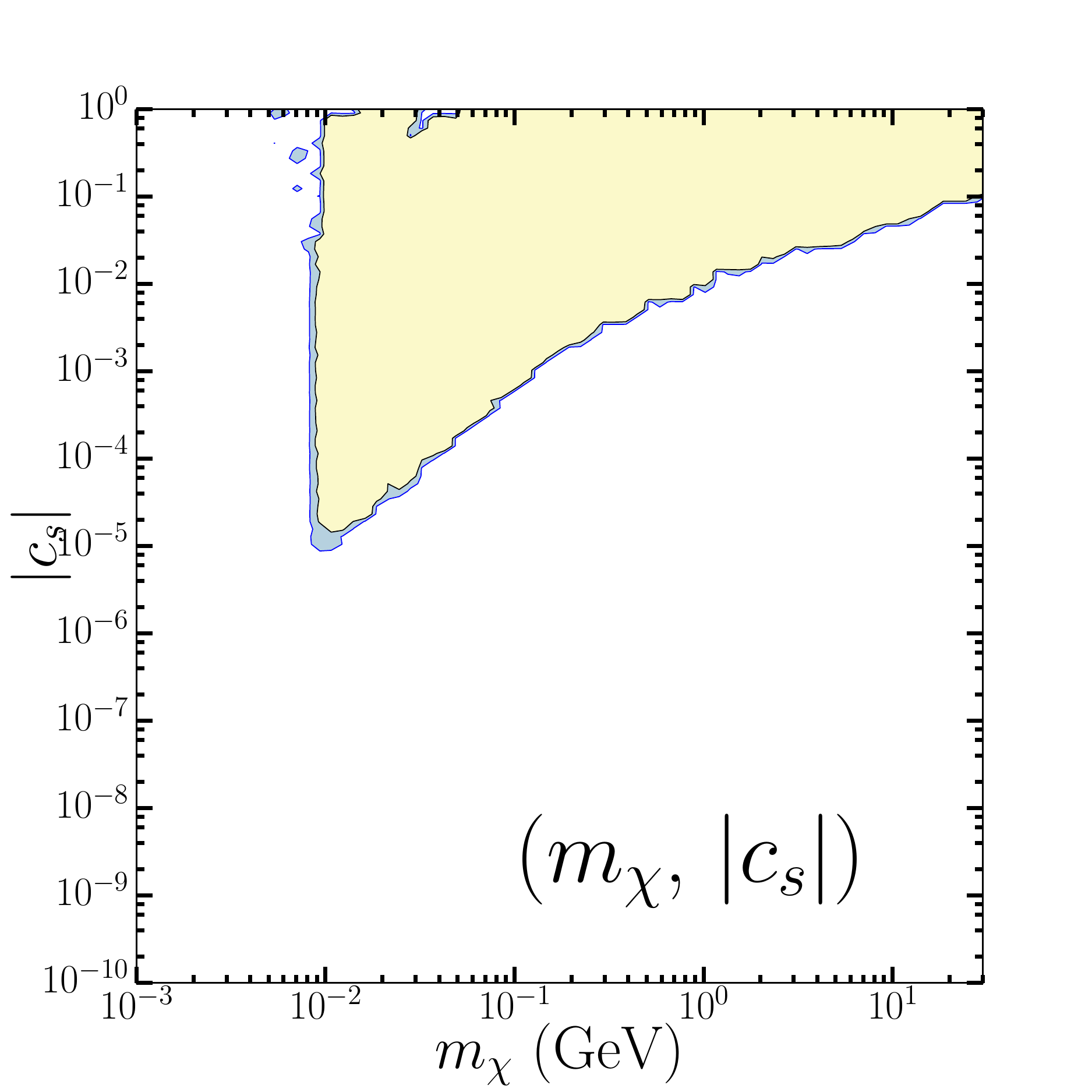}
	\includegraphics[height=1.52in,angle=0]{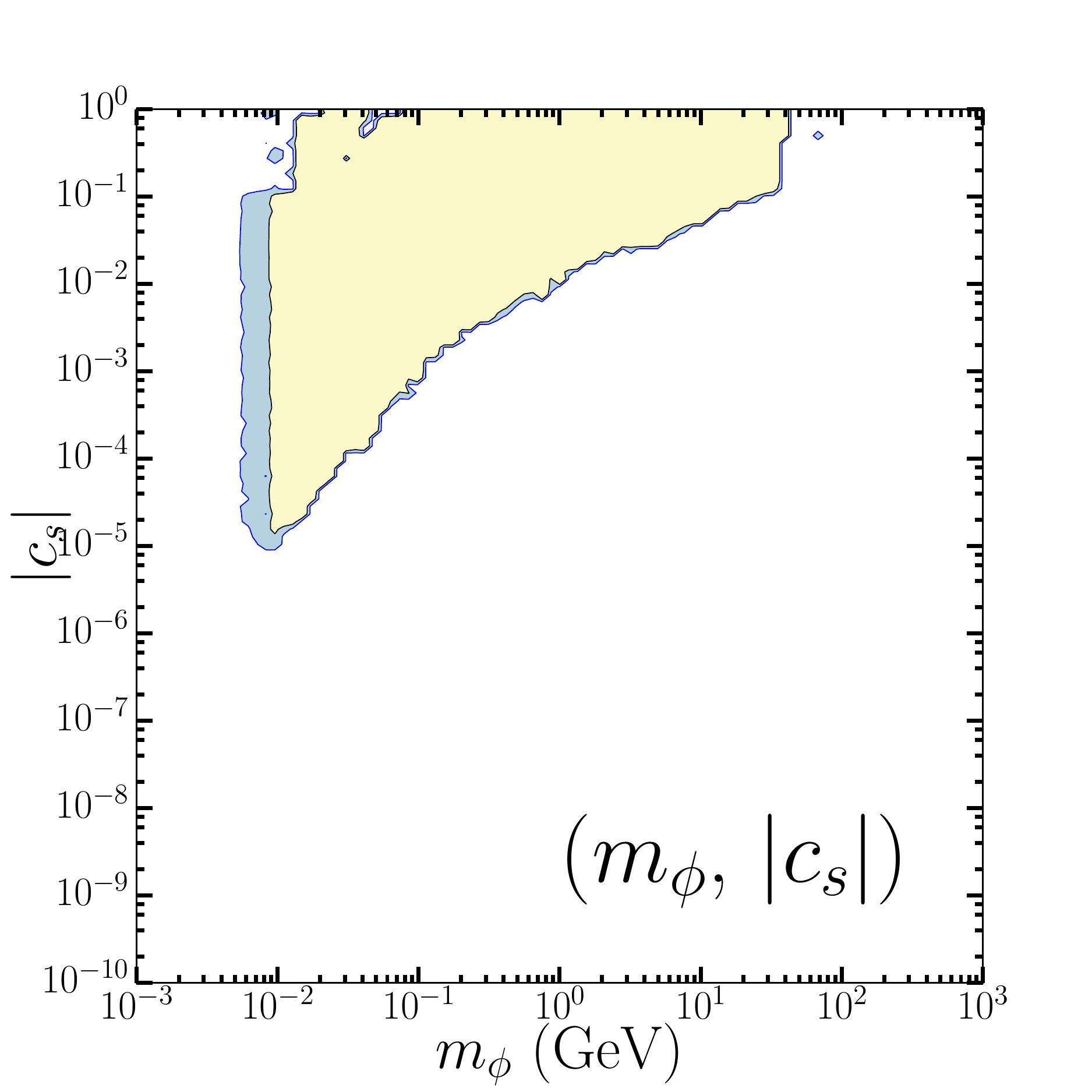}
	\includegraphics[height=1.52in,angle=0]{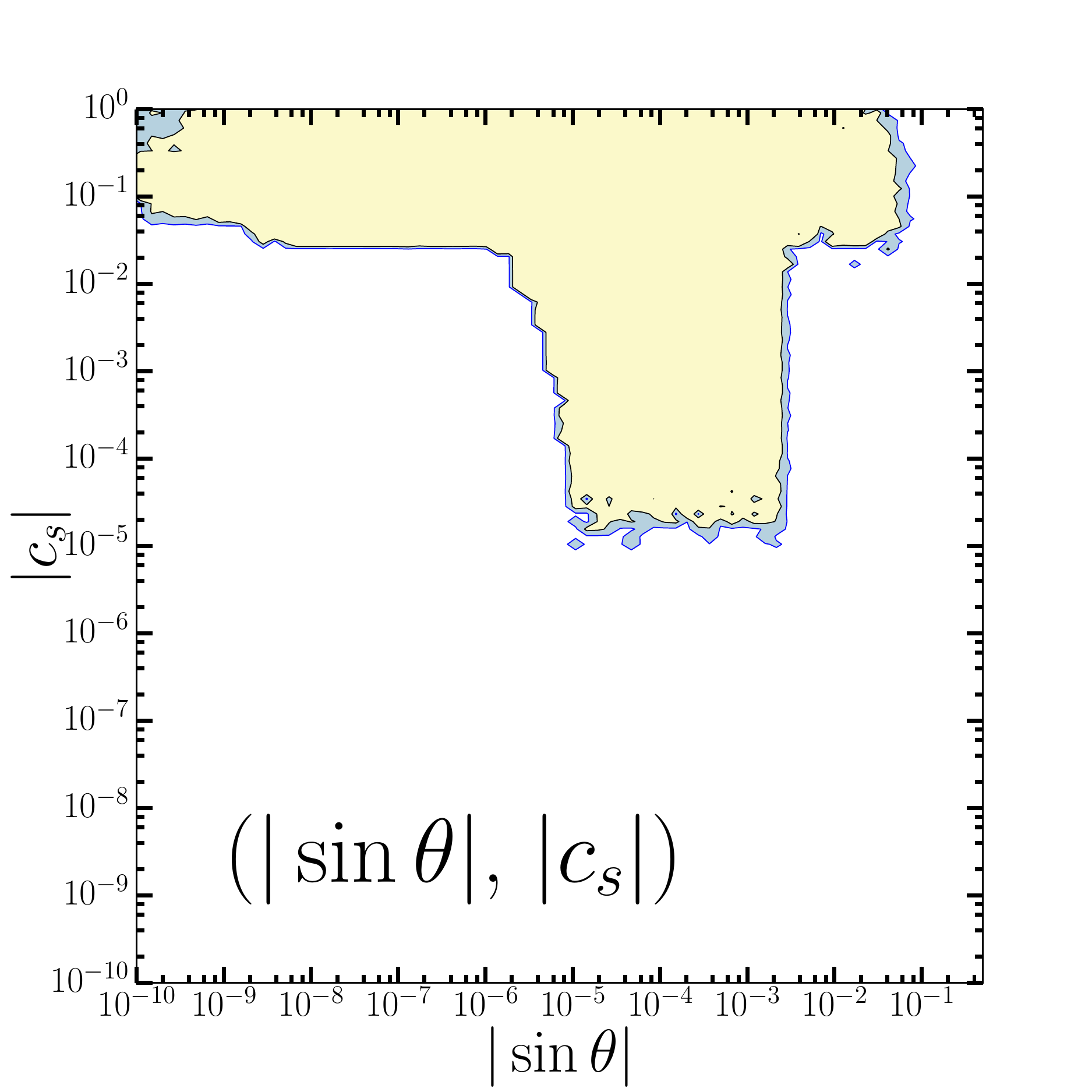}
	\includegraphics[height=1.52in,angle=0]{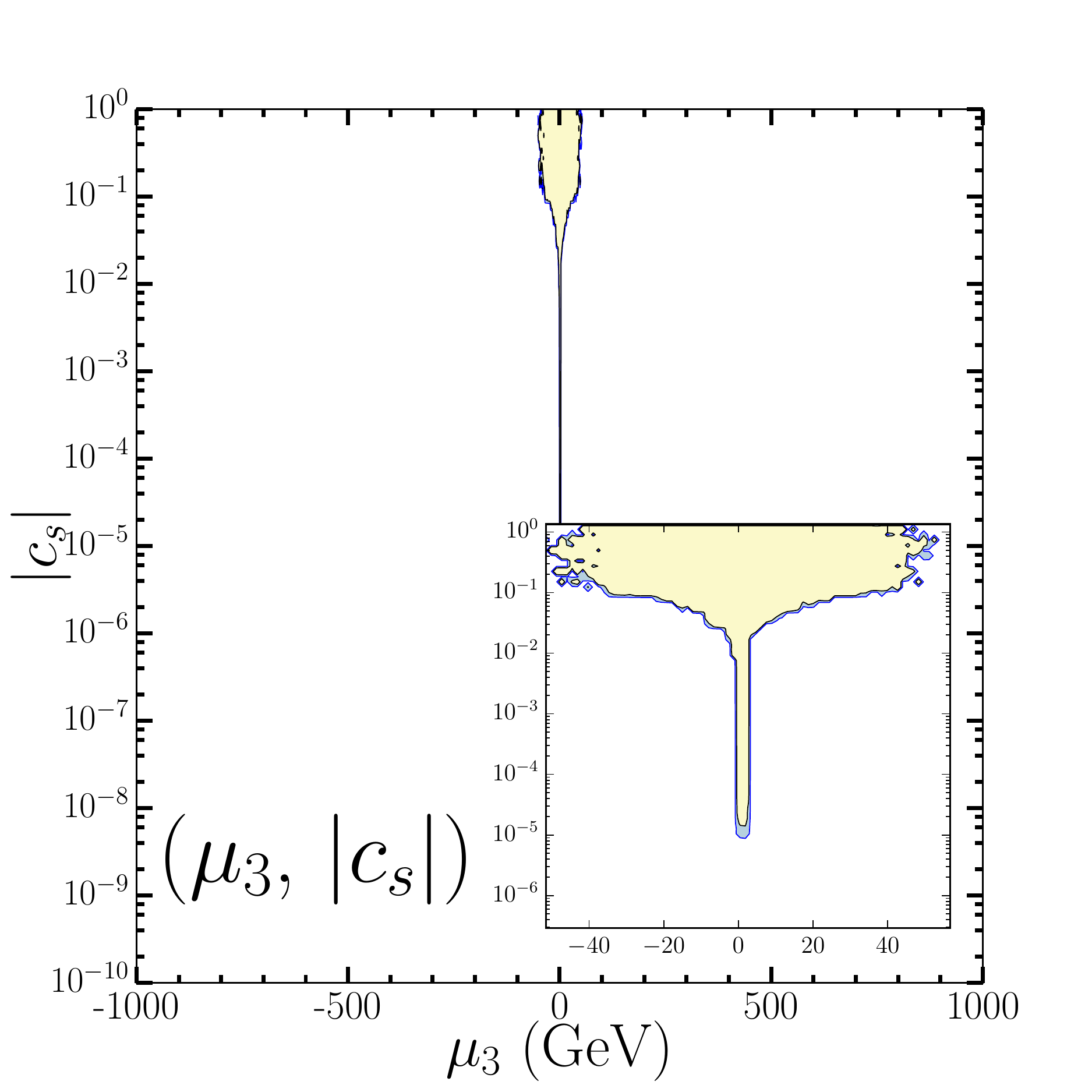}
	\caption{\small \sl Present status of the minimal WIMP model projected on all planes of input parameters.}
	\label{fig: present status all}
\end{figure}

Results of our analysis for the present status of the minimal WIMP model projected on all planes of input parameters ($m_{\chi}$, $c_s$, $m_\phi$, $\theta$, $\mu_3$) are shown in Fig.\,\ref{fig: present status all}. Since results on the $(m_\chi, m_\phi)$- and $(m_\phi, |\sin \theta|)$-planes are already discussed in section\,\ref{subsec: status}, we will focus mainly on those on other planes, spanned by different combinations of the input parameters.

First, the result on the $(m_\chi, |\sin \theta|)$-plane can be understood from the discussion in section\,\ref{subsec: status}, because the relic abundance condition requires $m_\chi \gtrsim m_\phi$ in most of parameter region. In fact, the lower limit on $m_\chi$, the lower limit on $|\sin \theta|$ in the range of $m_\chi \gtrsim 2$\,GeV and the upper limit on $|\sin \theta|$ are understood in this manner. Only the exception is about the lower limit on $|\sin \theta|$ in the range of $m_\chi \lesssim 2$\,GeV, where this region is excluded by the kinematical equilibrium condition, as already discussed in the previous appendix\,\ref{app: equilibrium condition}.

Next, as seen from the result on the $(m_\phi, \mu_3)$-plane, the tri-linear coupling $\mu_3$ is more constrained than the one in the corresponding panel of Fig.\,\ref{fig: After AP RE KE constraints} when $m_\phi \lesssim 1$\,GeV. We have confirmed that this is due to constraints from meson decay experiments as well as the vacuum stability condition which was imposed as one of preselection criteria: The mixing angle $|\sin \theta|$ is suppressed less than ${\cal O}(10^{-2})$ because of the collider constraints, so that $\mu_3$ is required to be small enough when $m_\phi \lesssim 1$\,GeV in order to stabilize our vacuum. In a similar manner, results on the $(m_\chi, \mu_3)$- and $(|\sin \theta|, \mu_3)$-planes can be understood as well.

Third, concerning the result on the $(m_\phi, |c_s|)$-panel, the lower limit on $m_\phi$ comes mainly from the $\Delta N_{\rm eff}$ constraint, though the upper-left corner with $|\sin\theta| \gtrsim 5 \times 10^{-2}$ is further constrained by the direct dark matter detection. The lower bound on the coupling $|c_s|$ in the bulk region is from the relic abundance condition, where it is satisfied by the $\chi \chi \to \phi \phi$ annihilation. The result on the $(m_\chi, |c_s|)$-plane is also understood in the same manner.

Fourth, the result on the $(\mu_3, |c_s|)$-plane can be understood by those on the $(m_\chi, |c_s|)$- and $(m_\chi, \mu_3)$-planes. When $|c_s|$ is smaller, a larger $m_\chi$ is not allowed due to the relic abundance condition. On the other hand, a smaller $m_\chi$ leads to a very restricted tri-linear coupling $\mu_3$.

Finally, the result on the $(|\sin \theta|, |c_s|)$-plane can be understood as follows: The region of $|\sin \theta| \lesssim 10^{-3}$ is not very much different from that in the corresponding panel of Fig.\,\ref{fig: After AP RE KE constraints}. On the other hand, the region of $10^{-3} \lesssim |\sin \theta| \lesssim 10^{-1}$ is excluded by meson decay experiments, because a smaller $|c_s|$ indicates a smaller $m_\phi$, as seen on the $(m_\phi, |c_s|)$-plane. The region of $|\sin \theta| \gtrsim 10^{-1}$ is excluded by collider experiments as well as the direct dark matter detection.

\subsection{Future prospects}
\label{app: future prospects}

\begin{figure}[t!]
	\includegraphics[height=1.52in,angle=0]{future_mx_mphi_like}
	\newline
	\includegraphics[height=1.52in,angle=0]{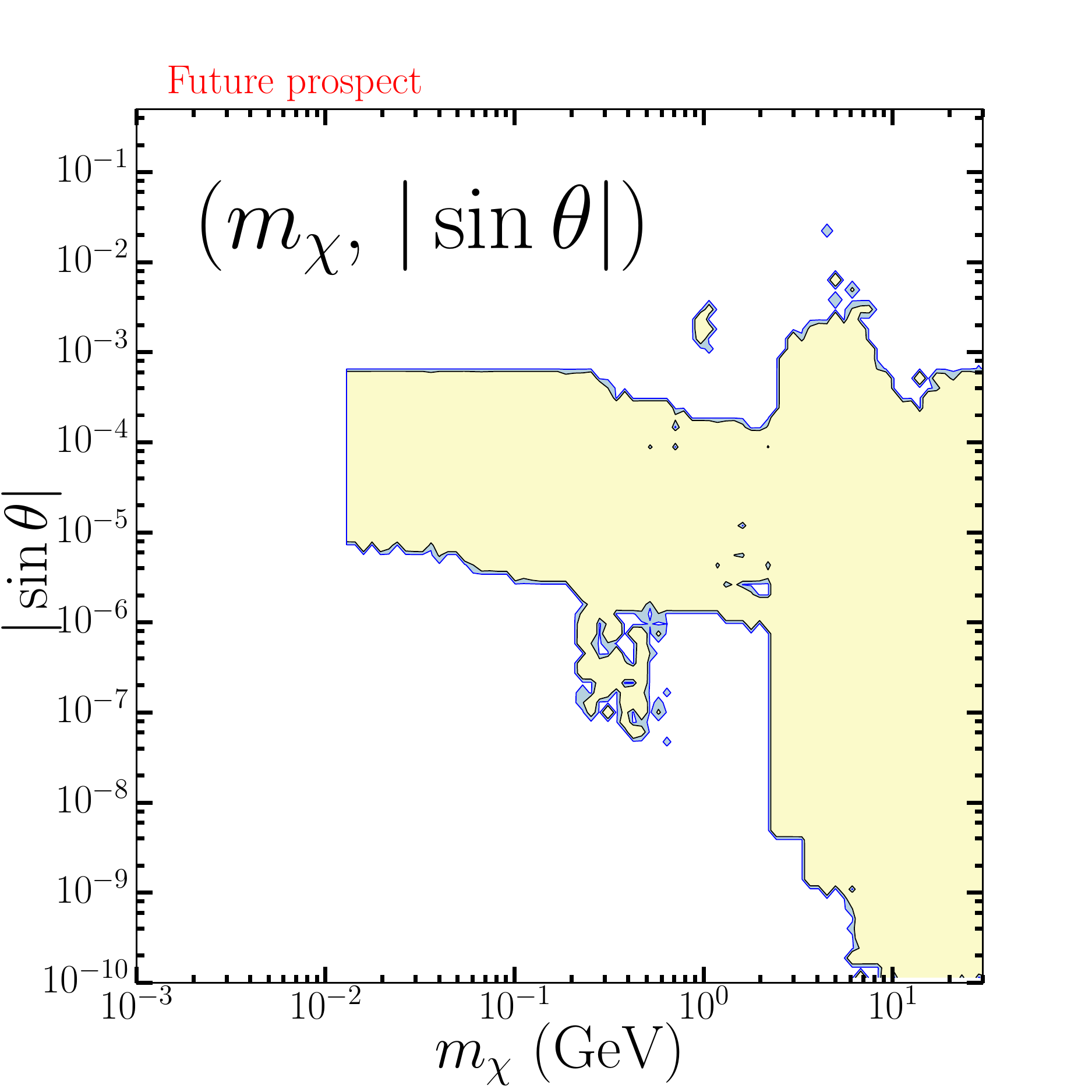}
	\includegraphics[height=1.52in,angle=0]{future_mphi_logsth_like}
	\newline
	\includegraphics[height=1.52in,angle=0]{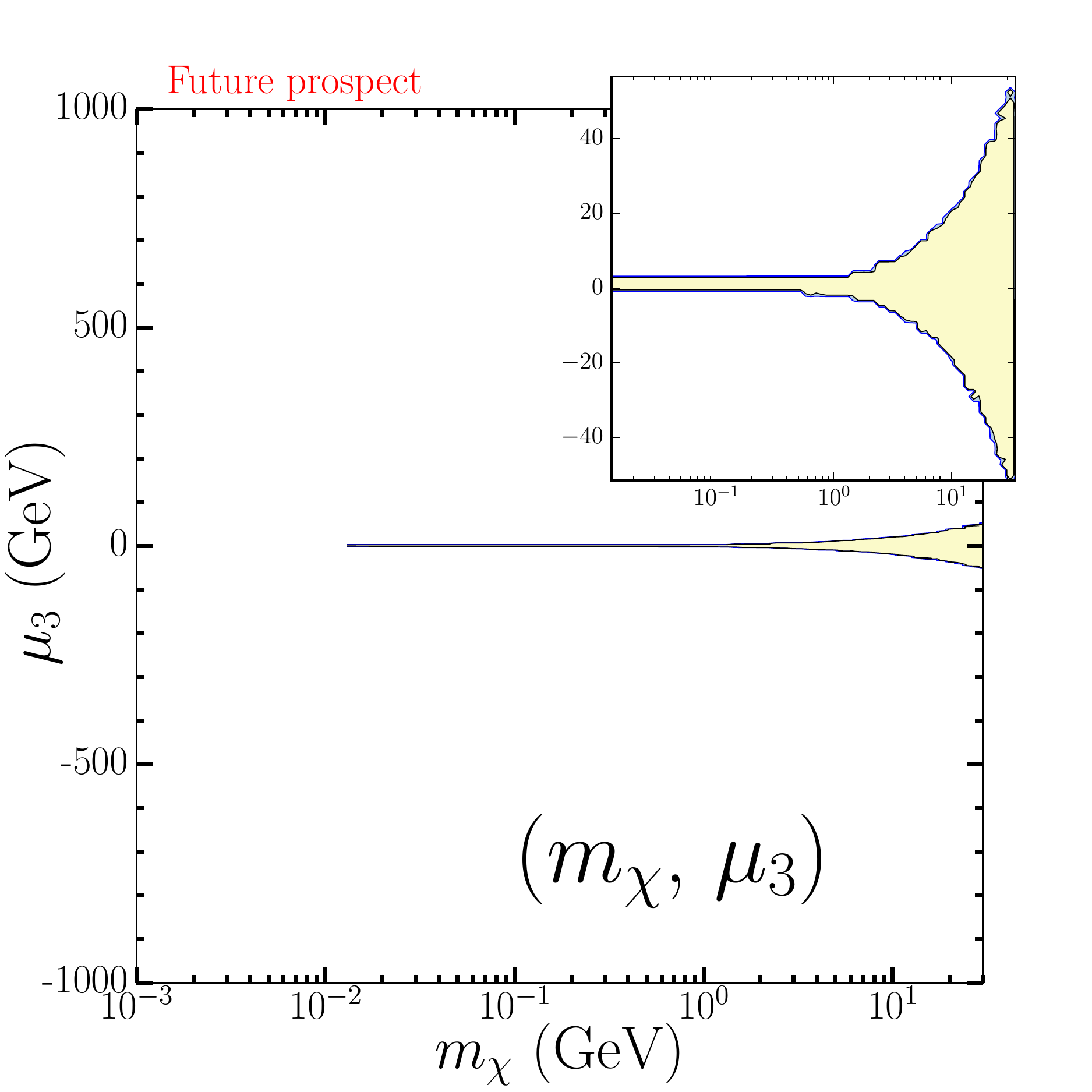}
	\includegraphics[height=1.52in,angle=0]{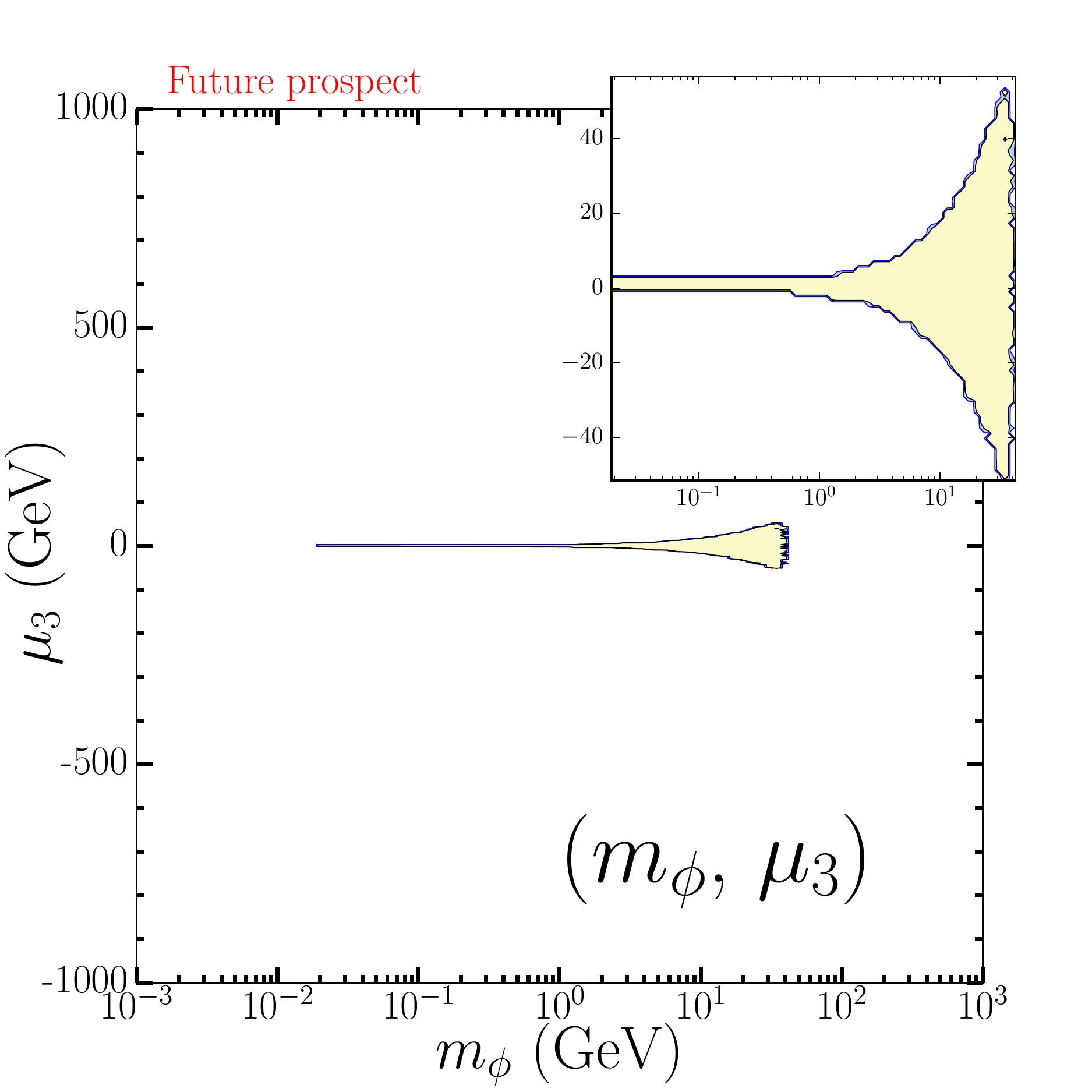}
	\includegraphics[height=1.52in,angle=0]{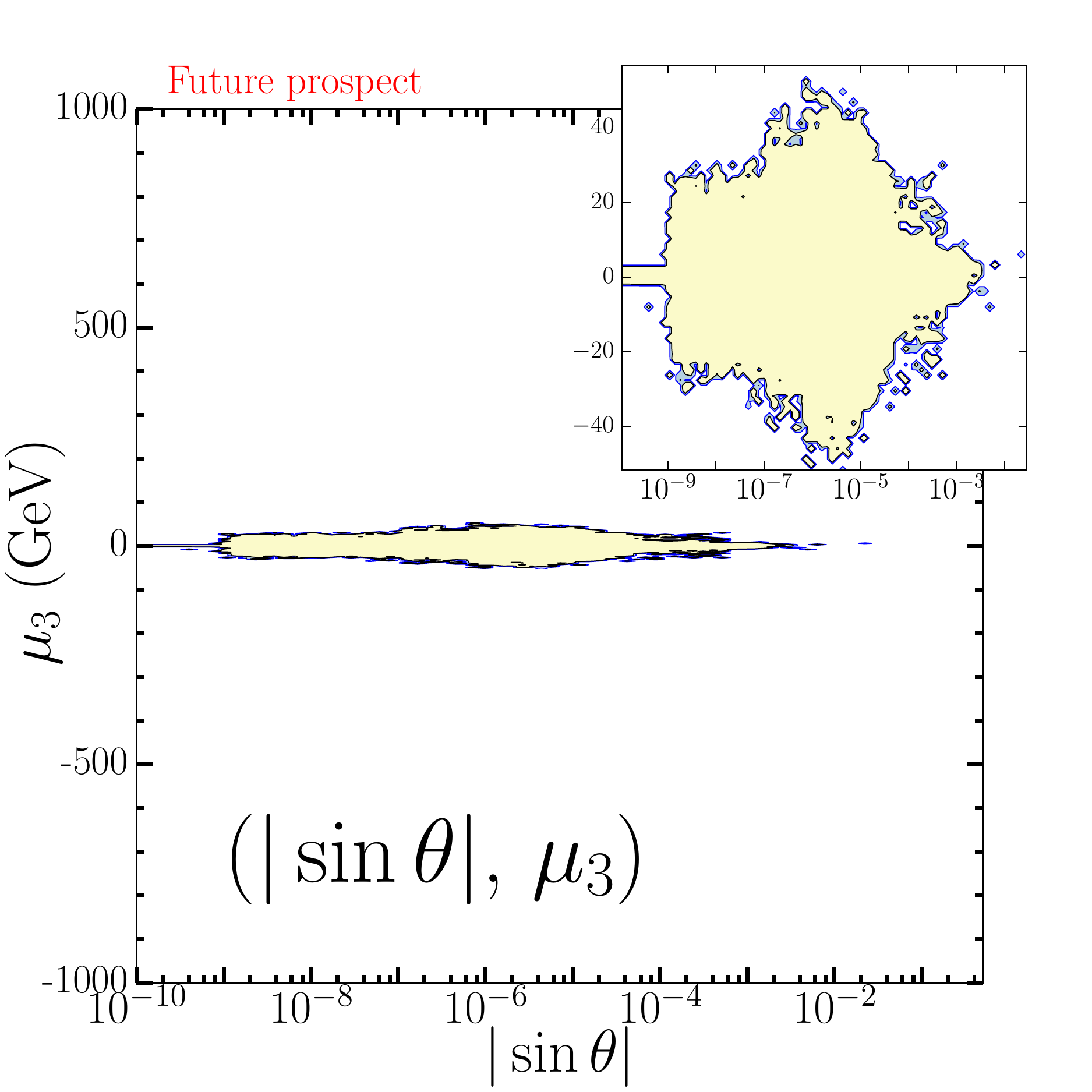}
	\newline
	\includegraphics[height=1.52in,angle=0]{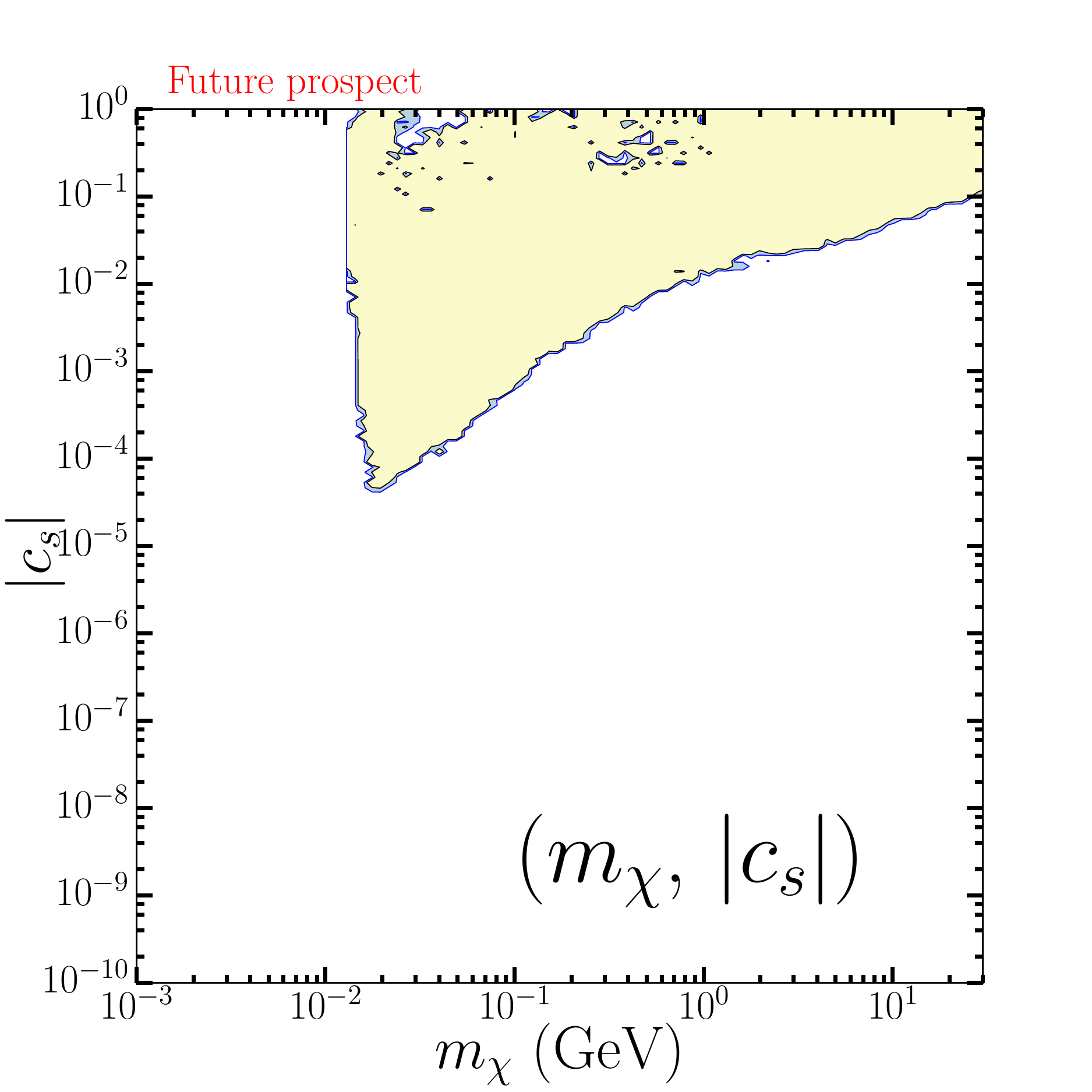}
	\includegraphics[height=1.52in,angle=0]{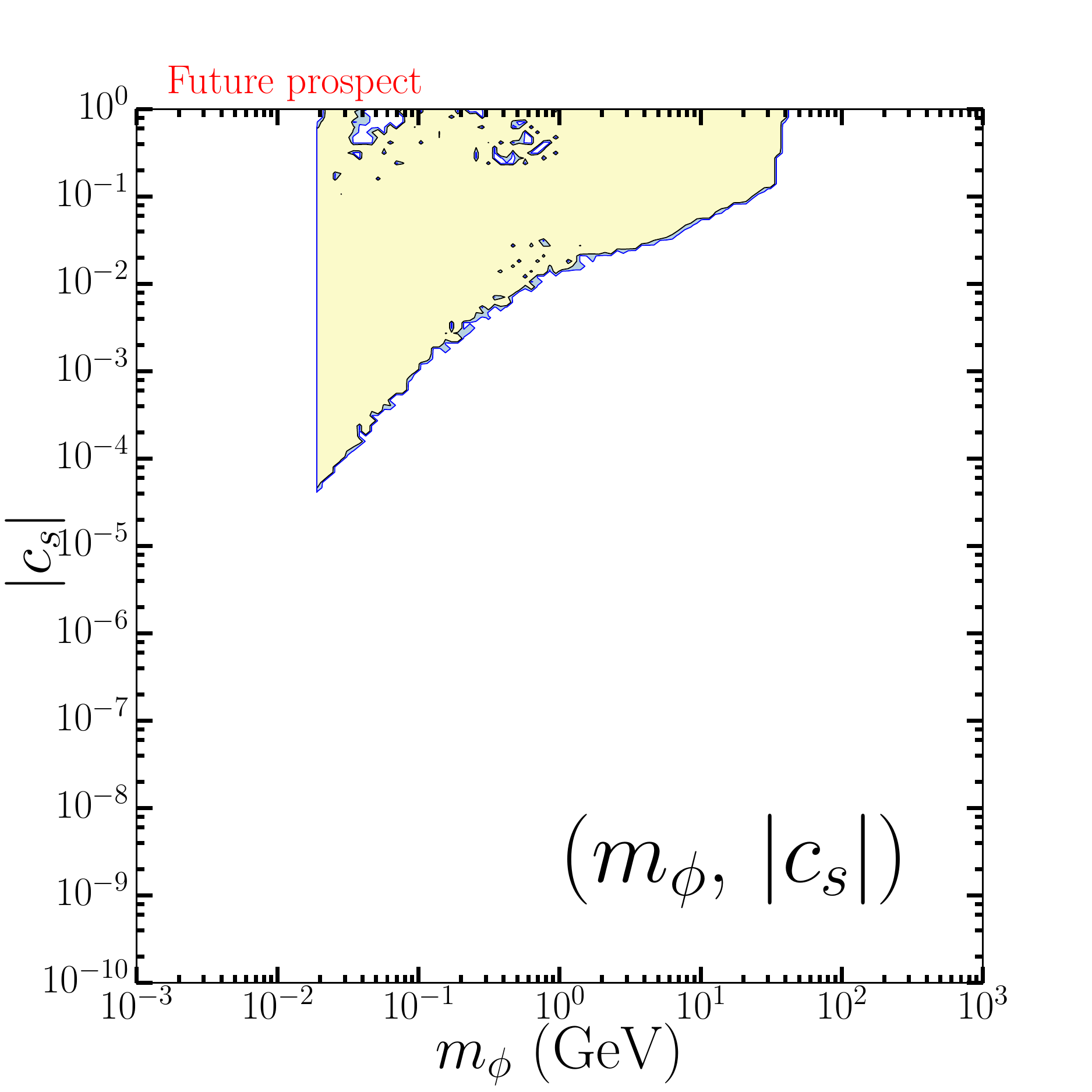}
	\includegraphics[height=1.52in,angle=0]{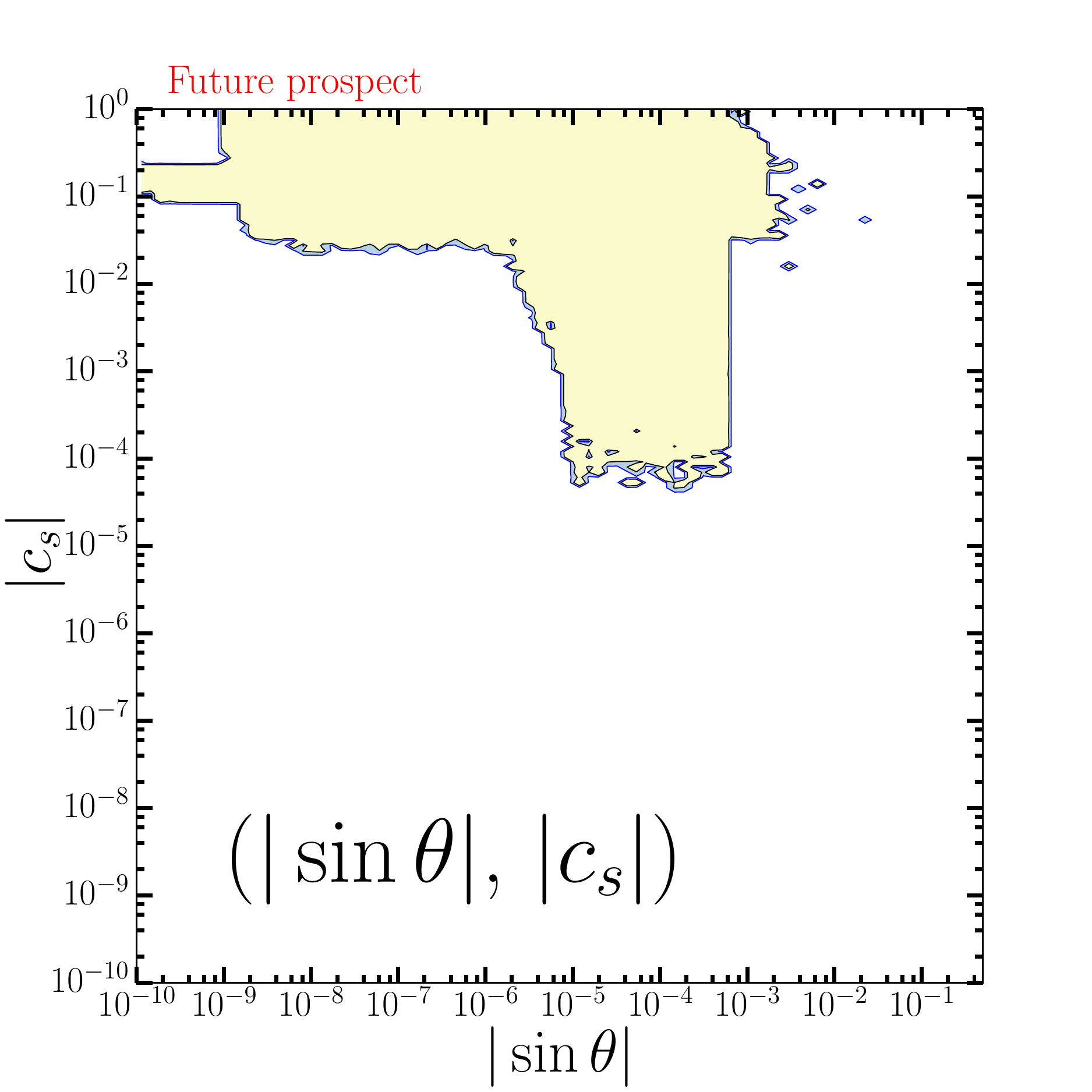}
	\includegraphics[height=1.52in,angle=0]{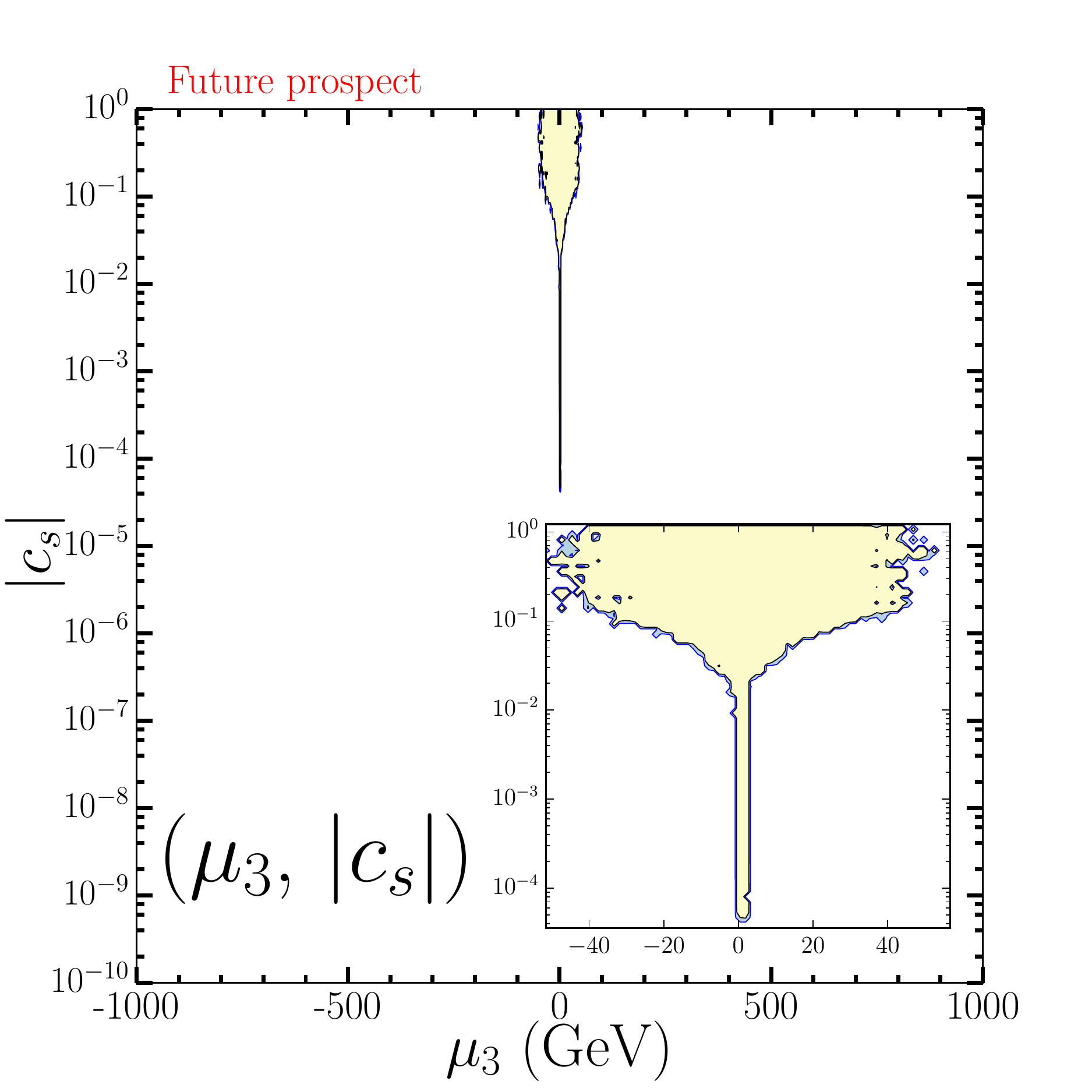}
	\caption{\small \sl Future prospects of the minimal WIMP model projected on all planes of input parameters.}
	\label{fig: future prospects all}
\end{figure}

Results of our analysis for the future prospects of the minimal WIMP model projected on all planes of input parameters are shown in Fig.\,\ref{fig: future prospects all}, assuming that no dark matter and mediator signals are detected even in the near future. Since results on the $(m_\chi, m_\phi)$- and $(m_\phi, |\sin \theta|)$-planes are already discussed in section\,\ref{subsec: prospects}, we will focus on those on other planes spanned by different combinations of the input parameters, as in the previous subsection\,\ref{app: present status}.

First, the result on the $(m_\chi, |\sin \theta|)$-plane can be understood from the discussion in section\,\ref{subsec: prospects}: the allowed region shrank compared to the corresponding panel in Fig.\,\ref{fig: present status all} because of the provisional future-update on $\Delta N_{\rm eff}$, direct detection and collider constraints.

Next, the results on $(m_\phi, \mu_3)$-, $(m_\chi, \mu_3)$- and $(|\sin \theta|, \mu_3)$-planes show that the constraint on the tri-linear coupling $\mu_3$ becomes severer than those of Fig.\,\ref{fig: present status all} even at $m_\phi \lesssim m_\chi \lesssim$ a few GeV, because future collider experiments will put a severer constraint on these mass region, if no signal is detected there. Moreover, $\mu_3$ is highly restricted in the range of $m_\phi \gtrsim$ a few ten GeV or $|\sin \theta| \gtrsim 10^{-3}$, 
but the resonant annihilation region is severely constrained by the provisional direct dark matter detection in the near future, as discussed in section\,\ref{subsec: prospects}.

Third, allowed regions on $(m_\chi, |c_s|)$- and  $(m_\phi, |c_s|)$-planes are, overall, shrunk compared to those in Fig.\,\ref{fig: present status all}. Those lower limits on $m_\phi$ and $m_\chi$ come from the $\Delta N_{\rm eff}$ measurement, while the lower limit on $|c_s|$ is from the 
relic abundance condition. The small void region at 0.3\,GeV $\lesssim m_\chi \lesssim$ 1\,GeV (0.3\,GeV $\lesssim m_\phi \lesssim$ 1\,GeV) and $|c_s| \gtrsim 0.1$ is due to the constraint from the SHiP experiment with the relic abundance condition being satisfied by the $\chi \chi \to \phi \phi$ annihilation at the threshold $m_\phi \sim m_\chi$. The other void region at 30\,MeV $\lesssim m_\chi \lesssim$ 80\,MeV (30\,MeV $\lesssim m_\phi \lesssim$ 80\,MeV) and $|c_s| \gtrsim 0.1$ is from the direct dark matter detection. 

Moreover, the allowed region on the $(\mu_3, |c_s|)$-plane remains similar as that of the present status in Fig.\,\ref{fig: present status all}, except the resonant annihilation region at $|c_s| \sim 10^{-2}$ is shrinking.

Last but not least, on the $(|\sin\theta|,|c_s|)$-plane, it is seen that the lower limit on the coupling constant $|c_s|$ is more or less the same as that in Fig.\,\ref{fig: present status all}. It is again from the relic abundance condition as seen in the $(m_\phi, |c_s|)$- and $(m_\chi, |c_s|)$-planes. On the other hand, future meson decay experiments and direct dark matter detection make the parameter region of $|\sin\theta| \gtrsim 10^{-3}$ excluded. On the other hand, the shape of the contour at the region of $|\sin \theta| \lesssim 10^{-9}$ is mainly due to the uncertainty of the mediator decay width as seen in the $(m_\phi, |\sin \theta|)$-plan. Since the dark matter mass is almost fixed to be around 10\,GeV as seen in the $(m_\chi, |\sin \theta|)$-plan, it requires a specific value of $c_s$ to satisfy the relic abundance condition.

\section{Relaxing Kinematic Equilibrium Condition}
\label{app: relaxing KEC}


\begin{figure}[t!]
	\centering
	\includegraphics[height=3.0in, angle=0]{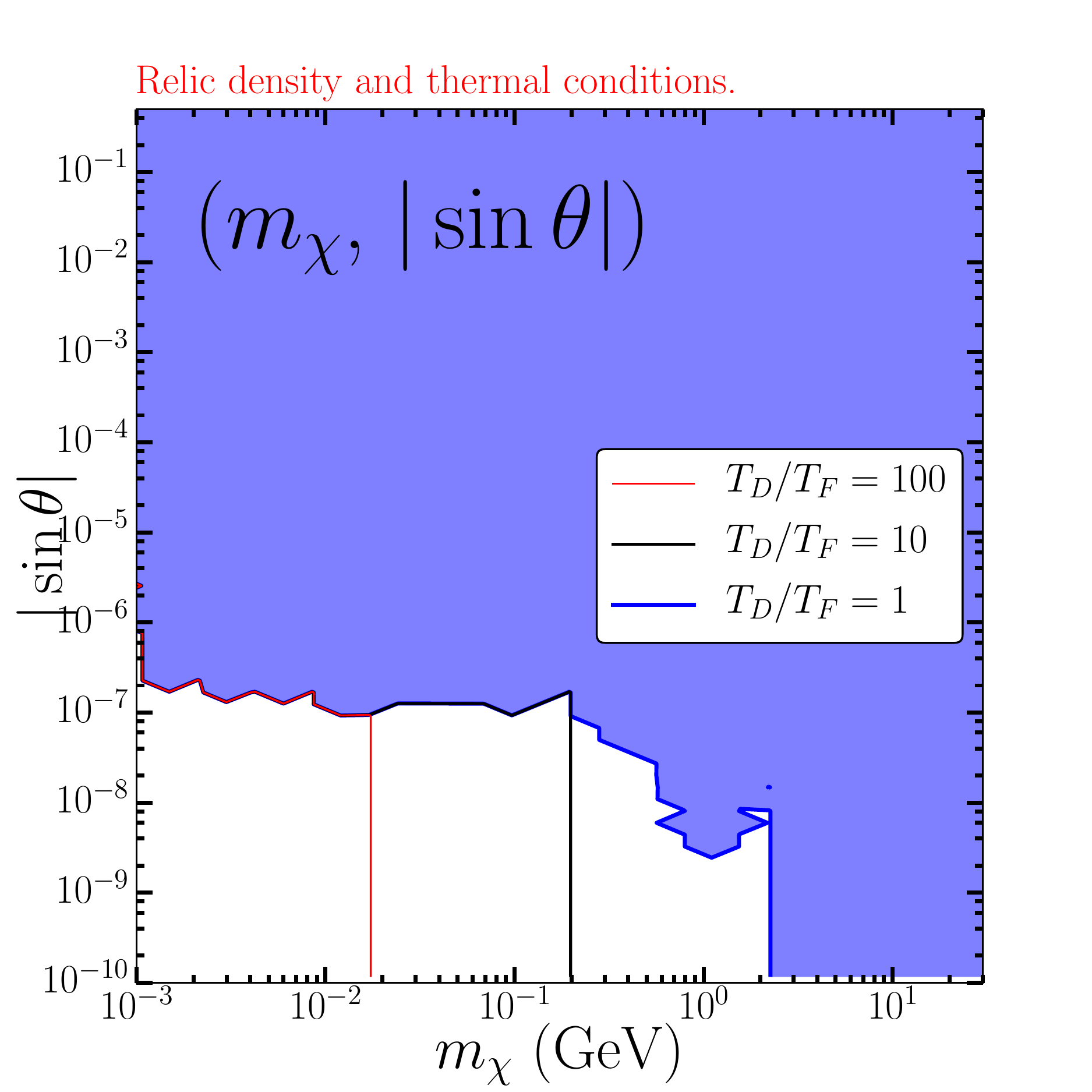}
	\includegraphics[height=3.0in, angle=0]{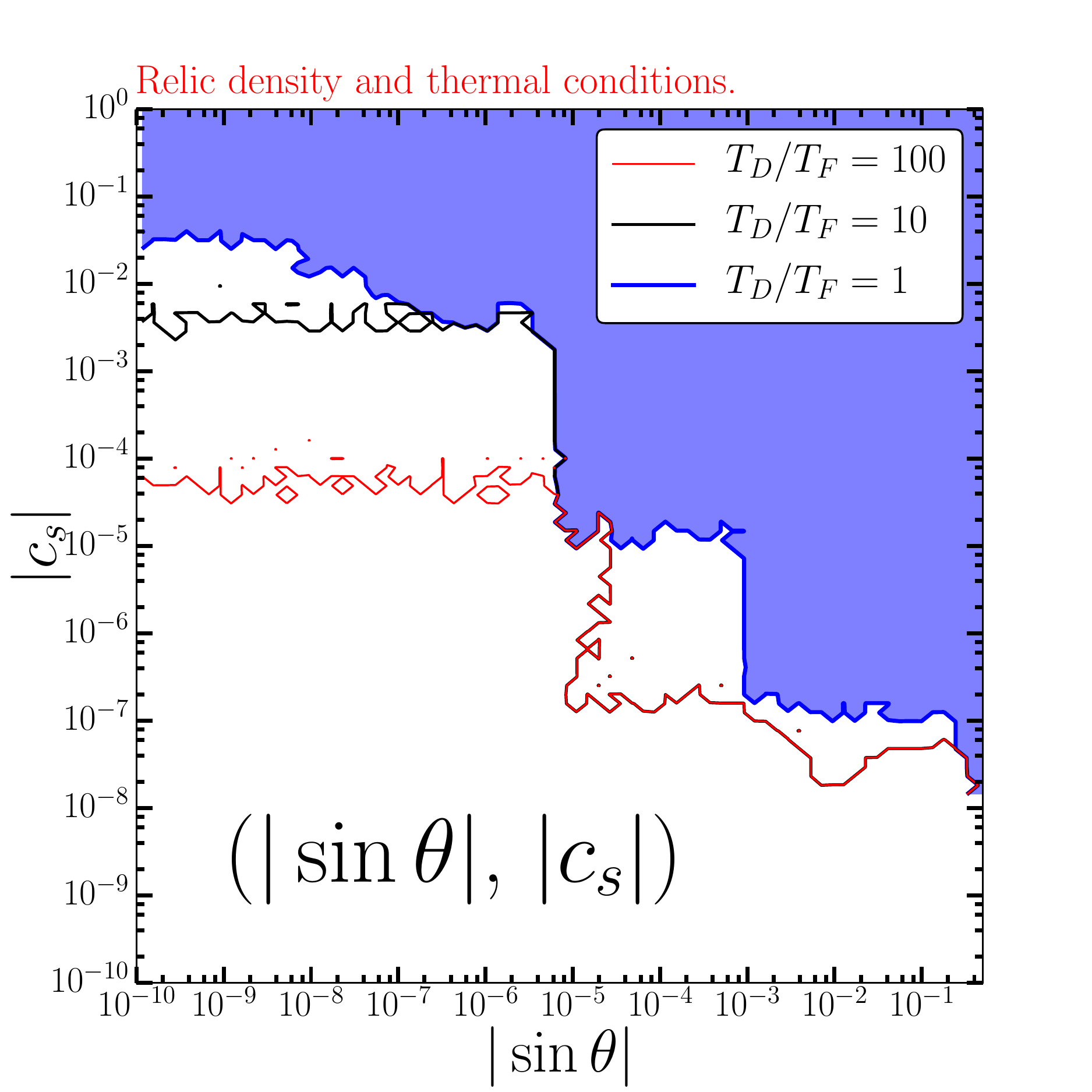}
	\caption{\small \sl 
	Model parameter regions survived after applying the relic abundance condition and the relaxed kinematical equilibrium condition. The black (red) line shows how the parameter regions extends when the kinematical equilibrium condition is relaxed to be $T_D/T_F = 10$ ($T_D/T_F = 100$) instead of $T_D/T_F = 10$. The blue shaded regions in both panels are the same as those in Fig.\,\ref{fig: After AP RE KE constraints}.
	}
	\label{fig: relaxing kec}
\end{figure}

In the main text, we imposed the kinematic equilibrium condition at around the freeze-out temperature $T_D \sim T_f$ to figure out the conventional WIMP parameter 
region. However, it is also acceptable that the dark sector and the SM sector 
kinematically decoupled at higher temperature above the freeze-out, and then two 
sectors evolve independently\,\cite{Evans:2017kti}.

Note that we adopt the condition $T_D=T_f$ to figure out a very conventional WIMP parameter region in our setup. On the other hand, the condition can be relaxed by requiring that the WIMP is in the equilibrium at some temperature of the universe before the freeze-out, because it still allows us to make a quantitative prediction on its abundance. 
In this section, we will discuss how the result of our analysis alters by relaxing the condition.

In order to understand how the kinematic equilibrium condition affects the parameter region,
we relax the condition by requiring 
the decouple temperature $T_D$ above the freeze-out temperature such as
$T_D=10T_f$ or $T_D=100T_f$.
Then, we show how the thermal DM parameter region is expanded
in the $(m_\chi,|\sin\theta|)$- and $(|\sin\theta|,|c_s|)$-planes
in Fig.\,\ref{fig: relaxing kec}.

The expansion of the parameter region can be understood as that
the higher decoupling temperature $T_D$ increases the light degree of freedom from heavier SM particles,
which have larger Yukawa couplings to help maintaining the kinematic equilibrium.
From another point of view, 
because the freeze-out condition fixes the relation $T_f\simeq m_\chi/20$,
relaxing the decoupling temperature $T_D$ to higher temperature than $T_f$ 
allows a lighter DM mass region still maintaining the same degree of freedom 
and keeping the thermal equilibrium.

In the $(m_\chi,|\sin\theta|)$-plane in Fig.\,\ref{fig: relaxing kec}, 
for the $T_D=T_f$ case, the vertical edge of $m_\chi\simeq 2$ GeV is from the pion threshold,
which shifts to $m_\chi\simeq 0.2$ GeV and $\simeq 0.02$ GeV for the $T_D=10T_f$ and $T_D=100T_f$ cases, respectively. We can clearly see how the parameter region is extended to the one with a smaller value of $m_\chi$.
The same behavior can be seen in the $(|\sin\theta|,|c_s|)$-plane in Fig.\,\ref{fig: relaxing kec},
where the parameter region extends to a smaller value of $|c_s|$ due to the correlation between $m_\chi$ and $|c_s|$ seen in the $(m_\chi,|c_s|)$-plane in Fig.\,\ref{fig: After AP RE KE constraints}.

It is worth pointing out that, once the kinematic equilibrium condition is relaxed, the temperature of the dark sector (both dark matter and mediator) during the freeze-out could be different from the temperature of the SM thermal bath. Since the result in Fig.\,\ref{fig: relaxing kec} is obtained assuming both the temperatures are (almost) equal, above discussions are validated only in such a case.\footnote{Such a case is realized when the mediator is relativistic in between the kinematical decoupling and the freeze-out, because the expansion histories of the two sectors are similar though they are independent.} On the other hand, when the temperatures are very different, the relic abundance condition as well as BBN and $N_{\rm{eff}}$ constraints have to be altered and the survived parameter space will be changed accordingly. A comprehensive study of early decoupled scenarios are indeed interesting but beyond the scope of our current study.


\end{document}